\documentclass[superscriptaddress,preprintnumbers,nofootinbib,10pt]{revtex4}
\usepackage[dvipdf,dvips,dvipdfmx]{graphicx}
\usepackage{amsmath}
\usepackage{amssymb}
\usepackage{comment}
\usepackage{float}
\usepackage{wrapfig}
\usepackage{bm}
\usepackage{tabularx}
\usepackage{color}
\catcode`\@=11
\def\lsim{\mathrel{\mathpalette\@versim<}}
\def\gsim{\mathrel{\mathpalette\@versim>}}
\def\@versim#1#2{\vcenter{\offinterlineskip
\ialign{$\m@th#1\hfil##\hfil$\crcr#2\crcr\sim\crcr } }}
\catcode`\@=12
 
\newcommand{\vect}[1]{\bm{#1}}

\newcommand{\bvec}[1]{\mbox{\boldmath $#1$}}

\newcommand{\p}{\partial}

\newcommand{\al}[1]{\begin{align}#1\end{align}}
\newcommand{\bp}{\begin{pmatrix}}
\newcommand{\ep}{\end{pmatrix}}
\newcommand{\nn}{\nonumber\\}

\newcommand{\bs}[1]{\boldsymbol}

\newcommand{\Tr}{{\rm Tr}\,}
\newcommand{\pmat}[1]{\begin{pmatrix}#1\end{pmatrix}}

\newcommand{\fn}[1]{\!\left(#1\right)}

\usepackage{etoolbox}
\let\bbordermatrix\bordermatrix
\patchcmd{\bbordermatrix}{8.75}{4.75}{}{}
\patchcmd{\bbordermatrix}{\left(}{\left[}{}{}
\patchcmd{\bbordermatrix}{\right)}{\right]}{}{}

\begin{document}

\title{Scale and confinement phase transitions in scale invariant $SU(N)$ scalar gauge theory}

\author{Jisuke \surname{Kubo}}
\email{jik@hep.s.kanazawa-u.ac.jp}
\affiliation{Max-Planck-Institut f\"ur Kernphysik, 69117 Heidelberg, Germany}
\affiliation{Department of Physics, University of Toyama, 3190 Gofuku, Toyama 930-8555, Japan}

\author{Masatoshi \surname{Yamada}}
\email{m.yamada@thphys.uni-heidelberg.de}
\affiliation{Institut f\"ur Theoretische Physik, Universit\"at Heidelberg, Philosophenweg 16, 69120 Heidelberg, Germany}

\preprint{UT-HET-127}

\begin{abstract}
We consider scalegenesis, spontaneous scale symmetry breaking, by the scalar-bilinear condensation
in $SU(N)$ scalar gauge theory.
In an effective field theory approach to the scalar-bilinear condensation at finite temperature, we include the Polyakov loop to take into account  the confinement effect.
The theory with $N=3,4,5$ and $6$ is investigated, and we find that  in all these cases the scale phase transition is a first-order phase transition.
We also calculate the latent heat  at and slightly below the critical temperature.
Comparing the results with those obtained without the Polyakov loop effect, we find that the Polyakov effect can considerably increase the latent heat in some cases,  which would mean a large  increase 
 in the energy density of  the gravitational waves background, if it  were produced by the scale phase transition.
\end{abstract}
\maketitle

\section{Introduction}
The understanding of the nature of the electroweak (EW) symmetry breaking is an important subject in elementary particle physics.
It is expected that the future experiments could elucidate the details of the Higgs sector.
There have been many attempts to 
simultaneously address
 other important issues in the standard mode (SM)
 such as neutrino mass and mixing, the Baryon number asymmetry and dark matter in the Universe and  also 
 the question of what the origin of the EW scale is.

As a guide to an extension of the SM, the fact that only the Higgs mass term is dimensionful in the SM might be a hint for new physics.
This question is strongly related to the so-called gauge hierarchy problem~\cite{Gildener:1976ai,Weinberg:1978ym} which states why the Higgs mass is so much smaller than the Planck scale.
For this issue, the scale invariance could play an important role.
The mass term is set exactly  equal to zero in 
the bare action of the SM if the scale invariance is imposed.\footnote{
Although the origin of the scale invariance in the SM is still unclear, asymptotically safe gravity could explain it~\cite{Oda:2015sma,Wetterich:2016uxm,Hamada:2017rvn,Eichhorn:2017als}.
See \cite{Niedermaier:2006wt,Niedermaier:2006ns,Percacci:2007sz,Reuter:2012id,Codello:2008vh,Eichhorn:2017egq,Percacci:2017fkn} as reviews of asymptotically safe gravity.
}
The important fact here is that the mass term keeps vanishing along its renormalization group flow~\cite{Wetterich:1983bi,Bardeen:1995kv,Aoki:2012xs}.
A mass scale corresponding to the EW symmetry breaking has to be generated 
by quantum effects, which we call  ``scalegenesis".
There are two possible ways of scalegenesis: One is known as the Coleman--Weinberg mechanism which is based on improved perturbation theory~\cite{Coleman:1973jx}.
The other is the spontaneous scale 
symmetry breaking due to non-perturbative dynamics such as Quantum Chromodynamics (QCD).

Several possibilities of scalegenesis due to strong dynamics have been suggested~\cite{Hur:2011sv,Heikinheimo:2013fta,Holthausen:2013ota,Kubo:2014ida,Heikinheimo:2014xza,Carone:2015jra,Ametani:2015jla,Kubo:2015cna,Haba:2015qbz,Hatanaka:2016rek,Ishida:2017ehu,Haba:2017wwn,Haba:2017quk,Tsumura:2017knk,Aoki:2017aws}.
One of them is based on the scalar gauge theory~\cite{Kubo:2014ida,Kubo:2015cna}, where complex scalar fields $S$ coupled to a hidden $SU(N)$ gauge fields are introduced. 
There 
we have considered a situation 
in which the condensate $\langle S^\dagger S\rangle \neq 0$
of  the complex scalars
is  formed by the strong  non-abelian gauge interaction and
 breaks  dynamically the scale invariance in the confining phase, 
 in a similar way as the chiral condensate in QCD does.
An obvious  interest is to see whether or not such a vacuum state is actually realized.
However, it is highly non-trivial, though 
not impossible \cite{Osterwalder:1977pc,Fradkin:1978dv},
 to investigate the vacua of the scalar gauge theory.
For phenomenological applications of the scalar-bilinear condensate
$\langle S^\dagger S\rangle \neq 0$, it is therefore highly
desired to describe it in an effective theory.
In the paper~\cite{Kubo:2015cna}, by mimicking the concept of the Nambu--Jona-Lasinio (NJL) model~\cite{Nambu:1961tp,Nambu:1961fr},
we have attempted to formulate an effective theory.
Using  the mean-field approximation, we  have found that 
the desirable vacuum structure is realized  in the effective theory
~\cite{Kubo:2015cna,Kubo:2017wbv}.
It has also turned out that  the theory involves a dark matter candidate if a flavor symmetry is imposed on the scalar fields.
Moreover, at finite temperature, the scale phase transition 
could be strongly first-order for a wide parameter space~\cite{Kubo:2015joa}.
Its signal can be observed as primordial gravitational waves~\cite{Kubo:2016kpb} in the future experiments such as DECIGO~\cite{Seto:2001qf,Kawamura:2006up,Kawamura:2011zz} and LISA~\cite{Seoane:2013qna}.

These previous works have focused on only  the scalar 
field dynamics:
The effective theory contains no gauge field,
and consequently  the confinement effects have been neglected.
Thus, it is important to investigate the impact 
of confinement on the scalar dynamics, especially, on phase transitions at finite temperature.
A key quantity representing the confinement is the Polyakov loop which is
an order parameter for spontaneous breaking of the center symmetry of $SU(N)$ in the pure Yang--Mills theory. 
Effective potentials of the Polyakov loop 
for the pure Yang--Mills theory have been suggested to investigate the confinement--deconfinement phase transition at finite temperature~\cite{Weiss:1980rj,Pisarski:2000eq,Dumitru:2000in,Dumitru:2001xa,Pisarski:2001pe,Sannino:2005sk,Marhauser:2008fz,Braun:2010cy}.
In the literature~\cite{Fukushima:2003fw}, the Polyakov--NJL model has been proposed in order to discuss the synergy between the chiral symmetry breaking and the confinement in QCD  (see also e.g. \cite{Ratti:2005jh,Roessner:2006xn,Fukushima:2008wg,Herbst:2010rf,Fukushima:2010bq,Fukushima:2013rx,Fukushima:2017csk}).

In this paper, following the Fukushima's work~\cite{Fukushima:2003fw},
we study the phase transition in the effective theory for the 
scalar-bilinear condensation with the Polyakov loop included.
In the next section 
we briefly explain the basic idea of the scalegenesis 
in  the scalar gauge theory, and in the beginning part of
section\,\ref{Polyakov scalar theory}, we review how 
the scalegenesis is described in the effective theory.
In most of the early works on the Polyakov potential  the $N=3$ case 
has been discussed from the obvious reason. 
Since there is no constraint on $N$ in phenomenological applications
of the scalar-bilinear condensate,
we  start with analyzing the pure Yang--Mills theory for $N=4$, $5$, $6$
in the following part of section\,\ref{Polyakov scalar theory}.
After that we include the scalar fields coupled to the Polyakov loop, which
 is the main part of this paper.
 We investigate the phase transition with the assumption
 that the deconfinement transition and the scale phase transition
 appear at the same temperature.
 We also calculate the latent heat that is released during the 
 first-order phase  transition in the effective theory
 both with the Polyakov loop included and suppressed. 
 Needless to say that the latent heat is an important quantity
that enters into the energy density of  the gravitational waves
background which is produced by a first-order phase transition in
the early Universe.
Section\,\ref{summary} is devoted to summarize this work.

\section{Brief overview on scalar-condensate model}\label{Brief overview on scalar-condensate model}
We briefly introduce the model suggested in \cite{Kubo:2015cna} and 
outline what has been investigated so far.
We consider a hidden sector which is governed
 by the following $SU(N)$ scalar-gauge theory,
 \al{
{\cal L}_{\rm H} &=-\frac{1}{4}F_{\mu\nu}^aF^{a\mu\nu}+
([D^\mu S_i]^\dag D_\mu S_i)-
\hat{\lambda}_{S}(S_i^\dag S_i) (S_j^\dag S_j)
-\hat{\lambda'}_{S}
(S_i^\dag S_j)(S_j^\dag S_i)
+\hat{\lambda}_{HS}(S_i^\dag S_i)H^\dag H,
\label{LH}
}
where $S_i^{a}~(a=1,\dots,N,~i=1,\dots,N_f)$ are scalar fields in the fundamental representation of $SU(N)$, $F^{a\mu\nu}$ is the field strength of the $SU(N)$ hidden gauge field $A_\mu^a$, $D_\mu S_i = \partial_\mu S_i -ig A_\mu^a S_i^a$ is the covariant derivative, and the SM Higgs doublet field is denoted by $H^T=(\chi_1+i\chi_2,h+i\chi_3)/\sqrt{2}$.
The total Lagrangian is the sum of ${\cal L}_{\rm H}$ and ${\cal L}_{\mathrm{SM}}$, where the scalar potential of the SM part is $V_{\mathrm{SM}}=\lambda_H ( H^\dag H)^2$.
Note that the Higgs mass term is forbidden because of  classical scale invariance.
We suppose in this model that below a certain energy scale the hidden gauge coupling $g$ becomes so large that the $SU(N)$ invariant 
scalar bilinear dynamically forms a $U(N_f)$ invariant condensate,
\al{
\langle S^\dag_i S_j\rangle &=
\left\langle \sum_{a=1}^{N} S^{a\dag}_i S^a_j\right\rangle\propto \delta_{ij}.
\label{condensate}
} 
This scalar-bilinear  condensate triggers the EW symmetry breaking via the Higgs portal coupling and the Higgs mass term is generated; $m_H=-\hat{\lambda}_{HS}\langle S_i^\dag S_i\rangle$.
In other words, the origin of the EW vacuum is generated by the spontaneous scale symmetry breaking.
Note here that the corresponding Nambu-Goldstone (NG) boson to the spontaneous scale symmetry breaking is dilaton. This NG boson is, however, massive since the scale symmetry is broken by the scale anomaly.\footnote{
In the absence of the quark fields in QCD, the scale symmetry is broken by the gluon condensate, and the glueball is the dilaton. If the quark fields are present, the chiral condensate forms and breaks the chiral symmetry spontaneously.
The chiral condensate also breaks the scale symmetry and is another origin of the spontaneous breaking of the scale symmetry.
Therefore, in a general situation, the dilaton will be a mixing of the glueball and the chiral partner of the pion.
If  the running of the gauge coupling constant is sufficiently slow, the explicit, hard breaking effect by  the trace anomaly can be weak compared with that of the spontaneous breaking by the chiral condensate~\cite{Bardeen:1985sm}.
Here we have  a similar  situation in mind.}

Although the idea of this model is simple, the actual analysis is highly complicated due to the non-perturbative dynamics.
In the paper \cite{Kubo:2015cna}, we have attempted to formulate an effective theory of \eqref{LH} with the Higgs quartic interaction included:
The classical scale invariance with the $U(N_f)$ flavor symmetry
uniquely singles out it to be
\al{
 {\cal L}_{\rm eff} &=  ([\partial^\mu S_i]^\dag \partial_\mu S_i)-\lambda_{S}(S_i^\dag S_i) (S_j^\dag S_j)
-\lambda'_{S}(S_i^\dag S_j)(S_j^\dag S_i)
+\lambda_{HS}(S_i^\dag S_i)H^\dag H-\lambda_H ( H^\dag H)^2,
\label{Leff}
}
and we have investigated the  vacuum structure 
in this effective theory by using the mean-field approximation.
Then the following facts have been emerged: 
The effective theory can describe  the scalegenesis in the hidden sector,
which  produces the Higgs mass term  in the expected way.
For  finite $N_f$ the model has a weakly interacting massive particle (WIMP), a dark matter candidate,
 as an excited state above the vacuum \eqref{condensate}, and it could be tested by the future experiments of the dark matter direct detection~\cite{Kubo:2015cna,Kubo:2017wbv}.
In the paper~\cite{Kubo:2015joa}, the restorations of the scale and EW symmetries at finite temperature have been investigated.
The scale phase transition becomes strongly first-order for a wide parameter space in the model.
This scale phase transition can induce a strong first-order EW phase transition
 for a certain parameter choice, 
although the EW phase transition is weak within the SM.
It has been moreover found in ~\cite{Kubo:2016kpb} that
the released energy at the strong first-order 
scale phase transition can produce 
primordial gravitational waves which could be observed by the future space gravitational wave antennas.

In these analyses mentioned above, however, 
the confinement effects has not been taken into account.
The purpose of the present work is to introduce 
the Polyakov loop into the effective theory and 
to investigate the impact of the confinement effects
on the phase transition.

\section{Spontaneous Scale Symmetry Breaking and Polyakov-corrected Scale Phase Transition}\label{Polyakov scalar theory}

\subsection{At zero temperature}

For a small $\lambda_{HS}  \lsim 0.1$, 
the scale phase transition occurs at a critical 
temperature which is much higher than that 
of the EW phase transition \cite{Kubo:2015joa}, which means
$\langle H \rangle=0$ at the scale phase transition 
for small values of $\lambda_{HS}$.
We therefore consider the theory with the SM sector decoupled, i.e. $\lambda_{HS}=0$. The effective Lagrangian for this case is ${\cal L}_{\rm eff}$ given in \eqref{Leff} with $\lambda_{HS}=\lambda_H=0$.
To obtain the effective potential in the mean-field approximation, we introduce the the auxiliary fields, $f$ and $\phi_0^a~(a=1,\dots, N_f^2-1)$,
and rewrite the Lagrangian  in such a way that the rewritten Lagrangian (the mean-field Lagrangian ${\cal L}_{\rm MFA}$) yields the equations of motion
\al{
f &= \frac{1}{N_f}(S^\dag_i S_i)&
&\mbox{and}& \phi_0^a&=2(S_i^\dag
t^a_{ij} S_j),&
}
where $t^a~(a = 1,...,N_f^2 -1)$ are the $SU(N_f)$ generators in the fundamental representation.
The desired mean-field Lagrangian is given by \cite{Kubo:2015cna}
\al{
{\mathcal L}_\text{MFA}&=
 ([\partial^\mu S_i]^\dagger \partial_\mu S_i) 
 -2(N_f\lambda_S+\lambda'_S)f(S_i^\dagger S_i) 
 + N_f (N_f\lambda_S+\lambda'_S) f^2
 +\frac{\lambda_S'}{2}(\phi^a_0)^2
-2\lambda_S' \phi^a_0(S_i^\dagger t^a_{ij} S_j).
\label{MFA2ap}
}
We integrate out 
 the fluctuations
$\delta S_i$ of $S_i=\bar{S}_i+\delta S_i$ around the 
background $\bar{S}_i$ in the $\overline{\mbox{MS}}$  
subtraction scheme to obtain  the effective potential 
\al{
V_\text{MFA}\fn{\bar S,f}
&= \tilde{M}^2 (\bar S_i^\dagger \bar S_i)  -N_f (N_f \lambda_S +\lambda'_S)f^2 +\frac{N N_f}{32\pi^2}\tilde{M}^4 \ln\frac{\tilde{M}^2}{\Lambda_H^2},
\label{effective potential MFAa}
}
where $\Lambda_H = \mu e^{3/4}$ with $\mu$ being the 't Hooft
renormalization scale, and
\al{
\tilde{M}^2 &=
2(N_f\lambda_S+\lambda'_S)f.
\label{con mass}
}
The absolute minimum of the potential $V_\text{MFA}\fn{\bar S,f}$
is found to be located at
\al{
\langle \bar S\rangle &=0,&
\langle f\rangle&=
\frac{\Lambda_H^2/2}{N_f\lambda_S+\lambda_S'}\exp \left(  \frac{8\pi^2}{N(N_f\lambda_S+\lambda_S')}-\frac{1}{2} \right)&
}
if  $N_f\lambda_S+\lambda_S' >0$ is satisfied, and
the minimum value of $V_\text{MFA}$ is given by
\al{
\langle V_\text{MFA} \rangle
&= -\frac{1}{16 \pi^2}N N_f (N_f\lambda_S+\lambda_S')^2
\langle f\rangle^2 <0.
}
Therefore, as long as $N_f\lambda_S+\lambda_S' >0$ is satisfied,
the scale symmetry is spontaneously broken by the 
scalar-bilinear condensate $\langle f\rangle=\langle (S_i^\dag S_i) \rangle$ in the effective theory.

\subsection{At finite temperature}
At high temperatures we expect that the scale symmetry
is restored (up to anomaly) and the color degrees of freedom
are no longer confined.
At finite temperature $T$ the theory is 
equivalent to the Euclidean theory, which is periodic in the Euclidean time 
$x_4=i x^0$ with the period of $\beta=1/T$.
The local gauge transformation has to respect this periodicity
at finite temperature.
In spite  of this the pure gluonic action
is invariant  under a non-periodic (singular) gauge transformation
(center symmetry \cite{Polyakov:1978vu,Susskind:1979up,Svetitsky:1982gs,Svetitsky:1985ye}) defined by the transformation matrix
\al{
U_k(x_4) &=\left({\bf 1}e^{2\pi i k/N}\right)^{x_4/\beta},&
k&=1,\dots,N-1,&
\label{Uk}
}
where  
${\bf 1}\exp(2\pi i k/N)$  
belongs to   the center of 
$SU(N)$, and  ${\bf 1}$ is the $N \times N$ unit matrix.\footnote{Since $(\exp\fn{ i \pi a})^b \neq \exp\fn{ i a b} $ for $ |a| \leq 1$,
$\det U_k =\left( (\exp 2 \pi k /N)^{x_4/\beta}\right)^{N}$ is not manifestly equal to one.}
Under this non-periodic gauge transformation, the traced Polyakov loop in the fundamental representation 
\al{
\ell &= \frac{1}{N}\mbox{Tr}\, L
\label{ell}
}
transforms as $\ell\to \ell'=e^{2\pi i k/N}\ell$,
where the Polyakov loop is defined as
\al{
L &={\mathcal P} \exp \left(i g\int_0^\beta A_4(x) d x_4\right)~.
\label{polyakov}
}
Here $\mathcal P$ is the path-ordering, $A_4 = A^a_4T^a$ is  the temporal component of the gauge field with $T^a$ being the generators 
of $SU(N)$, and $g$ is the gauge coupling constant. 
Since $\ell$  is a gauge invariant observable, it can be an exact  order parameter for the spontaneous breaking of the center symmetry
in the pure gluonic theory.

Furthermore, the vacuum expectation value (VEV) of the traced Polyakov loop $\langle \ell\rangle $ can be expressed as
$\langle \ell \rangle =\exp (-\beta f_q)$, where $f_q$ is the free energy of an isolated static massive quark at a spatial position.
 Therefore, in the confining phase  $f_q$ is infinite, and consequently the center symmetry is unbroken, i.e. $\langle \ell\rangle =0$, while in the deconfining phase $f_q$ is finite so  that $\langle \ell\rangle \neq 0$, implying that the center symmetry is  spontaneously broken in this phase.
Thus,  $\ell$ can be used as an order parameter for deconfinement transition as well (see for a review \cite{Greensite:2003bk} for instance).

Since the scalar field $S$ transforms as $S(x)\to S(x)'=U_k(x_4)S(x)$ under the center symmetry 
transformation \eqref{Uk}, the discrete center symmetry is explicitly broken
in the presence of the scalar field by the boundary condition.
Consequently, the traced Polyakov loop $\ell$ can not be 
an exact order parameter if the scalar field is dynamically active.
In fact it has been proven that there exists no exact order parameter
for deconfinement transition in the presence of the scalar field
in the fundamental representation of $SU(N)$ 
\cite{Osterwalder:1977pc,Fradkin:1978dv}.
Therefore, the VEV of the traced Polyakov loop
$\ell$ is  finite  in the presence of the scalar field
in the fundamental representation.

The situation is quite similar to QCD with massive dynamical quarks,
because the presence of a massive dynamical 
 quark breaks explicitly the center symmetry as well as the chiral symmetry;
so  there exists  no exact order parameter.
 Nevertheless, it has been observed that
 $\ell$ and the chiral condensate undergo
a crossover transition at the same pseudo-critical temperature~\cite{Fukugita:1986rr,Karsch:1994hm,Aoki:1998wg,Karsch:2000kv,Allton:2002zi}.
Fukushima \cite{Fukushima:2003fw} has proposed an effective theory to describe 
this behavior of $\ell$ and the chiral condensate 
in the mean field approximation. The effective theory
consists of two sectors; the effective potential for 
$\ell$, where the temperature independent part is based
 on the Haar measure of the group integration, and
 the NJL sector for the chiral condensate,
which is so constructed that  the finite temperature effect vanishes 
if $L=0$ is imposed by hand
($\ell$ does not imply $L=0$).

Following Fukushima \cite{Fukushima:2003fw}, we make a phenomenological ansatz for the effective potential:
\al{
V_\text{eff}(L,f,T)=V_{\rm gluon}(L,T)+V_{\rm matter}(L,f,T),
}
where $V_{\rm gluon}(L,T)$ is the purely gluonic part, while $V_{\rm matter}(L,f,T)$ is the matter part and satisfies that the temperature effect vanishes in $V_{\rm matter}(L,f,T)$ at $L=0$, i.e. $V_{\rm matter}(L=0,f,T)=V_{\rm matter}(L,f,T=0)$.
We further require that $T_\ell =T_f$, where $T_\ell$ is the critical temperature for the deconfinement transition, and $T_f$ is that for the scale transition.

In the following subsection we first consider the two sectors separately  and then discuss the phase transition in the combined system.

\subsubsection{$V_{\rm gluon}(L,T)$ and the Haar measure for $N=3, 4,5,6$}
The Polyakov loop $L$ defined in \eqref{polyakov} assumes a simple form in the Polyakov gauge:
It is independent of $x_4$ and diagonal, i.e.
\al{
L &= \mbox{diag}(e^{i\theta_1},
\dots, e^{i\theta_{N-1}},e^{i\theta_{N}})&
&\mbox{with}~~\sum_{n=1}^{N}\theta_{n}=0,&
\label{polyakov1}
}
which implies that
\al{
\ell &= \frac{1}{N}\left( e^{i\theta_1}+\cdots+e^{i\theta_{N}}  \right)
\label{ell1}
}
in the Polyakov gauge. Although $\langle \ell\rangle$ can be
complex valued in general, we assume here that 
$\ell$ is real valued, as it has been assumed in \cite{Fukushima:2003fw,Fukushima:2008wg}.
Clearly, for an arbitrarily chosen set of $\theta_n$,
 $\ell$ can not be real: It is  possible only if at least two angles are
 related, e.g. $\theta_1=-\theta_2$ etc.
 Therefore, this realty assumption reduces the number of degrees
of freedom  down to $(N-1)/2$ for the  odd $N$
 and $N/2$ for the even $N$. That is,
\al{
 \ell = \begin{cases}
\displaystyle \frac{2}{N}\left(\cos\theta_1+ 
 \cdots+\cos\theta_{(N-1)/2} +\frac{1}{2} \right)& 
 \mbox{for odd}~N\\[10pt]
\displaystyle \frac{2}{N}\left(\cos\theta_1+ 
 \cdots+\cos\theta_{N/2}  \right) & \mbox{for even} ~N.
\end{cases}
\label{traced polyakov}
}
 Note that $\theta$s are a function of ${\vect x}$ because
 the Polyakov loop $L$ defined in \eqref{polyakov} is a function
 of ${\vect x}$. However, we recall that in deriving the effective potential
 \eqref{MFA2ap} we have treated the mean field $f$ as a constant field
 independent of $x$, and therefore, we regard $\theta_n$, too,
 as a constant field in deriving its effective potential.
 
With these preparations we come to the potential part
$V_{{\rm gluon},N}(L,T)$.\footnote{From here on we add the subscript $N$ to the potential.} 
As for the temperature independent part
$V_{{\rm gluon},N}^0(L)$, we use the form which is motivated 
by  the Haar measure $\mu(\theta)$ as it has been assumed in \cite{Fukushima:2003fw,Fukushima:2008wg}.
The appearance of the Haar measure may be understood
as a consequence of the variable transformation
from the gauge invariant measure (integration of link variables 
in lattice gauge theory) to $\theta$s.
It is  defined as
\al{
d\mu_{N}(\theta) &=H_{N}(\theta) \prod_n d \theta_n,
}
where 
\al{
H_{N}(\theta) &= \det \pmat{
1 & z_1 &\cdots & z_1^{N-1}\\[5pt]
1 & z_2 &\cdots  & z_2^{N-1}\\
 & & \vdots  & \\
1 & z_{N} &\cdots & z_{N}^{N-1}
}&
&\mbox{with}& z_n&=\exp (i\theta_n),&
}
from which we obtain
\al{
V_{{\rm gluon}, N}^0(L)&= - \ln H_{N}(\theta).
\label{gluon potential at zero temp}
}
Since we assume that the traced Polyakov loop $\ell$ is real
(see (\ref{traced polyakov})), we have to impose 
\al{
\theta_1 &= -\theta_{(N+1)/2},&
\theta_2 &= -\theta_{(N+3)/2},& &\dots,&
\theta_{(N-1)/2} &= -\theta_{N-1}& &\mbox{for odd}~N,&\\
\theta_1 &= -\theta_{N/2+1},&
\theta_2 &= -\theta_{N/2+2},& &\dots,&
\theta_{N/2-1} &= -\theta_{N-1}& &\mbox{for even}~N.&
}
Then adding the kinetic term \cite{Fukushima:2003fw,Fukushima:2008wg} to $V_{{\rm gluon}, N}^0(L)$ we finally obtain
\al{
\frac{V_{{\rm gluon},N}(L,T)}{b_{N}T }
&=-6 \exp(-a/T)N^2 ~\ell^2-\ln H_{N}(\theta).
\label{VglN}
}
In the following, we investigate the phase transition at finite temperature.
Since we cannot specify the number of the hidden gauge group $N$ from experiments such as collider and cosmological observations, in this work, we investigate the cases $N=3,4,5,6$. 
\\

\noindent
{\underline{$\bvec{N=3}$}}\\

\indent In this case, the traced Polyakov loop is parametrized by an angle as follows:
\al{
\ell=\frac{1}{3}(1+2x),
}
where $x=\cos\theta$.
The effective potential coming from the Haar measure is
\al{
V_{{\rm gluon}, 3}^0(L)&= - \ln \left[(1-x)^3(1+x)\right].
}
This potential can be written in term of $\ell$,
\al{
V_{{\rm gluon}, 3}^0(L)&=-\ln \left[1-6\ell^2+8\ell^3-3\ell^4\right],
}
where we neglected a constant term.

At finite temperature, we analytically obtain a minimum of the Polyakov loop:
\al{
\langle \ell \rangle= \frac{1}{9}\left( 3+\sqrt{36-3e^{-a/T}}\right).
\label{minimum of pure Nc3 case}
}
At $a/T=2.48491$ this minimum (false vacuum) appears, and at $\tilde a_c=a/T_c=2.45483$ two vacua at \eqref{minimum of pure Nc3 case} and $\langle \ell \rangle=0$ degenerate.
Since the case $N=3$ has been studied in a lot of works, see e.g.~\cite{Marhauser:2008fz,Haas:2013qwp,Kondo:2015noa} for details.\\

\noindent
{\underline{$\bvec{N=4}$}}
\begin{figure}
\includegraphics[width=8cm]{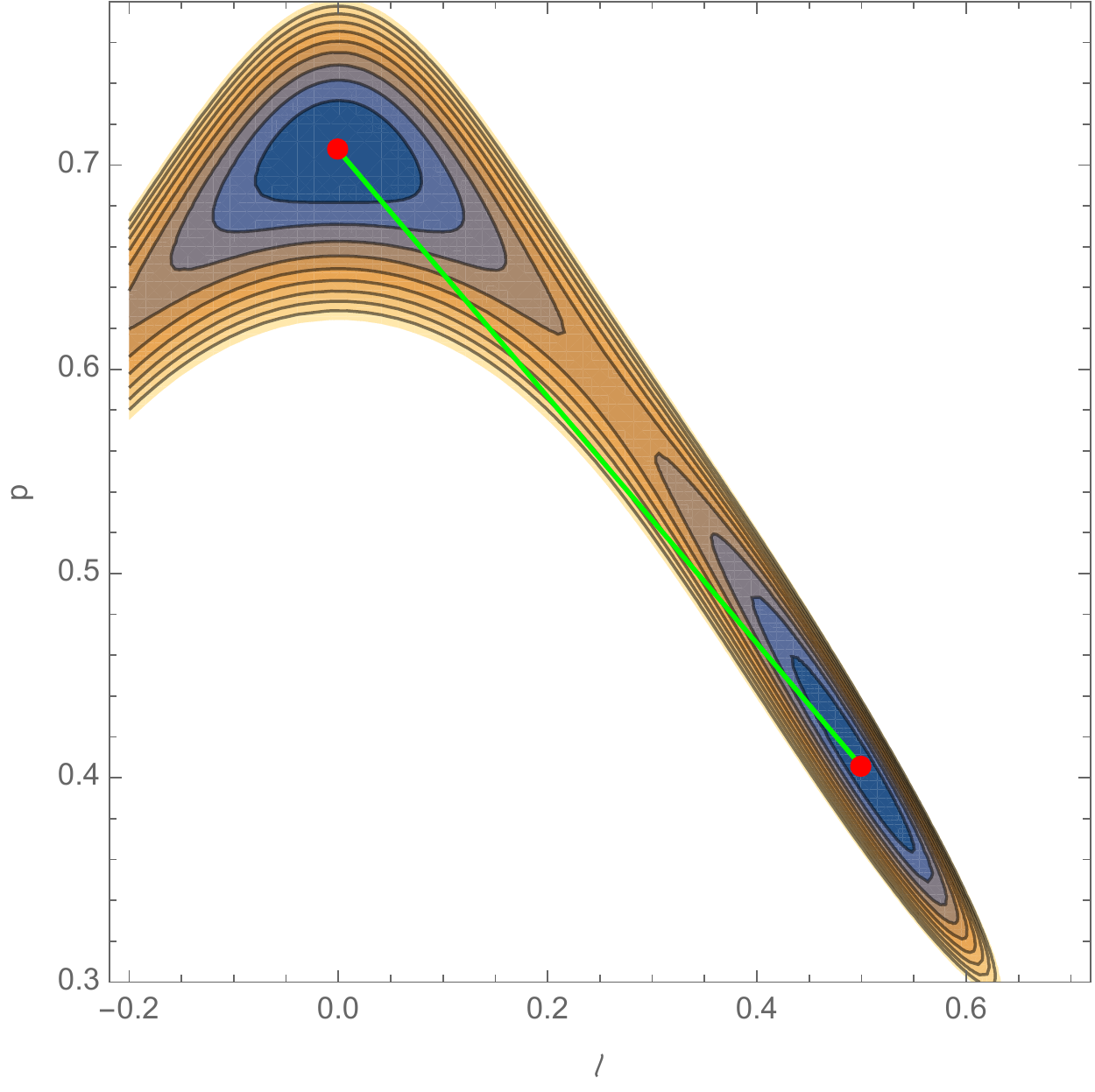}
\caption{
The contour plot of the potential $V_{{\rm gluon},4}(L,T)/b_4T$ at $\tilde{a}_c=a/T_c=2.237527$.
The green linear line links the two degenerated vacua \eqref{mini4-1} and \eqref{mini4-2} shown by the red points.
The shape of the potential as a function of $\ell$ or $p$ on the linear line corresponds to the black solid-line in Fig.\,\ref{fig:Vglue4}.
}
\label{fig:Vglue4ContourPlot} 
\end{figure}
\begin{figure}
\includegraphics[width=8.9cm]{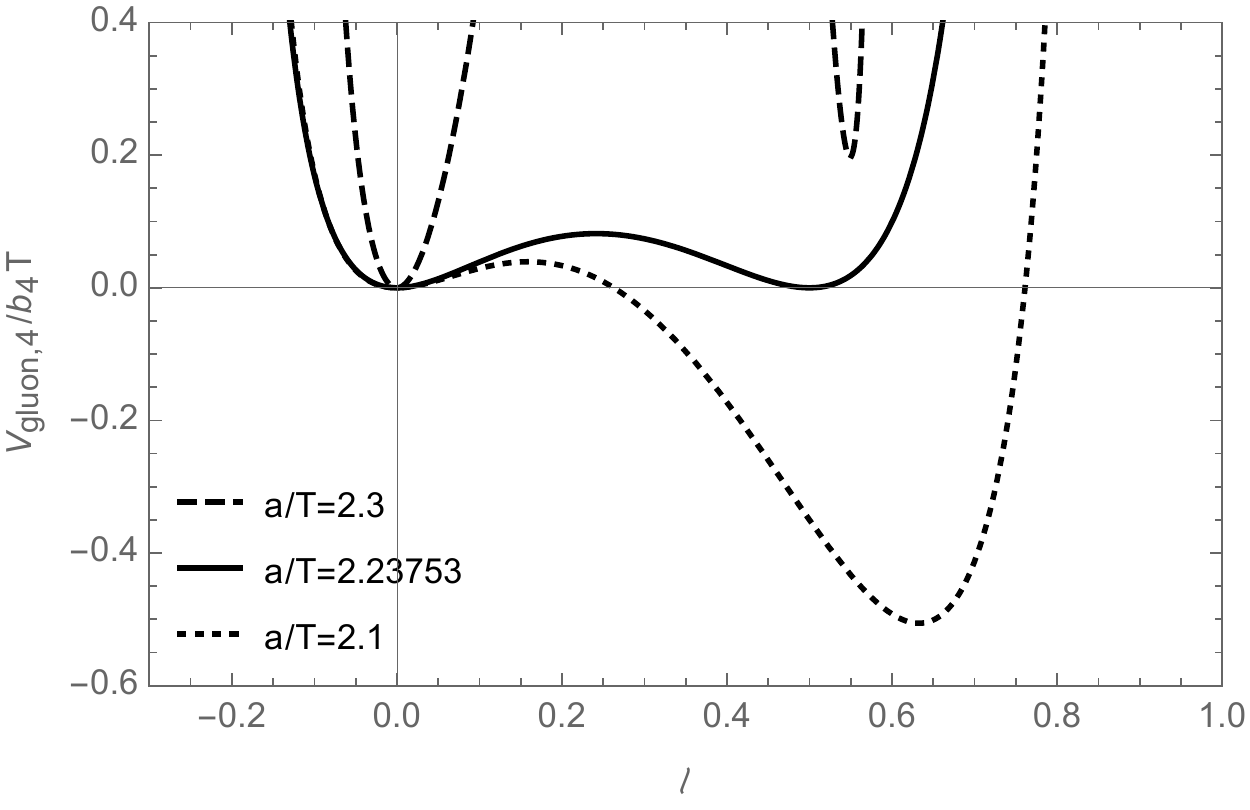}
\includegraphics[width=8.9cm]{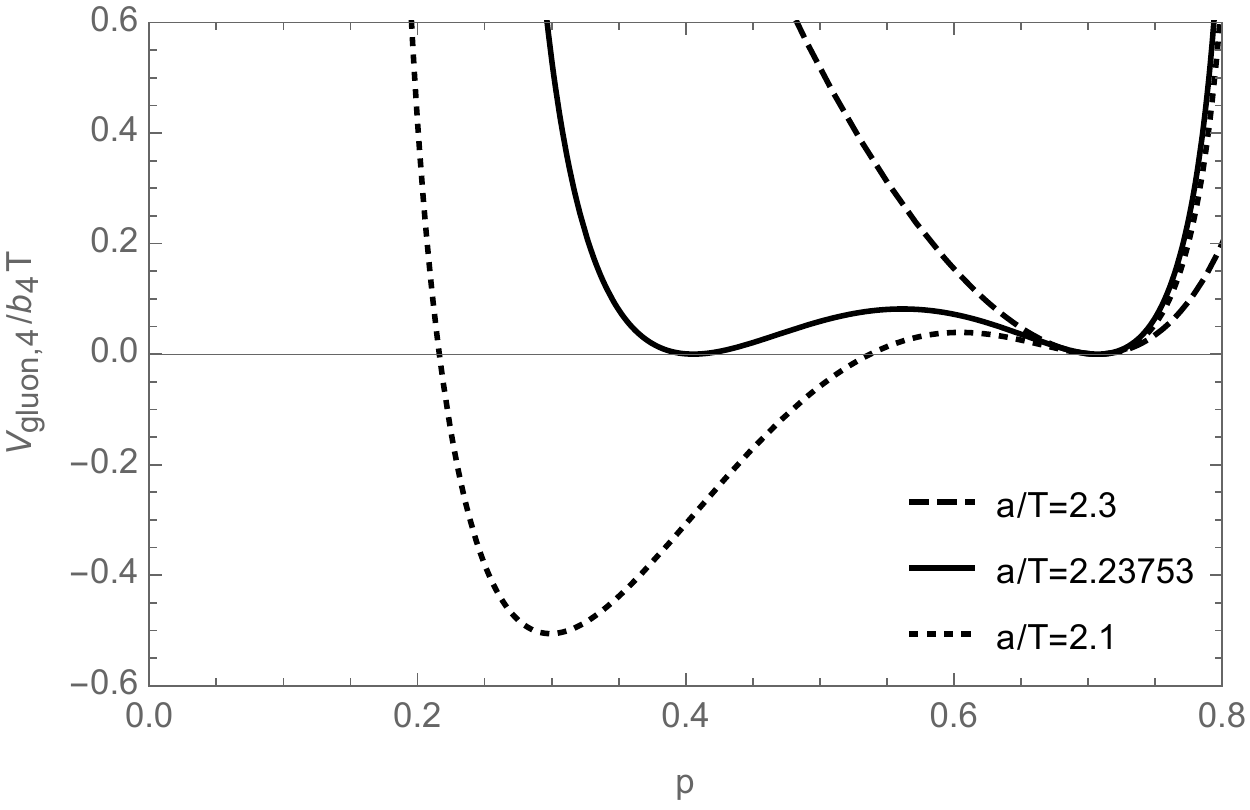}
\caption{
The potential $V_{{\rm gluon},4}(L,T)/b_4T$ as a function of $\ell$ (left) and of $p$ (right) on the straight line linking the two minimum points
given in \eqref{mini4-1} and \eqref{mini4-2} around the critical temperature $\tilde{a}_c=a/T_c=2.237527$.
Note that along the line $\ell$ increases while $p$ decreases.
}
 \label{fig:Vglue4} 
\end{figure}\\

\indent There are two independent  angles for $N=4$,
and the traced Polyakov loop $\ell$ assumes the form
\al{
\ell &= \frac{1}{2}\left(
x_1+x_2
\right),
}
where 
$x_1=\cos\theta_1$ and $x_2=\cos\theta_2$, and
$V_{{\rm gluon},4}^0(L)$ is given by
\al{
V_{{\rm gluon},4}^0(L)&=-\ln\left[ (x_1-x_2)^4(1-x_1^2)(1-x_2^2)\right]
}
up to a constant.
Note that $V_{\rm gluon}^0(L)$ has no longer
$Z_{N}$ center symmetry due to the reality assumption
on $\ell$. Instead, $V_{{\rm gluon},4}^0(L)$
is invariant under $S_2\times Z_2$, under which 
$x_1$ and $x_2$ transform as
\al{
S_2 &: x_1,x_2\to x_2,x_1,& Z_2&:~x_1,x_2\to -x_2,-x_1.&
\label{S2}
}
Since $\ell$ transforms as $\ell\to -\ell$
under $Z_2$, $\ell$ is an order parameter for
$Z_2$. An order parameter for $S_2$ is 
$p= (x_1-x_2)/2$. In terms of these order parameters 
$V_{{\rm gluon},4}^0(L)$ can be rewritten
as
\al{
V_{{\rm gluon},4}^0(L)
&=-\ln\left[ p^4 \left(1+\ell^2(\ell^2-2)+p^2(p^2-2)-2 \ell^2 p^2\right)
\right],
\label{V4g}
}
where we have suppressed the constant in (\ref{V4g}).
Note that $|\ell|, |p| \leq 1$, and one finds that the absolute
minimum of (\ref{V4g}) in this interval is located  at
\al{
\ell  &= 0,& p&=\cos(\pi/4)=0.7071\dots&
\label{mini4-1}
}
The two points $p=+ \cos(\pi/4)$ and $p=-\cos(\pi/4)$
are the physically same point because of the permutation symmetry
$S_2$ for $x_1$ and $x_2$ (and hence for $\theta_1$ and $\theta_2$).
Consequently, $Z_2$ is unbroken, while $S_2$ is spontaneously broken.

In the presence of the kinetic term (the first term on the rhs of \eqref{VglN})
the location of the absolute minimum changes. 
We find that
the critical value of $a/T$ denoted by $\tilde{a}_c$ is $2.237527$,
and at $\tilde{a}_c$ the total potential $V_{{\rm gluon},4}(L,T)$
is minimized at two points in the $\ell-p$ plane (up to the sign of $\ell$
and $p$ because of the permutation symmetry $S_2$):
The one is given in \eqref{mini4-1}, and the second point is located at
\al{
\ell  &= 0.4995,& 
p&=0.4056,&
\label{mini4-2}
}
implying that $Z_2$  and $S_2$ are both spontaneously broken
for $a/T < 2.237527$.
In Fig.\,\ref{fig:Vglue4ContourPlot} we show the contour plot of the potential $V_{{\rm gluon},4}(L,T)$
at $a/T=\tilde{a}_c=2.237527$ as a function of $\ell$ and of $p$.
The green straight-line links the two minimum points (red points).
Fig.\,\ref{fig:Vglue4} shows the shapes of the potential as a function of $\ell$ (left) or $p$ (right) on the straight line linking the two minimum points around $\tilde{a}_c$.
The black solid-line corresponds to the green line in Fig.\ref{fig:Vglue4ContourPlot}.
\\

\noindent
{\underline{$\bvec{N=5}$}}\\

\indent In this case there are also two independent  angles,
and the traced Polyakov loop $\ell$ assumes the form
\al{
\ell &= \frac{2}{5}\left( x_1+x_2+\frac{1}{2}
\right),
}
where 
$x_1=\cos\theta_1$ and $x_2=\cos\theta_2$
(as in the case for $N=4$), and
$V_{{\rm gluon},5}^0(L)$ is given by
\al{
V_{{\rm gluon},5}^0(L)&=-\ln\left[ (x_1-x_2)^4
(1-x_1^2)(1-x_2^2)(1-x_1)^2(1-x_2)^2\right]
}
up to a constant.
Note that $V_{{\rm gluon},5}^0(L)$ has no longer
$Z_{2}$ symmetry (\ref{S2}), but only the permutation symmetry
$S_2$ for $x_1$ and $x_2$. Therefore,
$\ell$ can no longer serve as an order parameter,
while $p= (x_1-x_2)/2$ is still
a good order parameter for $S_2$.

Note that $|x_1|, |x_2| \leq 1$, and one finds that the absolute
minimum  in this interval appears at
\al{
x_1  &= \cos(2 \pi/5)=0.30901\dots,& x_2&=\cos(4\pi/5)=-0.80901\dots,&
}
(up to the permutation of $x_1$ and $x_2$), at which $\ell$ and $p$ take the value 
\al{
\ell  &= 0,&
p&=\frac{\sqrt{5}}{4}=0.55901\dots&
\label{min5-1}
}
(up to the sign of $p$).
So, $S_2$ is spontaneously broken as in the case for $N=4$.
Although $\ell$ for $N=5$ is not related to any symmetry,
the absolute minimum of $V_{{\rm gluon},5}^0(L)$ appears at
$\ell=0$.

\begin{figure}
\centering
\includegraphics[width=8cm]{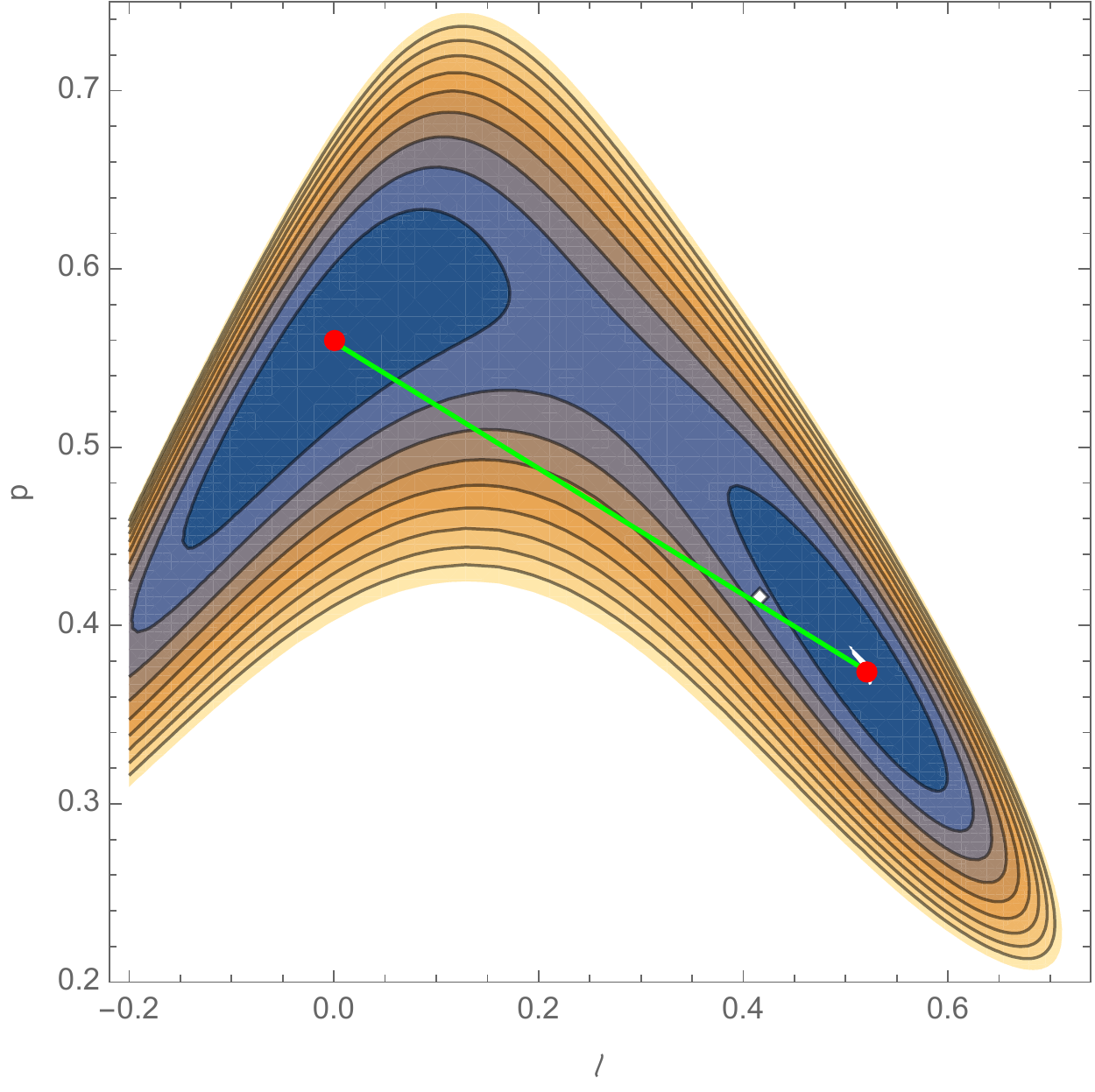}
\caption{
The contour plot of the potential $V_{{\rm gluon},5}(L,T)/b_5T$ at $\tilde{a}_c=a/T_c=2.12699$.
The green linear line links the two degenerated vacua \eqref{mini4-1} and \eqref{mini4-2} shown by the red points.
The shape of the potential as a function of $\ell$ or $p$ on the linear line corresponds to the black solid-line in Fig.\,\ref{fig:Vglue5}.
}
\label{fig:Vglue5ContourPlot} 
\end{figure}

\begin{figure}
\includegraphics[width=8.9cm]{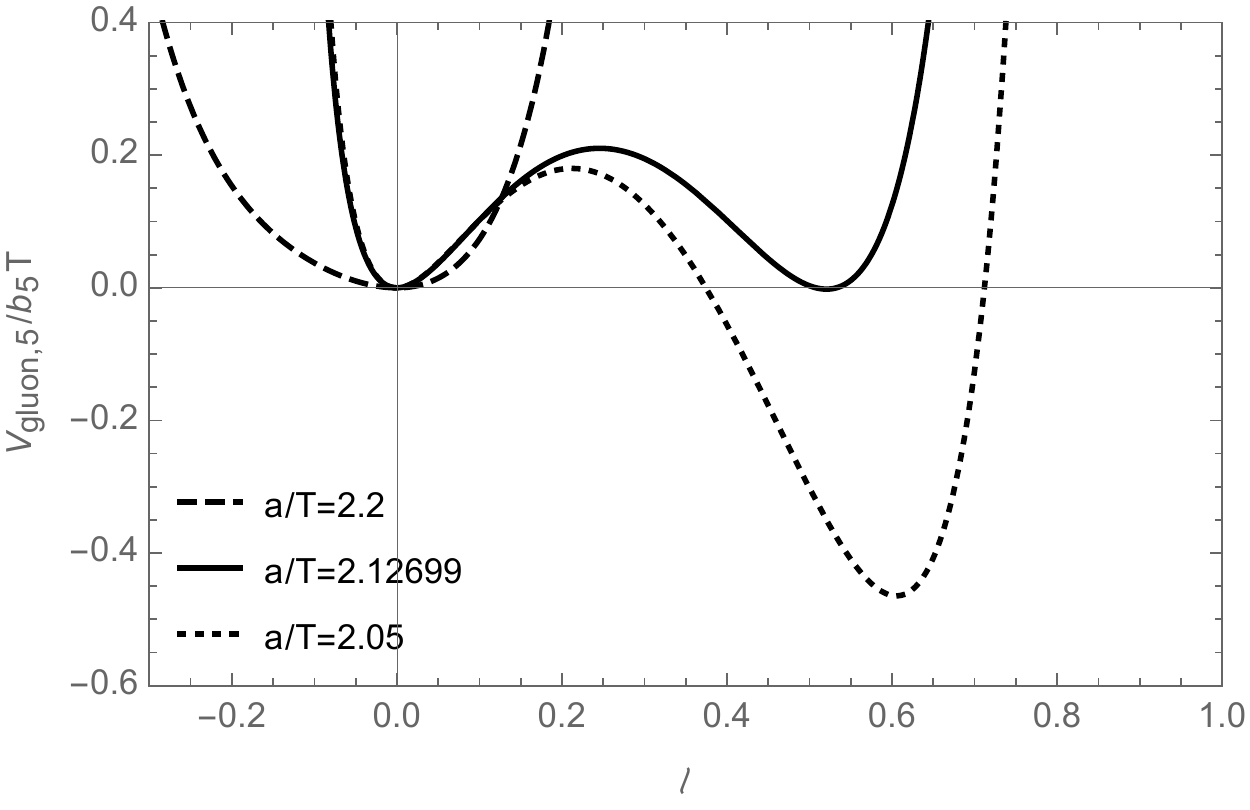}
\includegraphics[width=8.9cm]{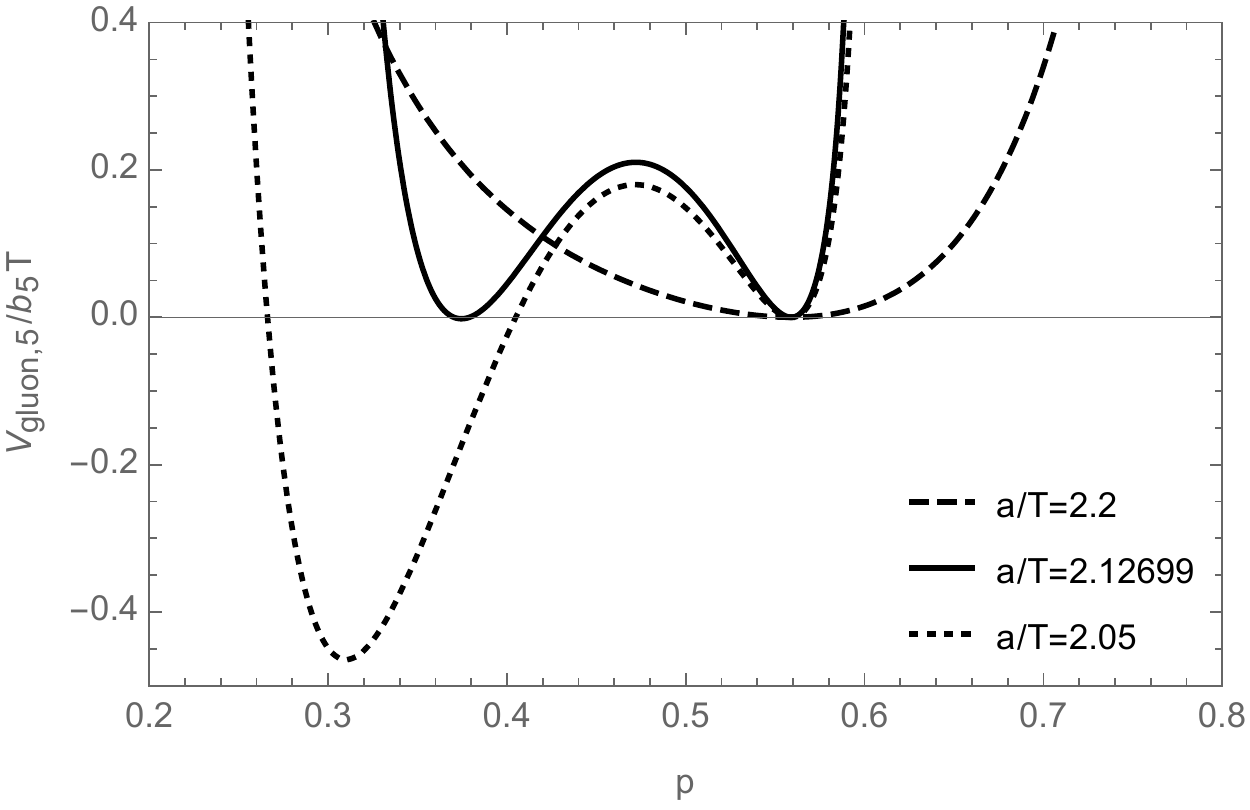}
\caption{
The potential $V_{{\rm gluon},5}(L,T)/b_5T$ as a function of $\ell$ (left) and of $p$ (right) on the straight line linking the two minimum points
given in \eqref{min5-1} and \eqref{min5-2} around the critical temperature $\tilde{a}_c=a/T_c=2.12699$.
}
 \label{fig:Vglue5} 
\end{figure}

The presence of the kinetic term in (\ref{VglN})
can change the location of the absolute minimum. We find that
the transition is a first-order phase transition, and that
the critical value of $ a/T$ is $\tilde{a}_c=2.12699$.
At $\tilde{a}_c$ the total potential $V_{{\rm gluon},5}(L,T)$
is minimized at two points in the $x_1-x_2$ plane (up to the 
the permutation of $x_1$ and $x_2$):
The one is given in (\ref{min5-1}), and the second point is located at
\al{
x_1  &= 0.7752,& x_2&=0.02621,&
\label{min5-w}
}
which means
\al{
\ell  &= 0.5206,&
p&=0.3745.&
\label{min5-2}
}
In Fig. \ref{fig:Vglue5ContourPlot} we plot the potential $V_{{\rm gluon},5}(L,T)$ at $a/T=\tilde{a}_c=2.12699$ on the $\ell$--$p$ plane.
The two degenerated vacua \eqref{min5-1} and \eqref{min5-2} shown by the red points are linked by the green straight line.
In Fig. \ref{fig:Vglue5}, we show the shapes of the potential as a function of $\ell$ (left) or $p$ (right) on the straight line around the critical temperature.
\\


\noindent
{\underline{$\bvec{N=6}$}}\\

\indent There are three independent  angles for $N=6$,
and the traced Polyakov loop $\ell$ is
\al{
\ell &= \frac{1}{3}\left(
x_1+x_2+x_3
\right),
}
where 
$x_i=\cos\theta_i~(i=1,2,3)$, and $V_{{\rm gluon},6}^0(L)$ is given by
\al{
V_{{\rm gluon},6}^0(L)&=-\ln\left[ (x_1-x_2)^4(x_2-x_3)^4(x_3-x_1)^4
(1-x_1^2)(1-x_2^2)(1-x_3^2)\right]
}
up to a constant.
$V_{{\rm gluon},6}^0(L)$
is invariant under $S_3\times Z_2$,
where $S_3$ consists of all the permutations of $x_1,x_2,x_3$,
and $x_1,x_2,x_3\to -x_2,-x_1,-x_3$ under $Z_2$.
$x_1$, $x_2$ and $x_3$ form a three dimensional reducible representation
of $S_3$ and can be decomposed into the irreducible representations
${\bf 1}$ and ${\bf 2}$:
\al{
&{\bf 1}:~\ell =\frac{1}{3}\left(x_1+x_2+x_3\right),&
&{\bf 2}:~\left( x_1-x_2~,~(x_1+x_2-2 x_3)/\sqrt{3}\right).&
}
The two dimensional representation can be further transformed to
a complex representation
\al{
z &= (x_1-x_2) +i (x_1+x_2-2 x_3)/\sqrt{3}.
}
Then using $\ell$ and $z$ we can now express the potential as
\al{
V_{{\rm gluon},6}^0(L)&=-\ln
\left((z^3+z^{*3})^4\left[ 1-3 \ell +3 \ell^2 -\ell^3-
z z^* (1-\ell)/4+\frac{i }{24\sqrt{6}}(z^3-z^{*3})\right]\right.\nn
&\qquad \left.\times \left[ 1+3 \ell +3 \ell^2 +\ell^3-
z z^* (1+\ell)/4-\frac{i }{24\sqrt{6}} (z^3-z^{*3})\right]\right)
}
(up to a constant),
which is invariant under $Z_2$:
\al{
\ell &\to  -\ell,&
z&\to z^*,&
}
and under  $Z_3$:
\al{
z &\to \exp(2\pi i k/3)z ,&
\ell&\to\ell&
}
with $k=1,2,3$.
Since $\ell$ transforms as $\ell\to -\ell$
under $Z_2$, $\ell$ is an order parameter for
 $Z_2$, and  the order parameter for $Z_3$ is 
$z$.
We find that the absolute minimum under $|\ell|\leq 1$ and $|z| \leq 4/\sqrt{3}$  is located at
\al{
\ell  &= 0,&
z&=z^*=\sqrt{3}.&
\label{mini6-1}
}
Therefore, $Z_2$ is unbroken, while $Z_3$ is spontaneously broken.
Note that this vacuum corresponds to
\al{
x_1&=\cos\fn{\pi/6}=0.8660,&
x_2&=\cos\fn{5\pi/6}=-0.8660,&
x_3&=\cos\fn{3\pi/6}=0.&
}

\begin{figure}
\includegraphics[width=8.9cm]{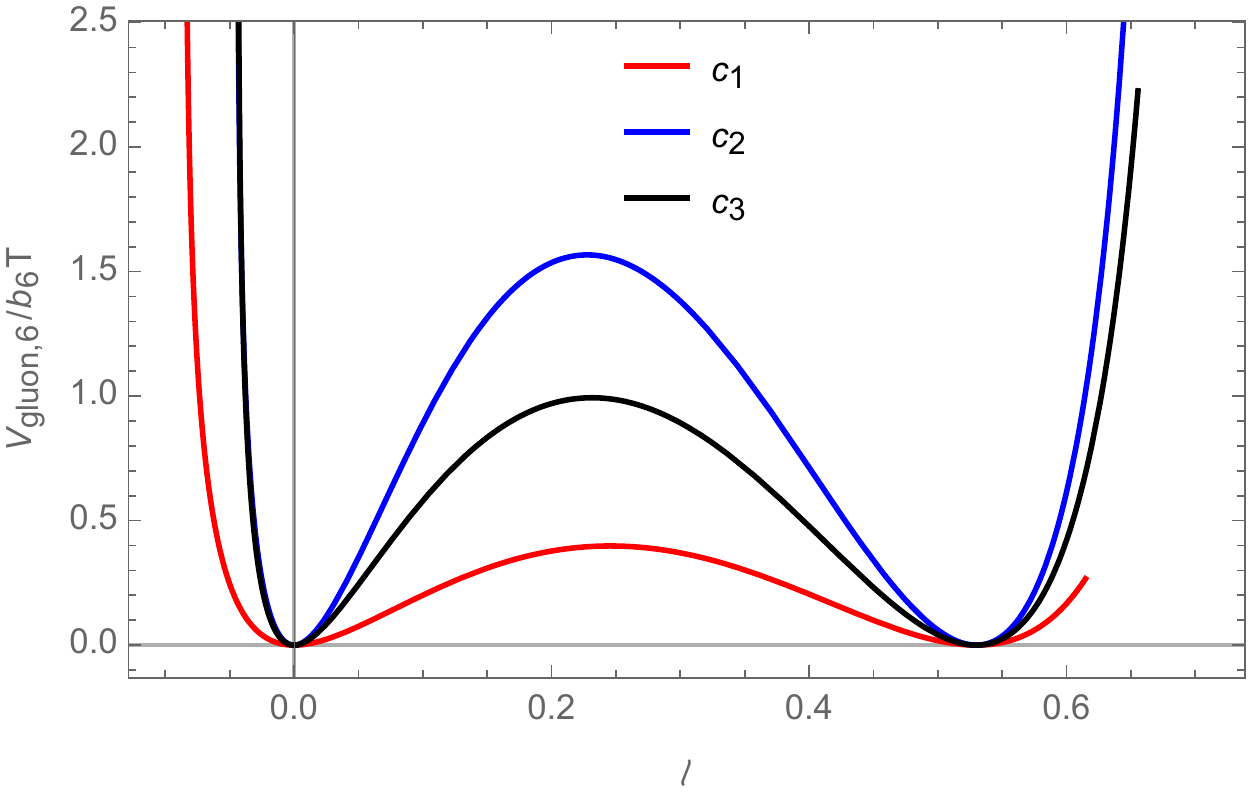}
\includegraphics[width=8.9cm]{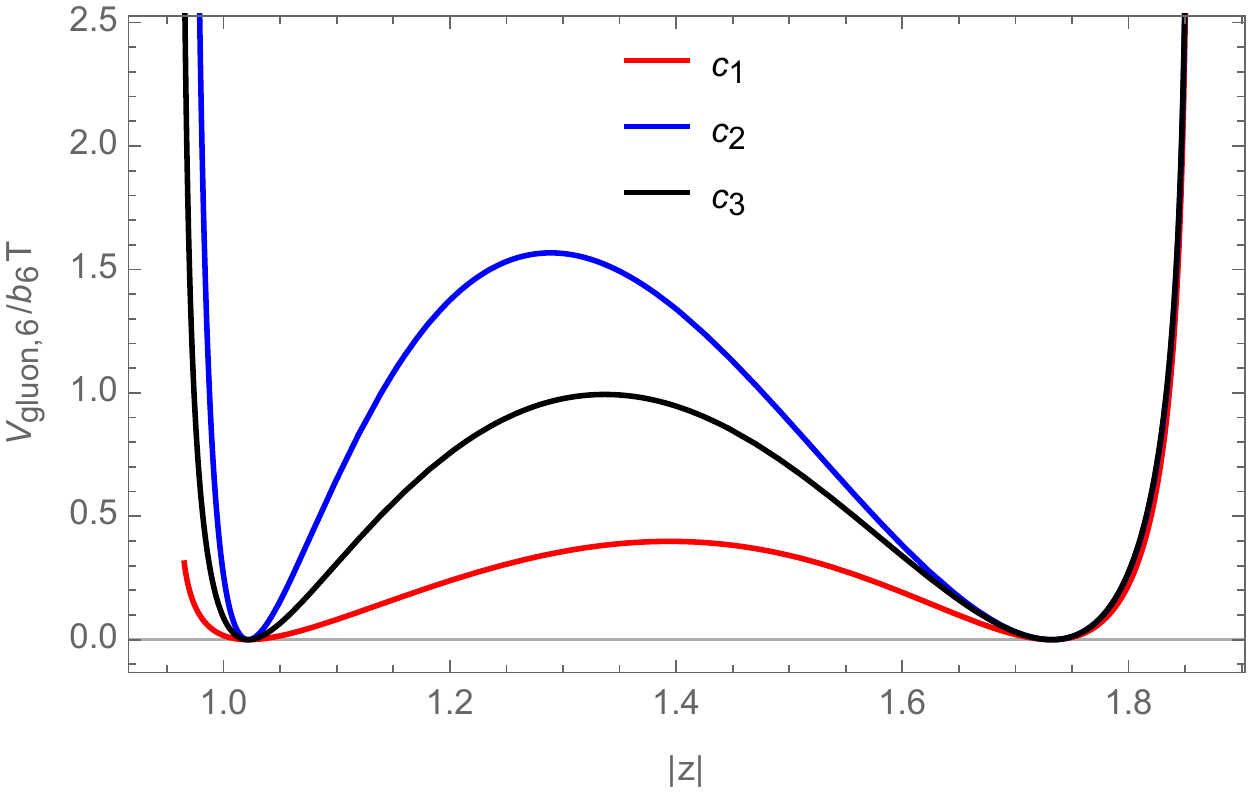}
\caption{ 
The potential $V_{{\rm gluon},6}(L,T)/b_6 T$ at $\tilde{a}_c=2.06064$ as a function of $\ell$ (left) and $|z|$ (right) on the three different lines $c_1$ (red), $c_2$ (blue), $c_3$ (black) linking the two minimum points given in \eqref{mini6-1} and \eqref{mini6-2}.
}
\label{fig:Vglue6} 
\end{figure}

\begin{figure}
\includegraphics[width=7.1cm]{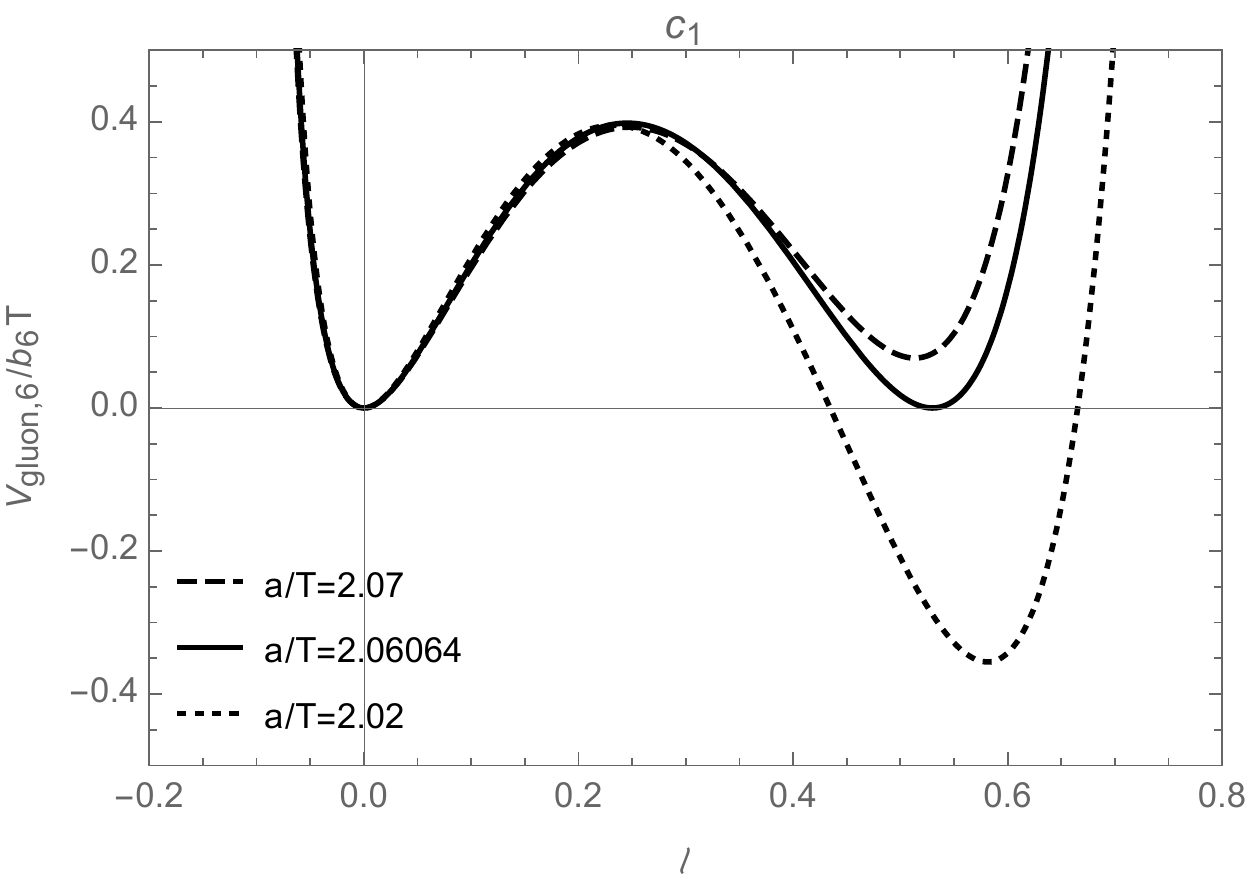}~~~~~
\includegraphics[width=7cm]{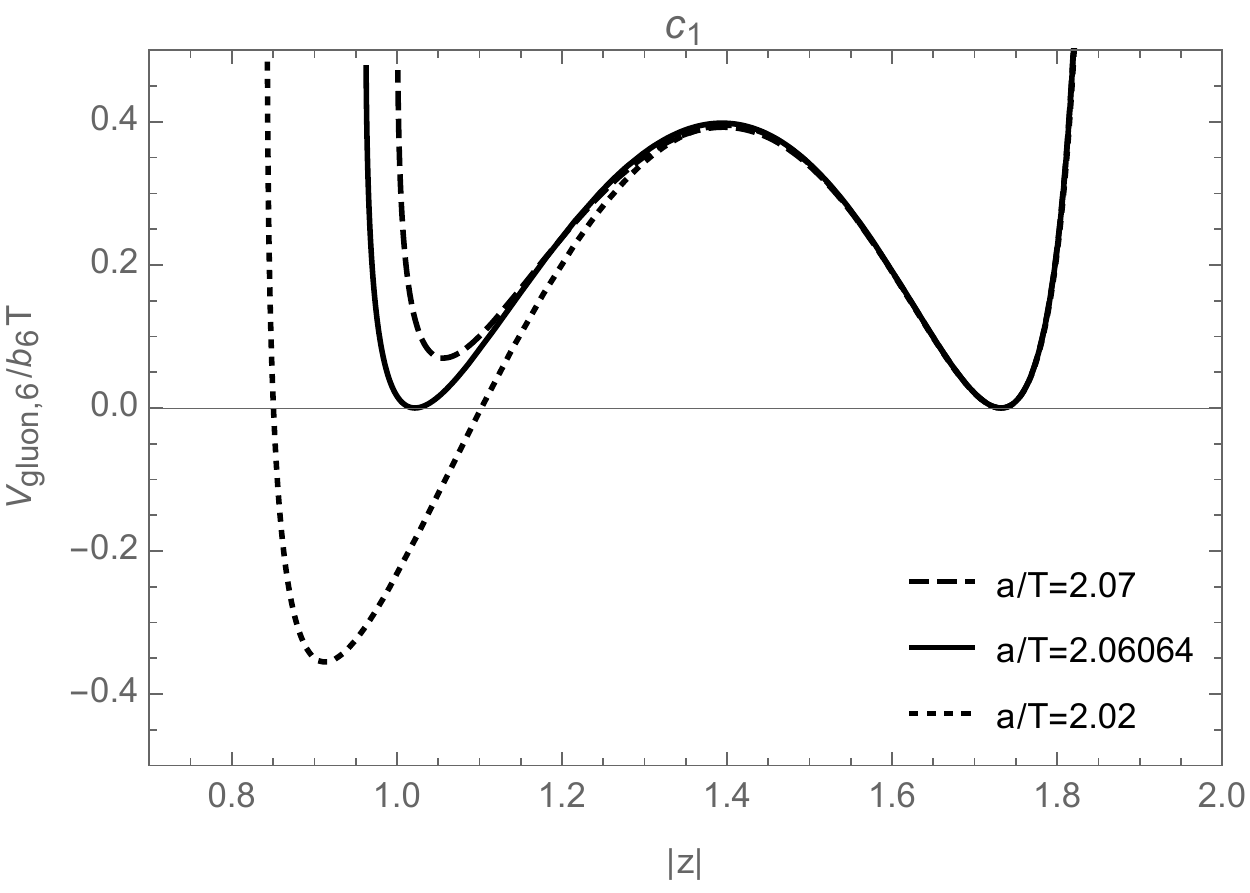}
\includegraphics[width=7.1cm]{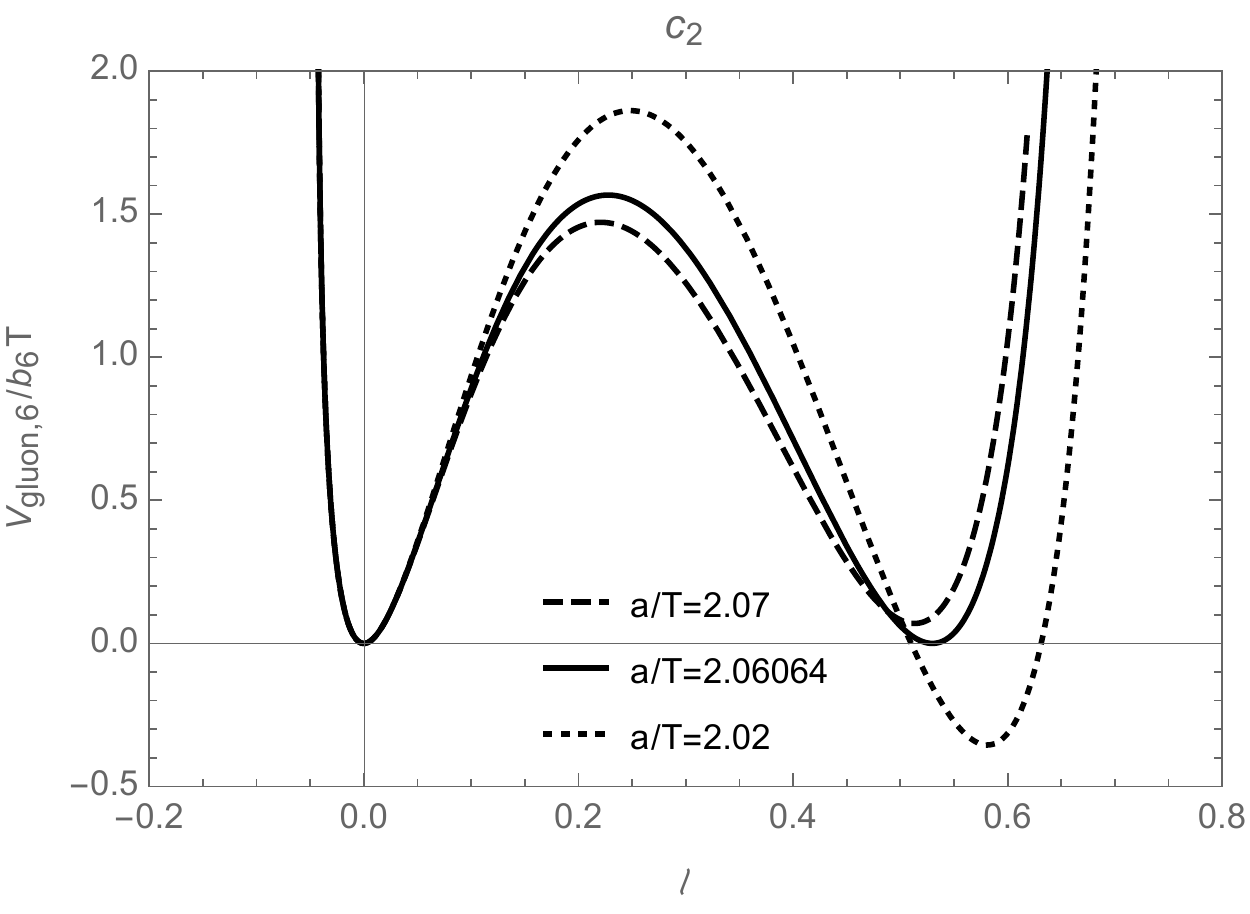}~~~~~
\includegraphics[width=7cm]{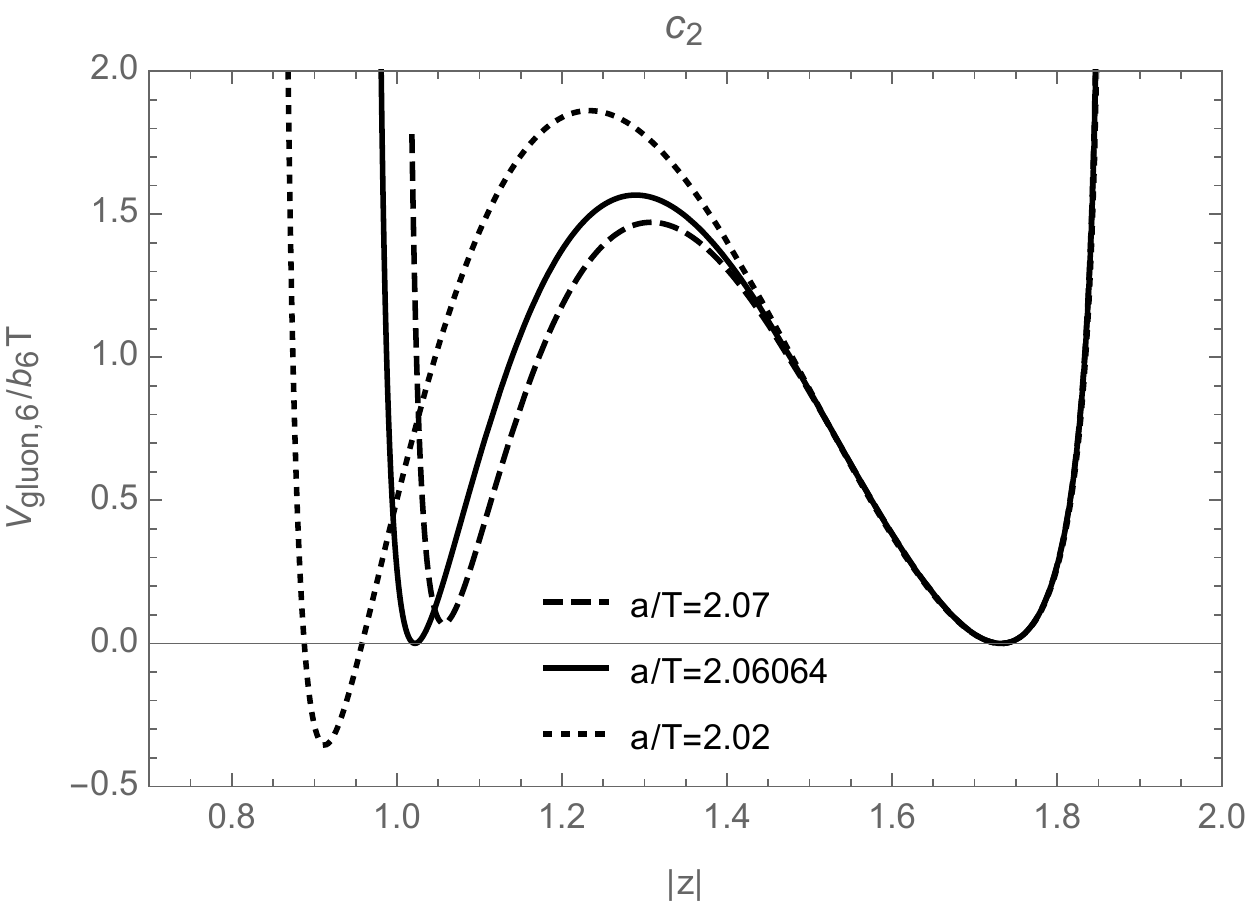}
\includegraphics[width=7.1cm]{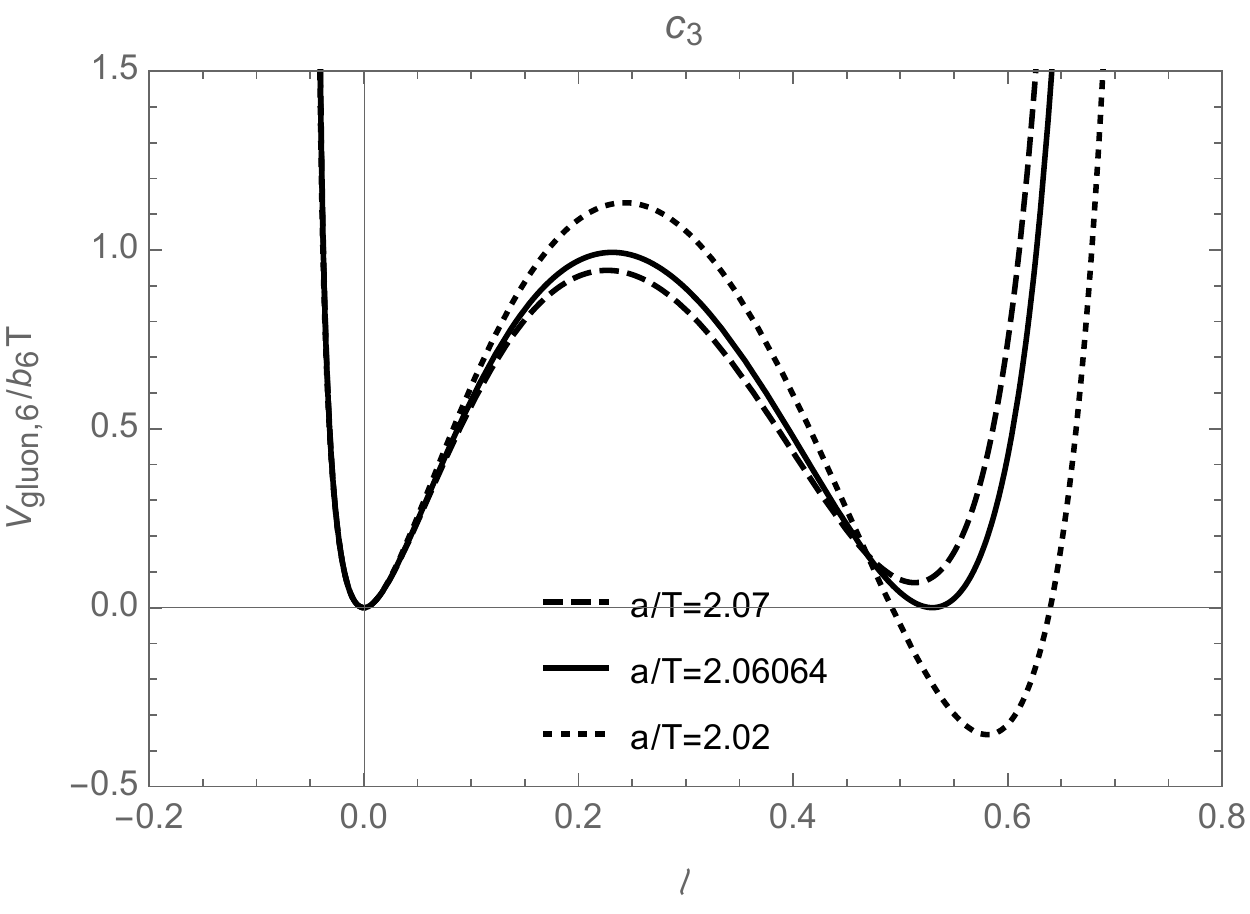}~~~~~
\includegraphics[width=7cm]{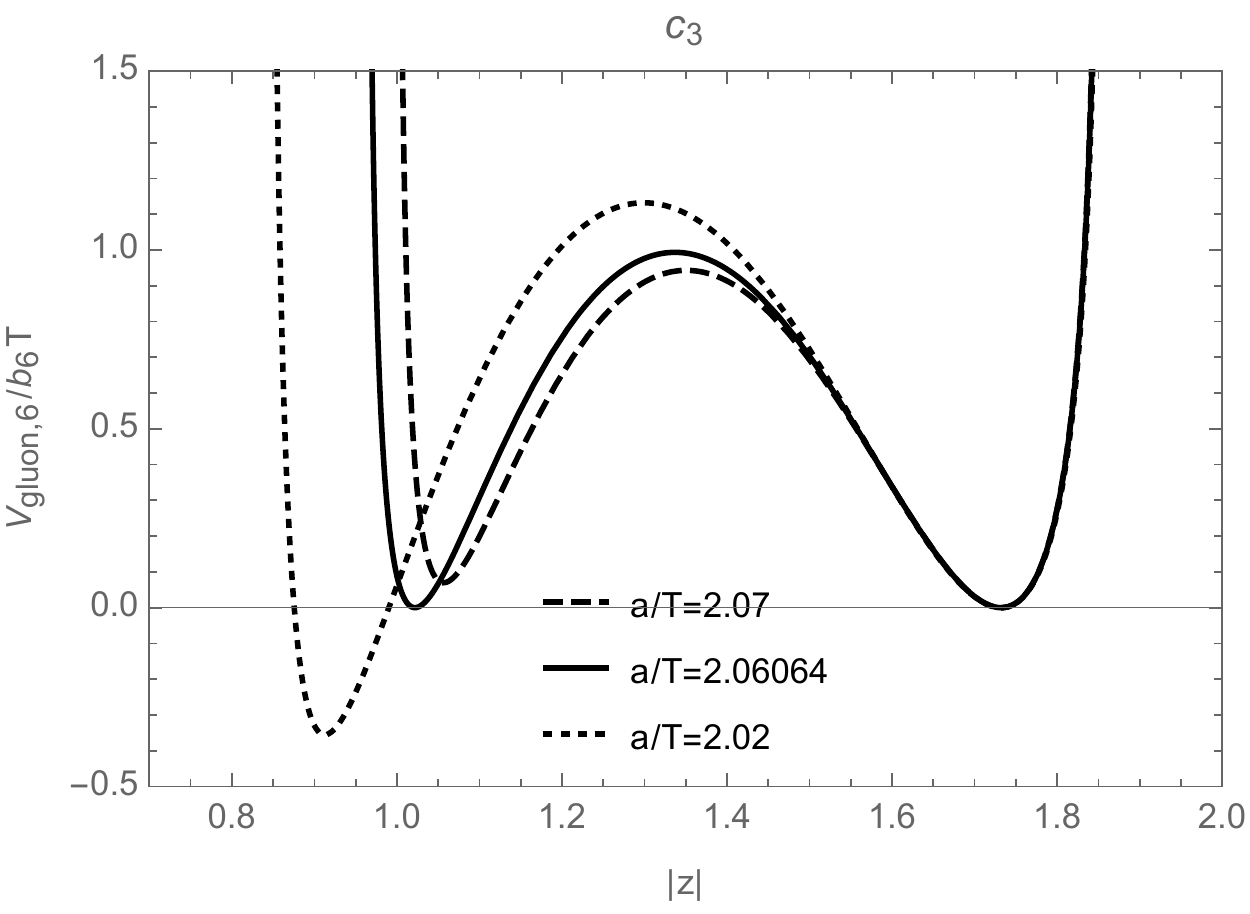}
\caption{
The potential $V_{{\rm gluon},6}(L,T)/b_6 T$ around the critical temperature $\tilde{a}_c=2.06064$ as a function of $\ell$ (left) and $|z|$ (right) on the three different lines $c_1$, $c_2$, $c_3$ linking the two minimum points given in \eqref{mini6-1} and \eqref{mini6-2}.
}
 \label{fig:Vglue6pass} 
\end{figure}

Now we include  the kinetic term and find that
the transition in this case  is also a first-order phase transition.
The critical value $\tilde{a}_c$ is $2.06064$,
and the second minimum point 
at $a/T=\tilde{a}_c$  is located at
\al{
\ell  &= 0.5299,&
z&=0.99689 -  0.224514i,&
z^*&=0.99689 + 0.224514i,&
\label{mini6-2}
}
or equivalently,
\al{
x_1&=0.9635,& 
x_2&=-0.03339,&
x_3&=0.6595,&
\label{mini6-22}
}
implying that $Z_2$  and $Z_3$ are both spontaneously broken
for $a/T < 2.06064$.
In Fig.\,\ref{fig:Vglue6} we plot the potential $V_{{\rm gluon},6}(L,T)$
at $a/T=\tilde{a}_c=2.06064$ as a function of $\ell$
(upper panel) and $|z|$ (lower panel)
varying 
along the three different  lines  $c_1~\mbox{(black)},c_2~\mbox{(red)}
$ and $c_3~\mbox{(blue)}$, which link the two minimum points using a parameter $t$:
\al{
&c_1:&
x_1 &= 0.0975 t +0.8660,&
x_2 &= 0.83261 t - 0.8660,&
x_3 &= 0.6595t &\\
&c_2 :&
x_1 &= 0.0975 t^2+0.8660,&
x_2 &= 0.83261 t - 0.8660, &
x_3 &= 0.6595 t^3&\\
&c_3 :&
x_1 &= 0.0975 t+0.8660,&
x_2 &= 0.83261 t - 0.8660,&
x_3 &=  0.6595t^2.&
}
We have \eqref{mini6-2} for $t=0$, while $t=1$ yields \eqref{mini6-22}.
In Fig.\,\ref{fig:Vglue6pass}, we show the temperature evolution of the potential around the critical temperature for the three different lines $c_1$, $c_2$ and $c_3$.



\subsubsection{Scale phase transition}
After we have analyzed the pure gluonic part $V_\text{gluon}\fn{L,T}$, we here study the scale phase transition including the Polyakov loop effect.
The matter part is $V_\text{matter}\fn{L,f,T}=V_\text{MFA}\fn{f}+V_{\rm FT}(L,f,T)$, where
\al{
V_\text{MFA}\fn{f}&=-N_f(N_f \lambda_S +\lambda_S')f^2+\frac{NN_f}{32\pi^2}M^4\ln\frac{M^2}{\Lambda_H^2},
\label{MFA potential}
\\
V_{\rm FT}(L,f,T)&=
2N_fT\int \frac{d^3 p}{(2\pi)^3}\,\Tr \ln\fn{1-L\, e^{-E_p/T}},
\label{FT potential}
}
and $E_p=\sqrt{\vec p^2+\tilde{M}^2\fn{T}}$ with the thermally dressed mass of $S$,
\al{
\tilde{M}^2\fn{T} 
&= \tilde M^2+\frac{T^2}{6}\Big ( (N N_f +1)\lambda_S +(N_f+N)\lambda'_S\Big).
}
Here, $V_\text{MFA}$ is given in \eqref{MFA potential} with $S$ suppressed, and $\tilde M^2$ is defined in \eqref{con mass}. $\Tr_c$ in \eqref{FT potential} stands for the trace in the $SU(N)$ color space.
Since the Polyakov loop is diagonal in the Polyakov gauge \eqref{polyakov1} and we assume that the angles $\theta$s are constants, the thermal effect part in \eqref{FT potential} can be written as
\al{
\Tr \ln\fn{1-L\, e^{-E_p/T}}
=\sum_{n=1}^{N}\ln\fn{1-\exp\fn{i\theta_n -E_p/T}}.
}
Then using the reality condition \eqref{traced polyakov}, we find that
\al{
V_{\rm FT}(L,f,T)&=-\frac{2N_f T^2 {\tilde M}^2}{2\pi^2}\sum_{n=1}^{N}\sum_{j=1}^\infty \frac{e^{i(j\theta_n)}}{j^2}K_2(j{\tilde M}/T)\nn
&=-\frac{2N_f T^2 {\tilde M}^2}{\pi^2}\sum_{j=1}^\infty \frac{K_2(j{\tilde M}/T)}{j^2}\nn
&\quad\times
\begin{cases}
\left( \cos\fn{j\theta_1}+\cdots +\cos\fn{j\theta_{N/2}} \right)\\[10pt]
\left( \cos\fn{j\theta_1}+\cdots +\cos\fn{j\theta_{(N-1)/2}}+1/2 \right)
\end{cases}
~~\text{for}~~
\begin{cases}
\text{even}~~N\\[10pt]
\text{odd }~~N
\end{cases},
\label{Finite temperature loop effect}
}
where $K_2\fn{x}$ is the modified Bessel function of the second kind of order two, and we will truncate the sum at $j=10$.
A useful identity is given by
\al{
\cos\fn{j\theta_i}=T_j\fn{x_i},
}
where $T_n\fn{x}$ is the Chebyshev polynomials of the first kind which satisfies the recurrence relation, $T_{n+1}\fn{x}=2xT_n\fn{x}-T_{n-1}\fn{x}$ with $T_0\fn{x}=1$ and $T_1\fn{x}=x$ for $n=1,2,\cdots$, and $x_i=\cos\theta_i$.\footnote{
More explicitly, the Chebyshev polynomials of the first kind is written as
\al{
T_j\fn{x_i}=\sum_{k=0}^j(-1)^k\pmat{j\\[2pt] 2k} x_i^{j-2k}(1-x_i^2)^k,\nonumber
}
where $\pmat{j\\[2pt] 2k}$ is the binomial coefficient.
}

Since in the $N=3$ case the volume factor $b_3T$ approximately satisfies $b_3T\simeq T^4$ at the critical temperature~\cite{Fukushima:2003fw}, we assume that this relation holds for other $N$, too.
Accordingly, we consider the dimensionless effective potential
\al{
V_\text{eff}\fn{L,f,T}/T^4=V_\text{matter}\fn{L,f,T}/T^4-6\exp\fn{-a_{N}/T}N^2\ell + V^0_{\text{gluon},N}\fn{L},
}
where the pure gluonic potential $V^0_{\text{gluon},N}\fn{L}$ is given in \eqref{gluon potential at zero temp}.
Further, since $f$ is a positive definite field with the canonical dimension two, we introduce a canonically defined field $\chi$ with the canonical dimension one:
\al{
f=\chi^2.
}
Note that the effective potential $V_\text{eff}\fn{L,f=\chi^2,T}$ is invariant under $\chi\to -\chi$.

For a given set of $\lambda_S$, $\lambda_S'$, $N_f$ and $N$, the effective potential $V_\text{eff}\fn{L,\chi,T}/T^4$ is controlled by $T$ and $a_{N}$. 
For an arbitrary choice of $T$ and $a_{N}$, the deconfinement and the scale transitions do not occur at the same temperature.
Since we assume that the both transitions occur at the same temperature $T_{c,N}$, we have to adjust $T$ and $a_{N}$, such that this happens.\footnote{
This condition, that both confinement and scale phase transition take place at the same temperature, is here nothing but an assumption. Therefore, it should be clarified by both analytical and numerical computations such as lattice Montecarlo simulations. 
}
We find that $T_{c,N}$ is not unique for a given set of $\lambda_S$ and $\lambda_S'$, $N_f$ and $N$ and varies slightly as $a_{N}$ varies. 
In the following we consider the cases with $N=3,4,5$ and $6$
for the set of the other  parameters given by
\al{
\lambda_S&=1,&
\lambda_S'&=2,&
N_f&=2.&
\label{parameter set}
}

\noindent
{\underline{$\bvec{N=3}$}}\\

\begin{figure}
\includegraphics[width=7cm]{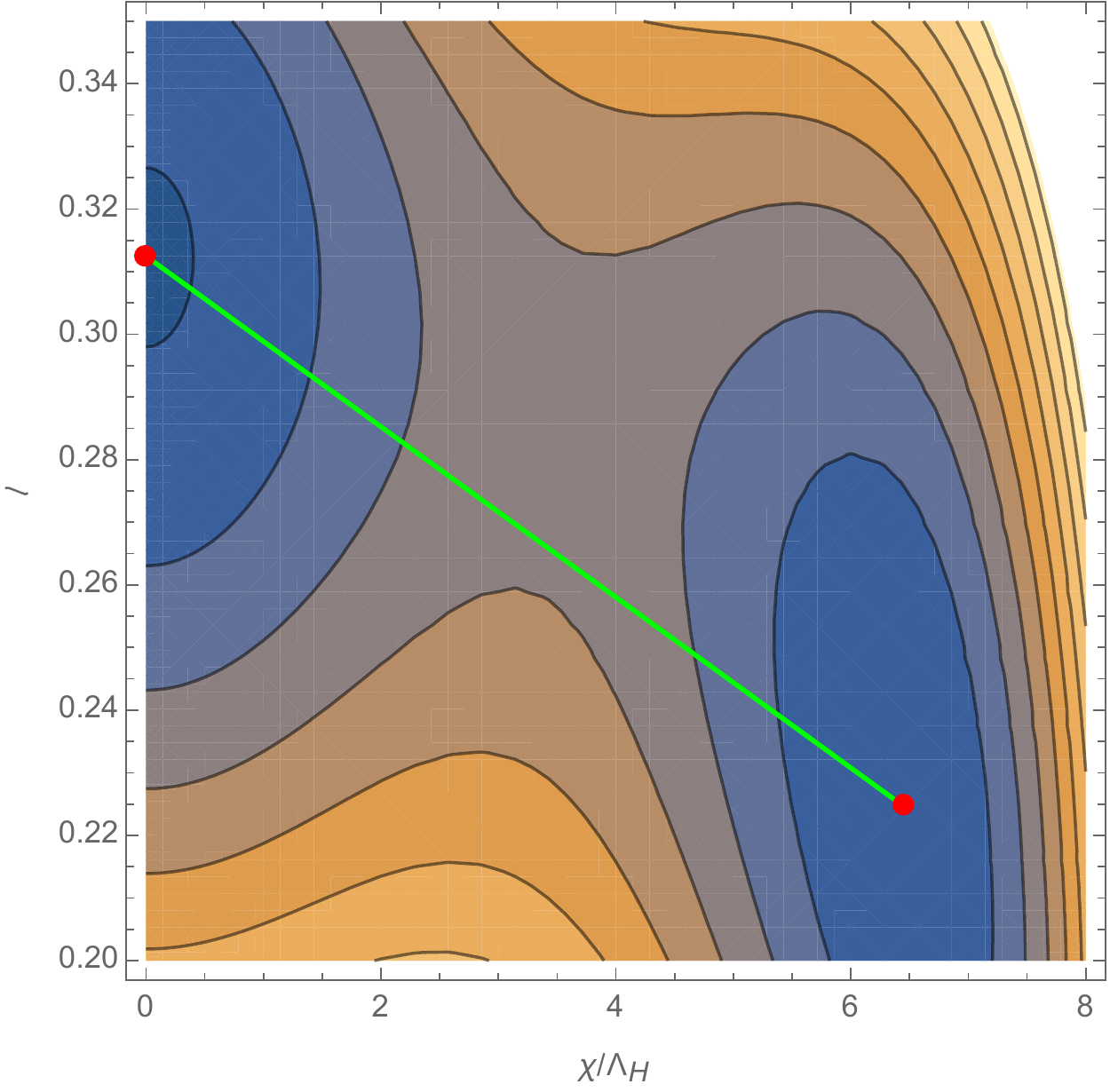}\\[5pt]
\includegraphics[width=8cm]{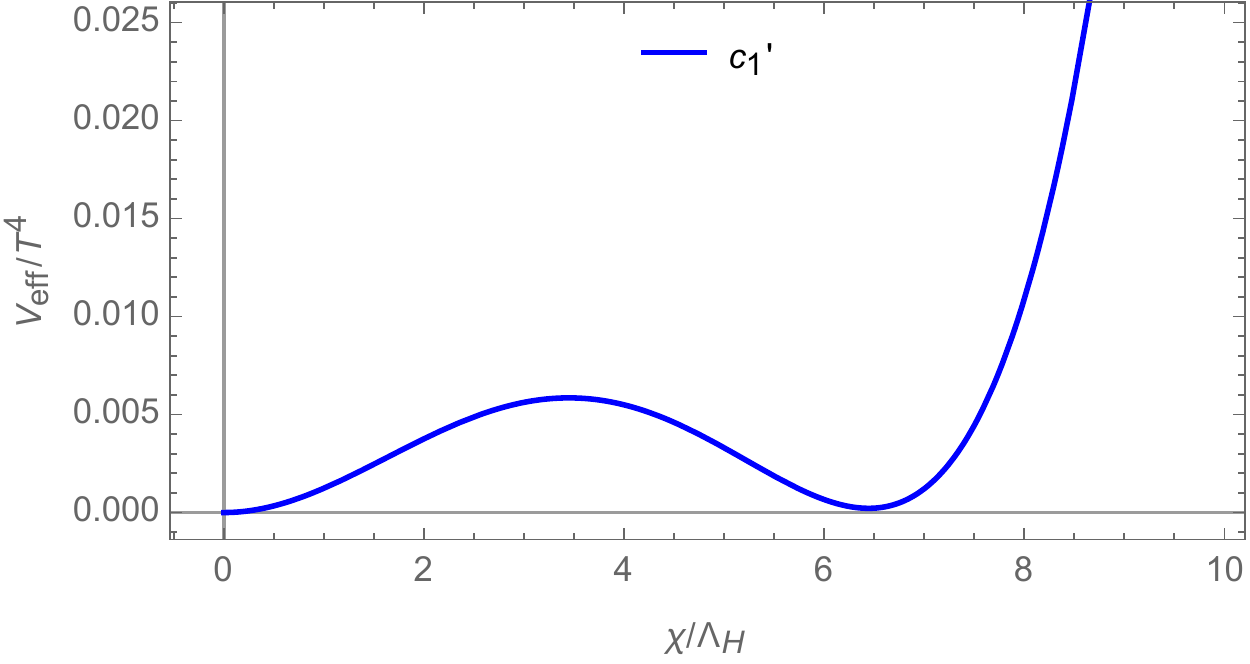}
\includegraphics[width=8cm]{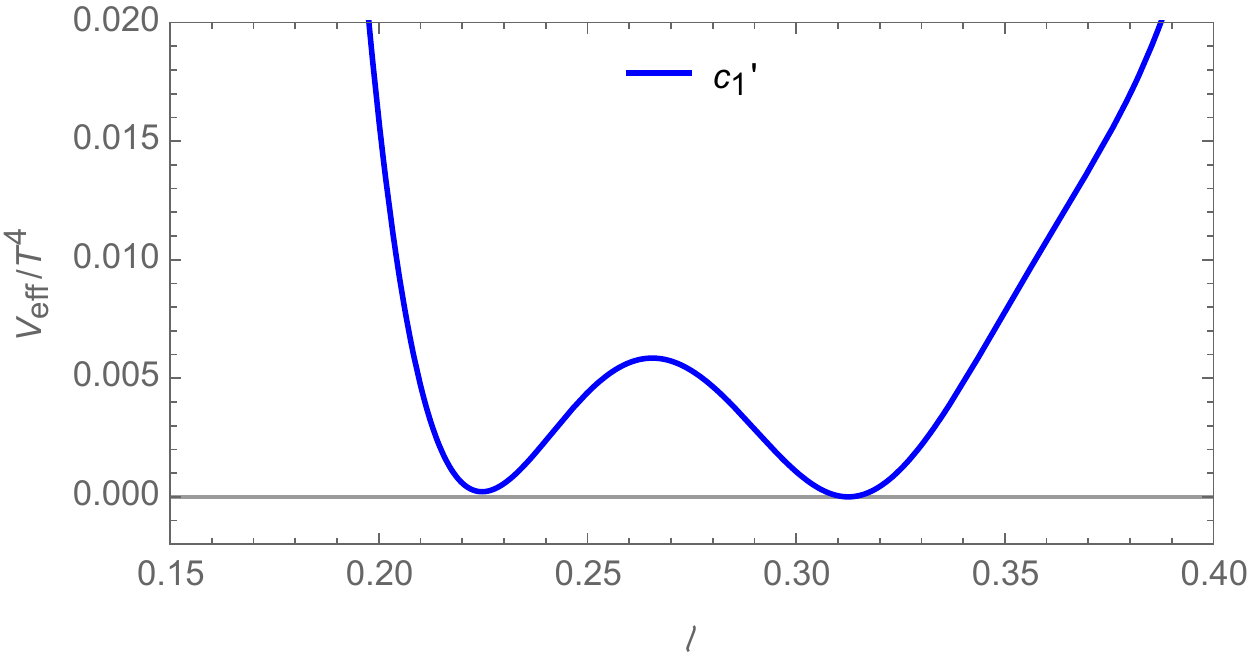}
\caption{ 
The potential $V_\text{eff}(L,\chi,T)/T^4$ at $T_{c,3}/\Lambda_H=14.38$ as a function of $\chi$ and $\ell$.
The top panel is a contour plot on $\chi$--$\ell$ plane. The green line ($c_1'$) links the degenerated vacua \eqref{min3-chiPol-1} and \eqref{min3-chiPol-2} which are shown by the red points.
The bottom panels show the effective potential along the line $c_1'$ as a function of $\chi$ (left) and $\ell$ (right).  
}
\label{fig:VSandglue3} 
\includegraphics[width=8cm]{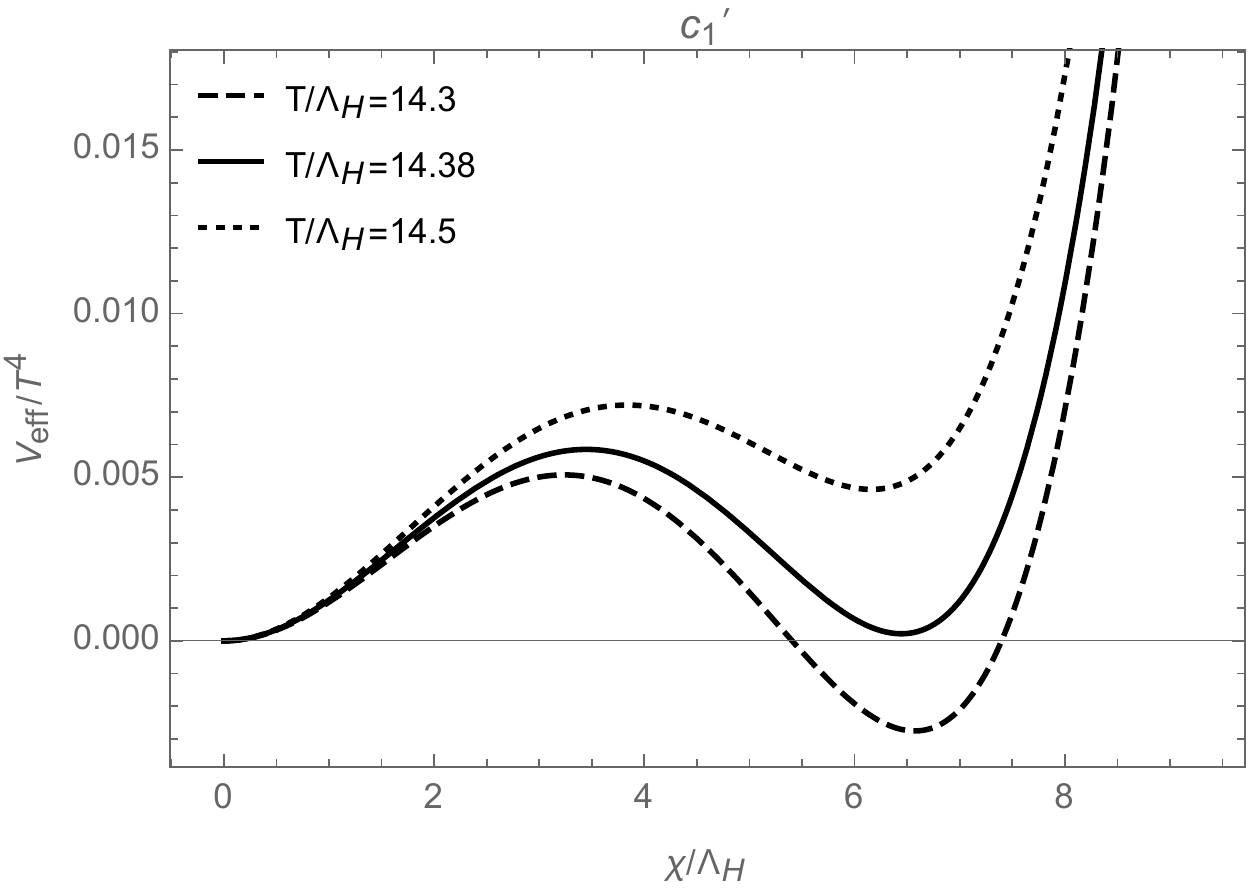}
\includegraphics[width=8cm]{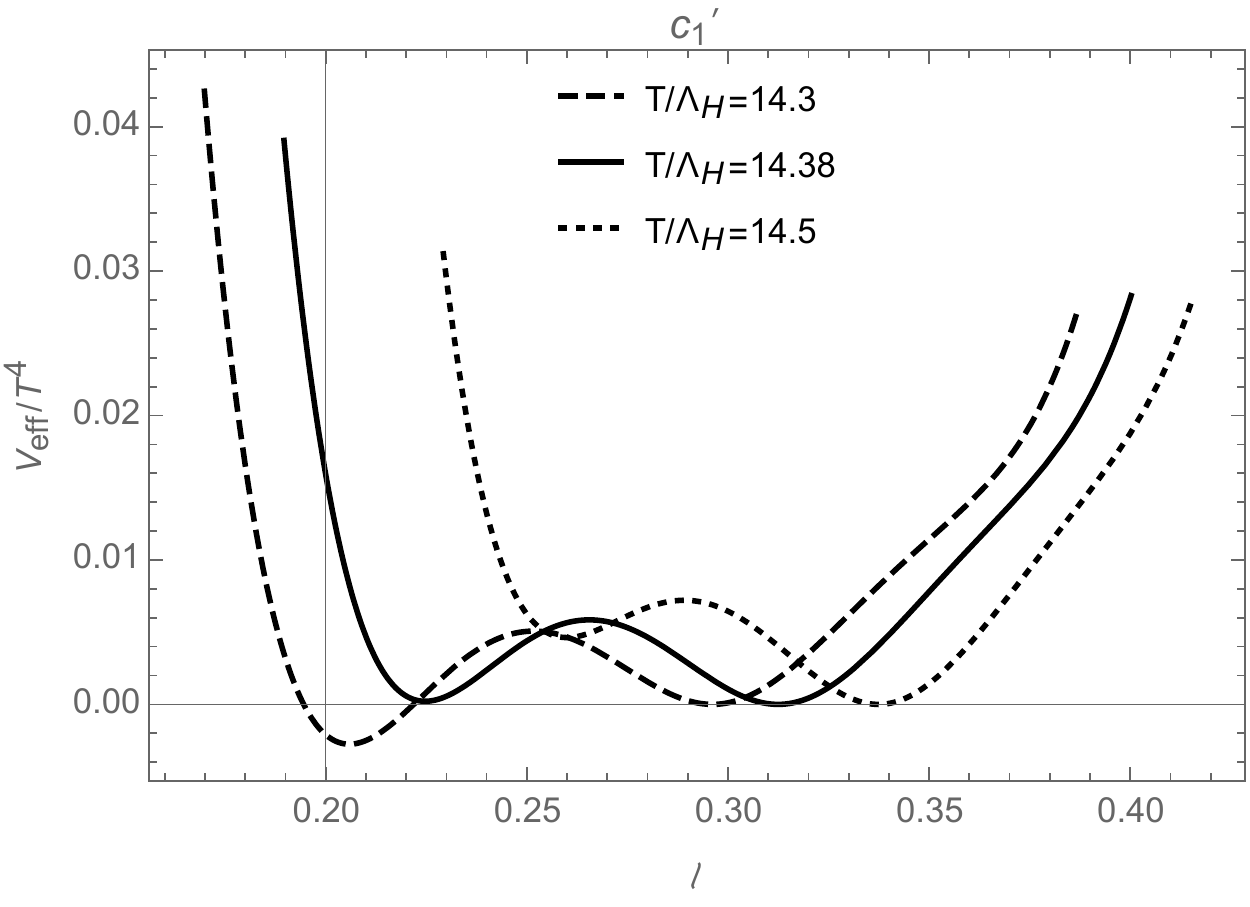}
\caption{ 
The potential $V_\text{eff}(L,\chi,T)/T^4$ around $T_{c,3}/\Lambda_H=14.38$ as a function of $\chi$ (left) and $\ell$ (right) on the line $c_1'$ linking the two minimum points given in \eqref{min3-chiPol-1} and \eqref{min3-chiPol-2}.
}
\label{fig:V3 change T} 
\end{figure}

To show behaviors of the effective potential, we here use \eqref{parameter set} and the choice
\al{
T_{c,4}/\Lambda_H&=14.38,&
a_4/\Lambda_H&=40,&
\label{temperature and a parameter for n3}
}
which yield the degenerated vacua
\al{
&\text{(i) broken phase}:&
&\chi/\Lambda_H=6.44464, &
&\ell=0.22473, &
\label{min3-chiPol-1}
\\
&&  && &(x_1=-0.16291)&\nn[10pt]
&\text{(ii) symmetric phase}: &
& \chi/\Lambda_H=0,&
&\ell=0.31247,&
\label{min3-chiPol-2}
\\
&&  && & (x_1=-0.03129).& \nonumber
}
In order to plot the effective potential, let us define two lines linking the vacua \eqref{min3-chiPol-1} and \eqref{min3-chiPol-2}:
\al{
&c_1':&
&\chi=6.4446 t,&
&x_1= -0.13162t-0.03129.&
}
The effective potential along these lines is shown in Fig.\,\ref{fig:VSandglue3}.
Fig.\,\ref{fig:V3 change T} shows the effective potential around the critical temperature \eqref{temperature and a parameter for n3}.\\

\noindent
{\underline{$\bvec{N=4}$}}\\

\begin{figure}
\includegraphics[width=5.9cm]{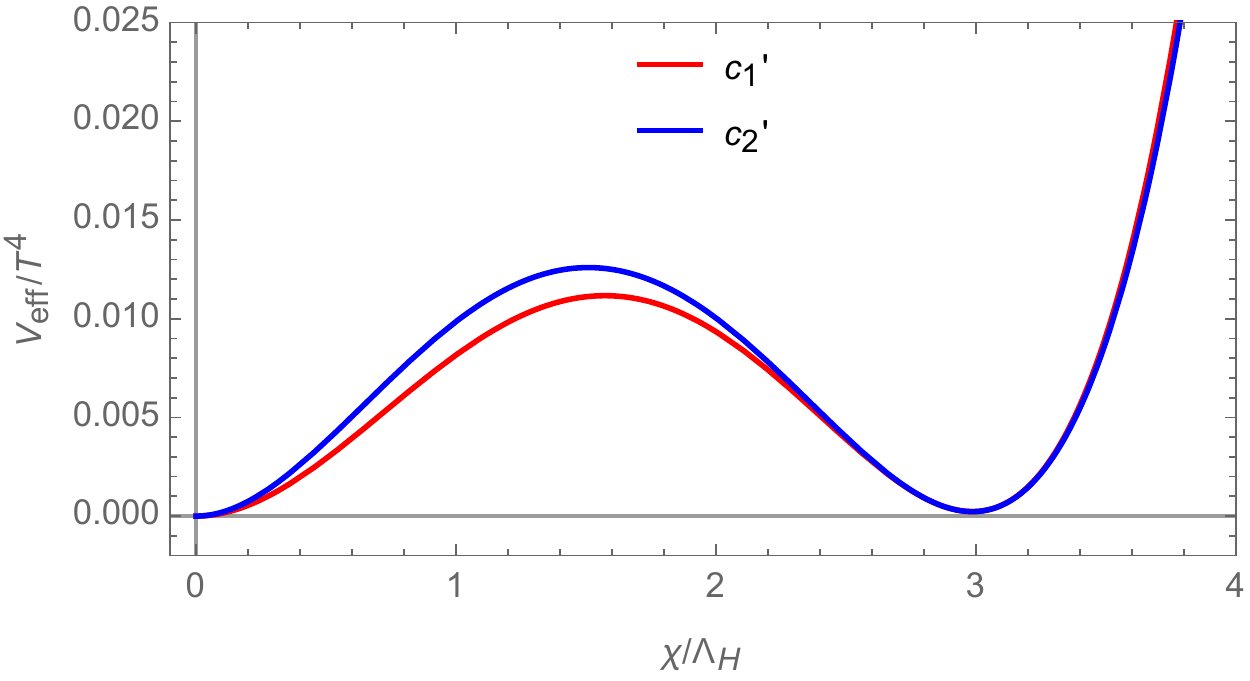}
\includegraphics[width=5.9cm]{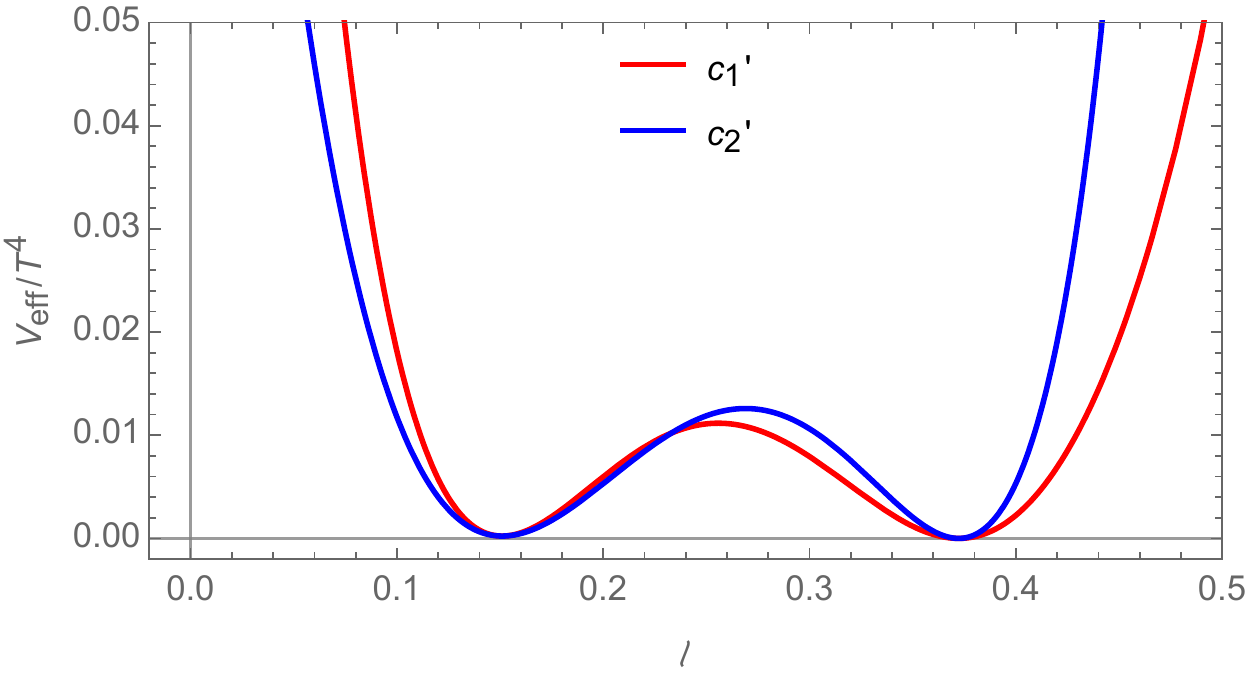}
\includegraphics[width=5.9cm]{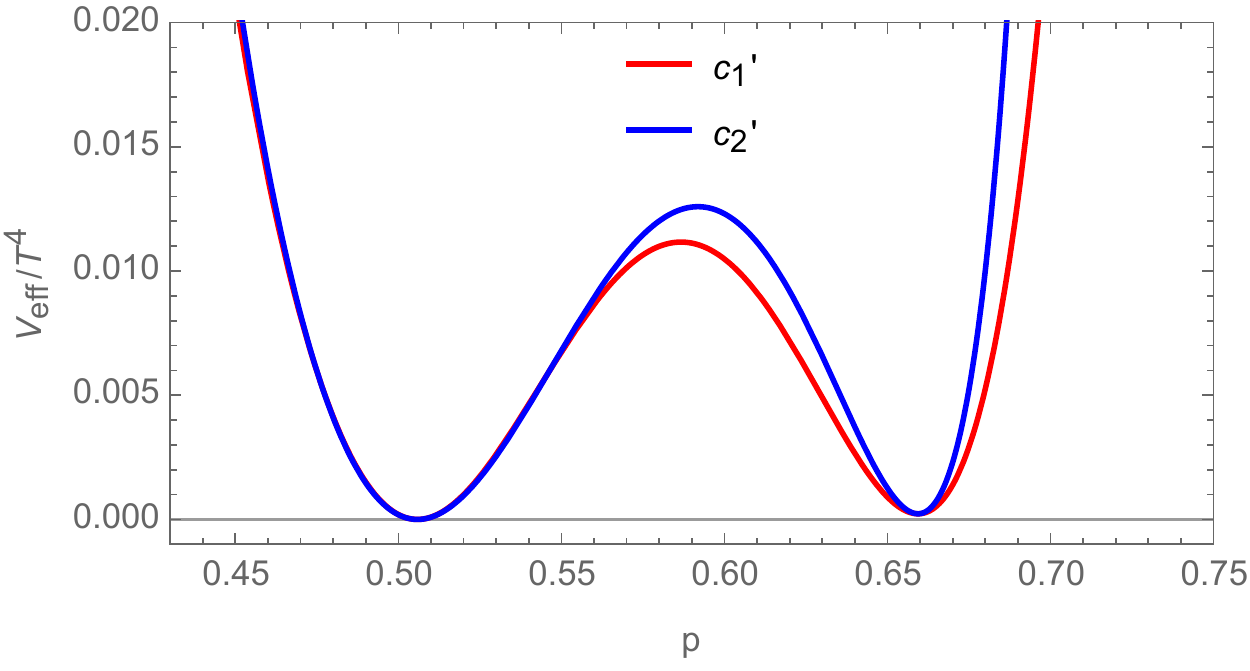}
\caption{ 
The potential $V_\text{eff}(L,\chi,T)/T^4$ at $T_{c,4}/\Lambda_H=6.670$ as a function of $\chi$ (left), $\ell$ (center) and $p$ (right) on the two different lines $c_1'$ (red) and $c_2'$ (blue) linking the two minimum points given in \eqref{min4-chiPol-1} and \eqref{min4-chiPol-2}.
}
\label{fig:VSandglue4} 
\includegraphics[width=5.9cm]{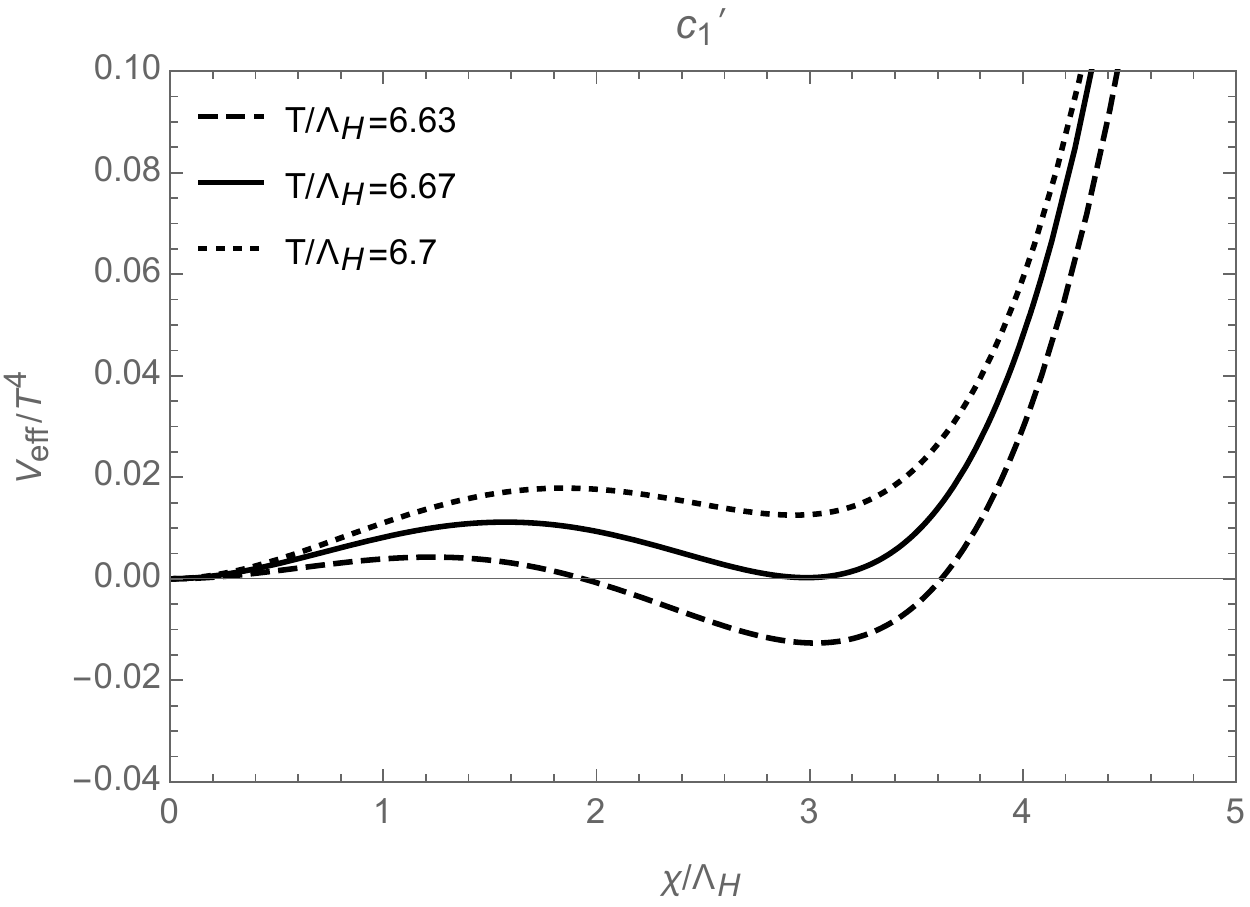}
\includegraphics[width=5.9cm]{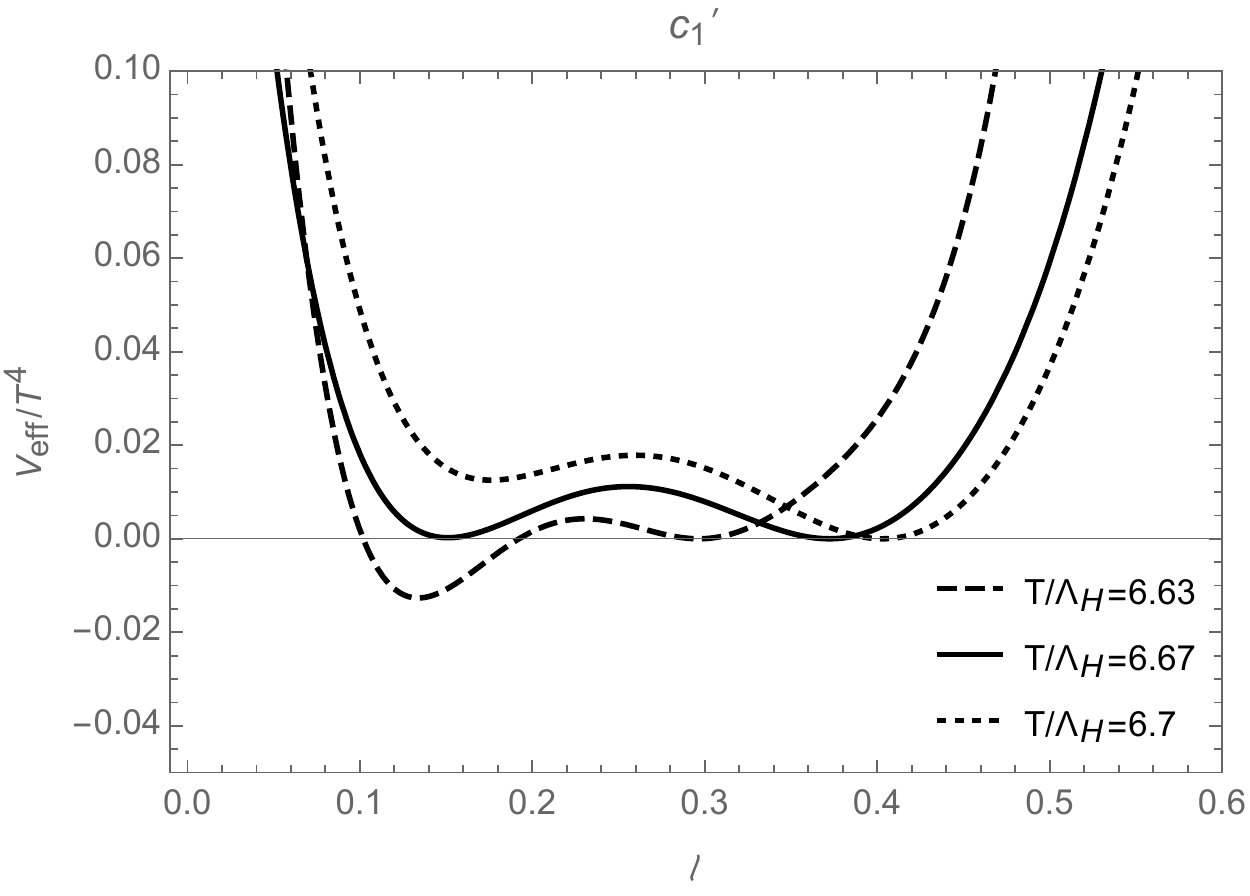}
\includegraphics[width=5.9cm]{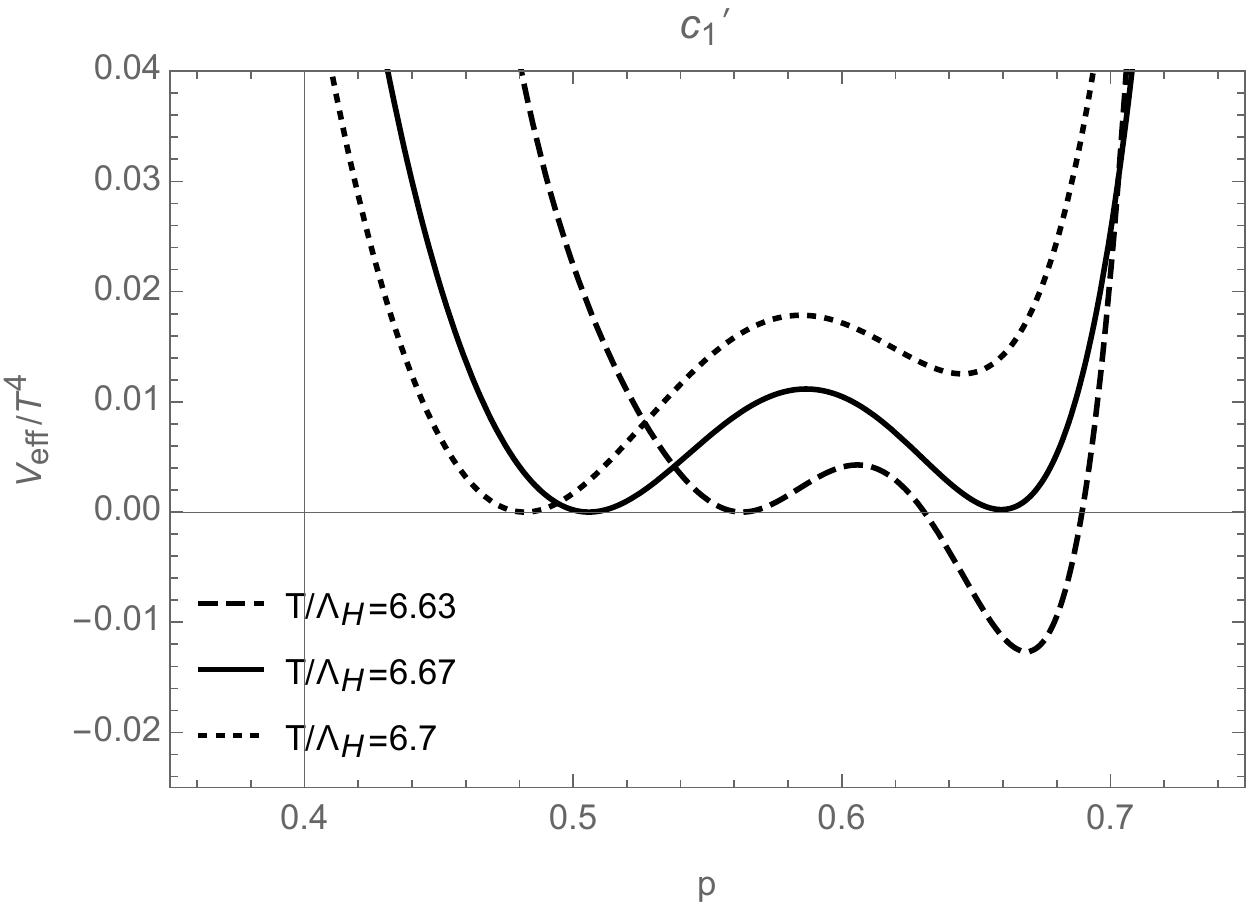}
\includegraphics[width=5.9cm]{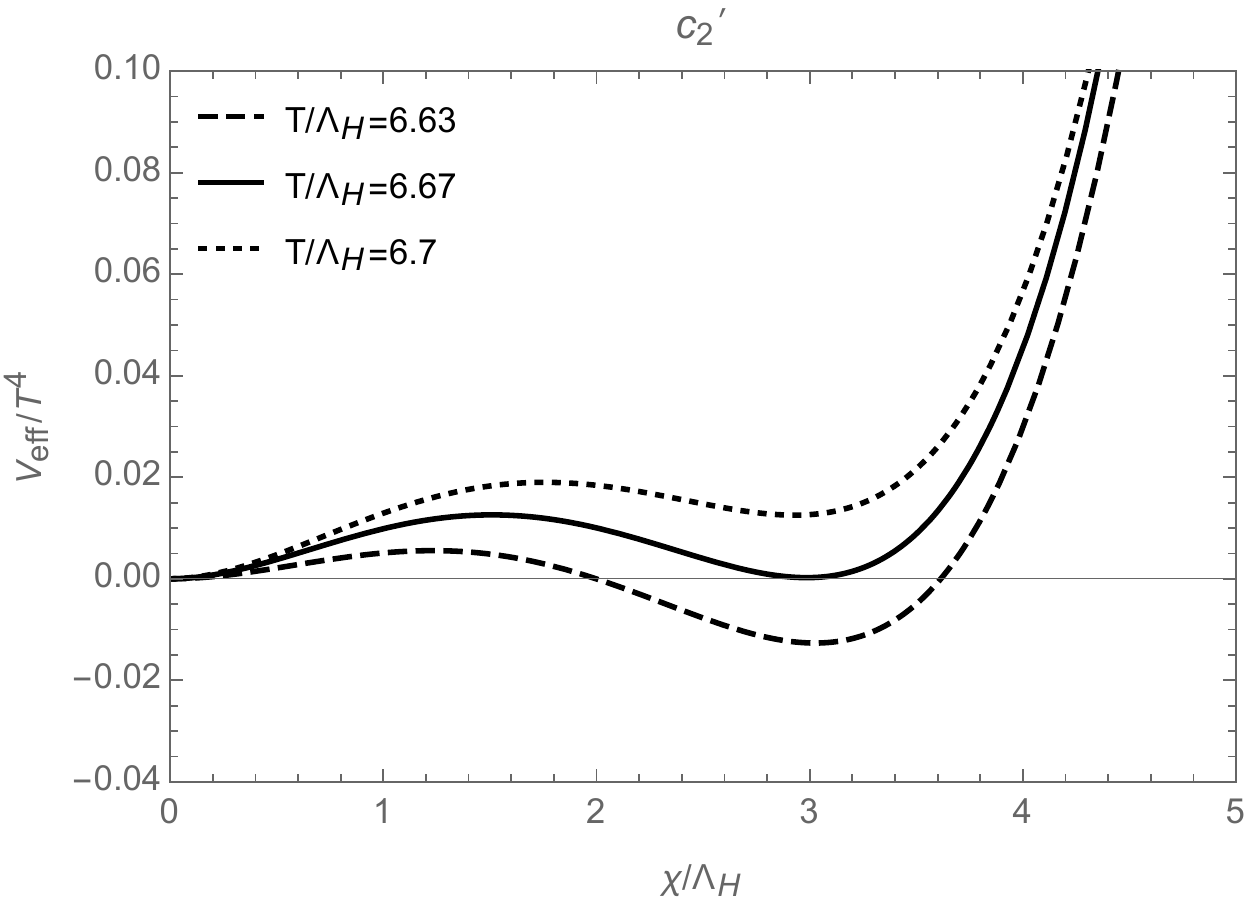}
\includegraphics[width=5.9cm]{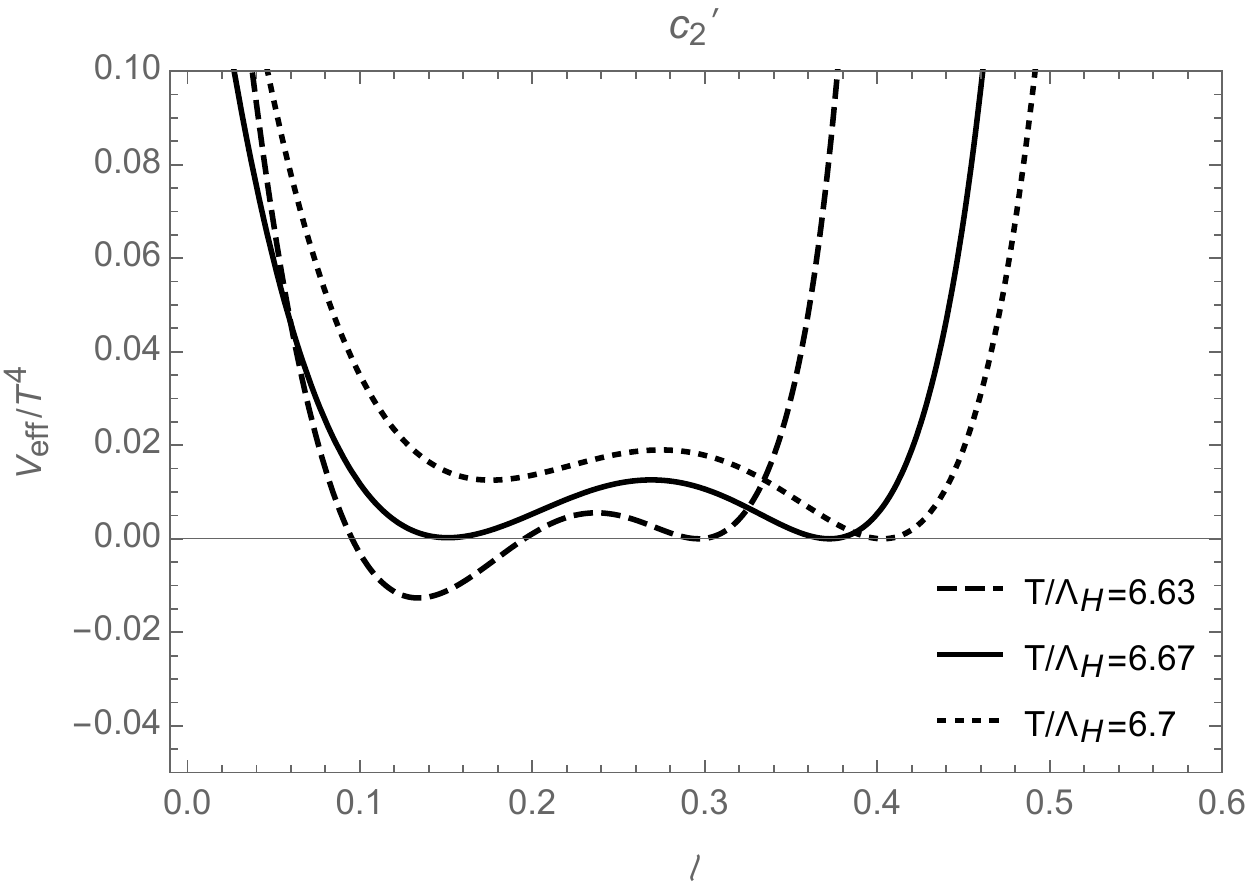}
\includegraphics[width=5.9cm]{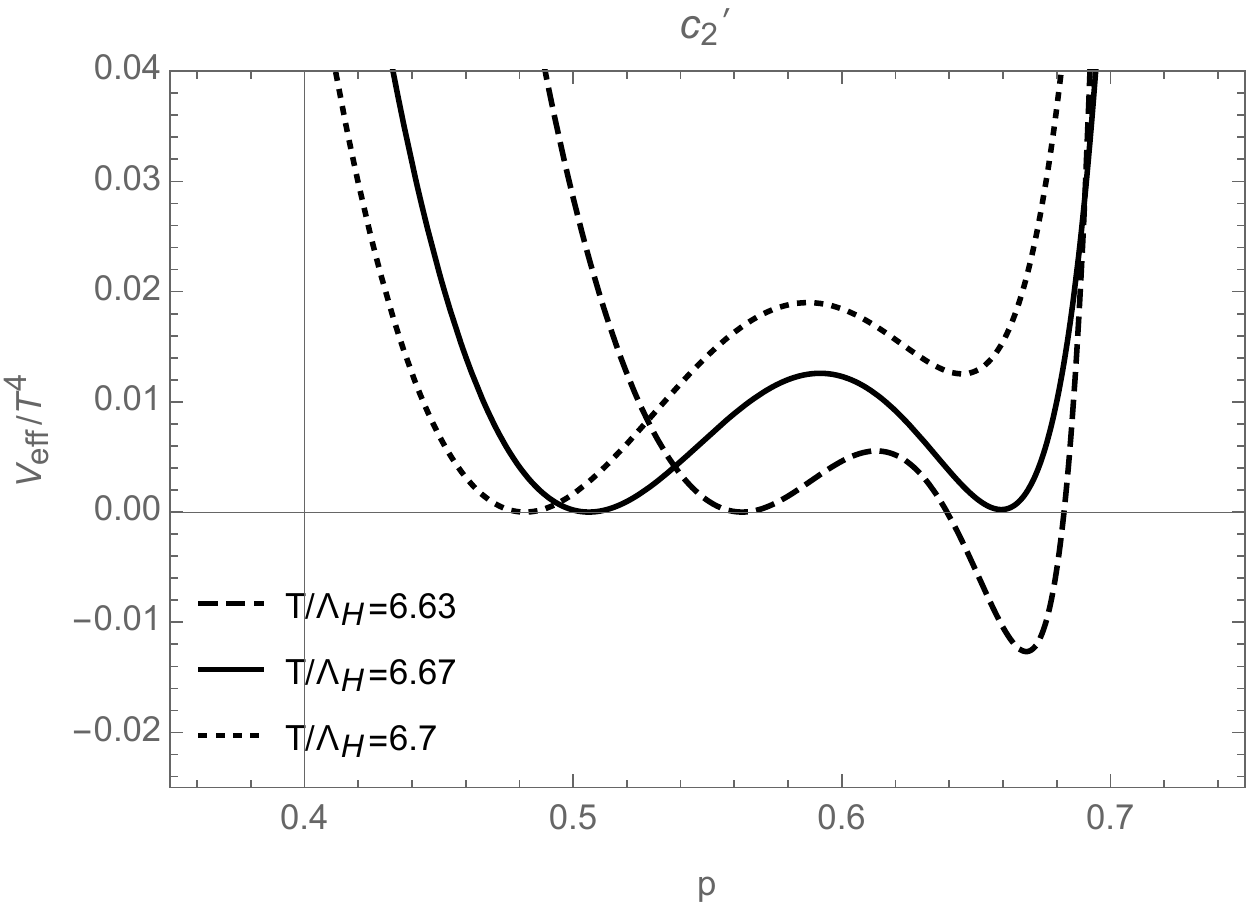}
\caption{ 
The potential $V_\text{eff}(L,\chi,T)/T^4$ around $T_{c,4}/\Lambda_H=6.670$ as a function of $\chi$ (left), $\ell$ (center) and $p$ (right) on the two different lines $c_1'$ and $c_2'$ linking the two minimum points given in \eqref{min4-chiPol-1} and \eqref{min4-chiPol-2}.
}
\label{fig:V4 change T} 
\end{figure}

As an example, we use \eqref{parameter set} and the choice
\al{
T_{c,4}/\Lambda_H&=6.670,&
a_4/\Lambda_H&=16.&
\label{temperature and a parameter for n4}
}
This set yields the following vacua:
\al{
&\text{(i) broken phase}:&
&\chi/\Lambda_H=2.9857, &
&\ell=0.1510, &
&p=0.6594, &
\label{min4-chiPol-1}
\\
&&  &(x_1=0.810,& &x_2=-0.5084)&\nn[10pt]
&\text{(ii) symmetric phase}: &
& \chi/\Lambda_H=0,&
&\ell=0.3723,&
&p=0.5059,&
\label{min4-chiPol-2}
\\
&&  & (x_1=0.8781,&  &x_2=-0.1336).& \nonumber
}
In order to plot the effective potential, let us define two lines linking the vacua \eqref{min4-chiPol-1} and \eqref{min4-chiPol-2}:
\al{
&c_1':&
&\chi=2.9857 t,&
&x_1= - 0.06775 t+0.8781,&
&x_2= - 0.3748 t-0.1336,&\\
&c_2':&
&\chi=2.9857 t,&
&x_1=  - 0.06775t^2 +0.8781,&
&x_2= - 0.3748 t-0.1336.&
}
The effective potential along these lines is shown in Fig.\,\ref{fig:VSandglue4}.
In Fig.\,\ref{fig:V4 change T} we plot the effective potential around the critical temperature \eqref{temperature and a parameter for n4}.\\

\noindent
{\underline{$\bvec{N=5}$}}\\

\begin{figure}
\includegraphics[width=5.9cm]{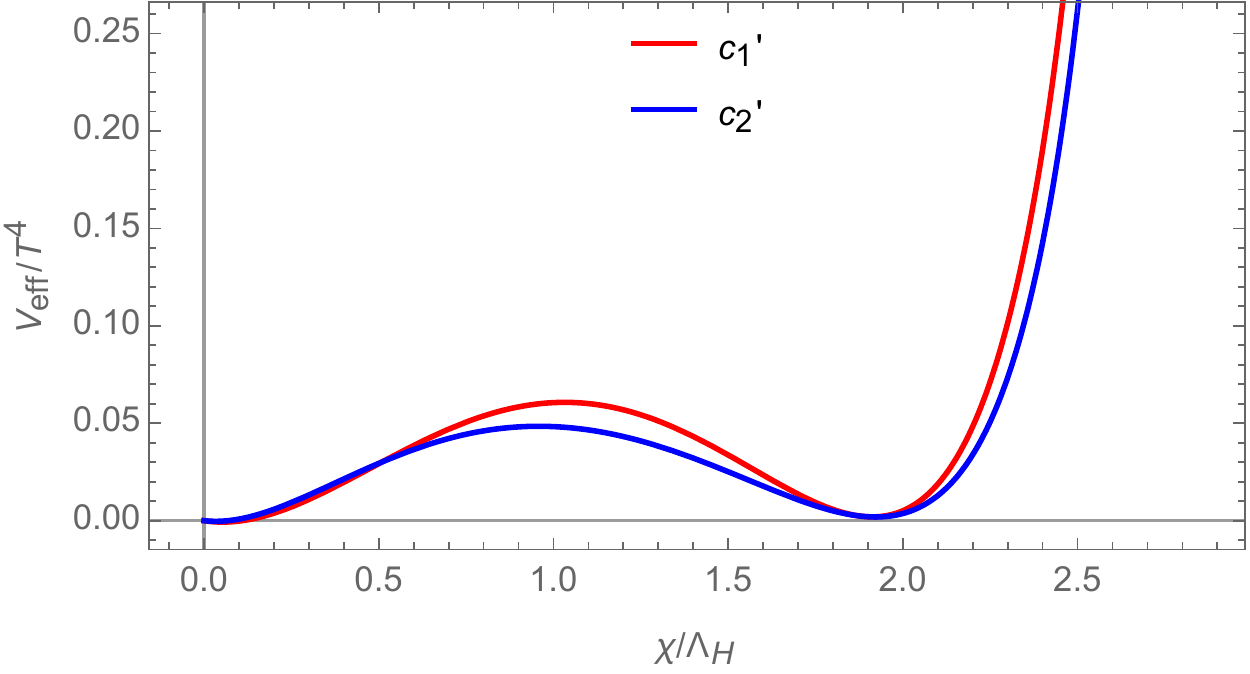}
\includegraphics[width=5.9cm]{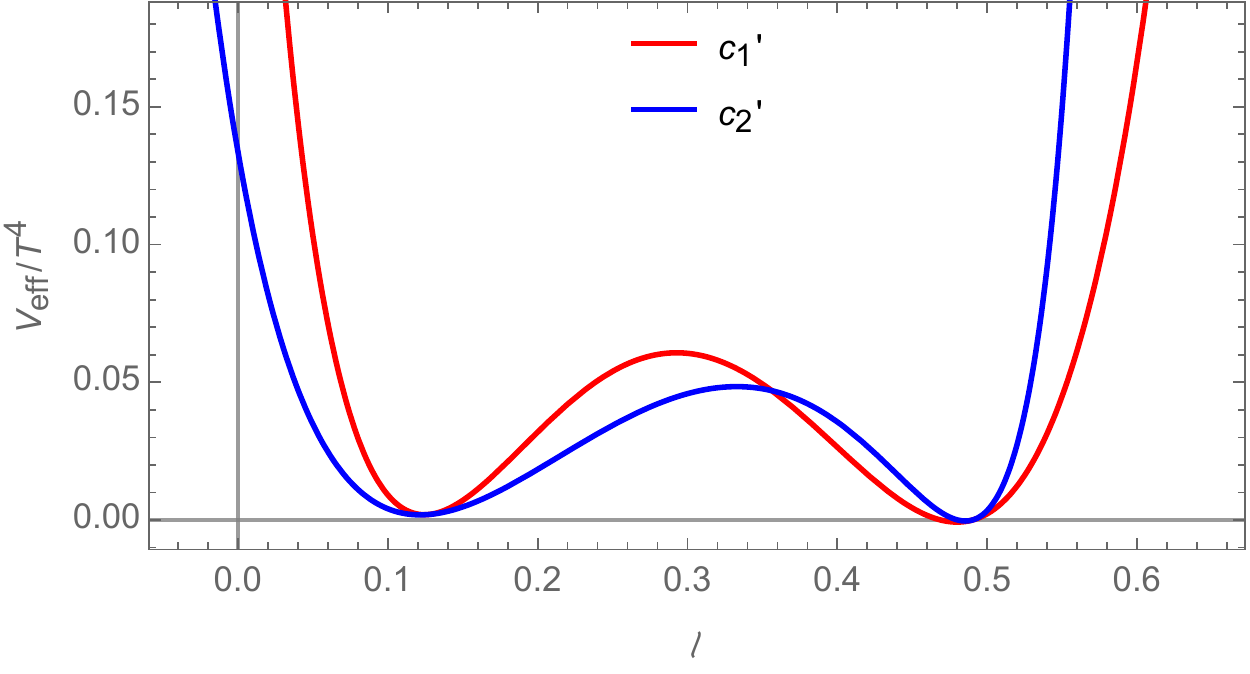}
\includegraphics[width=5.9cm]{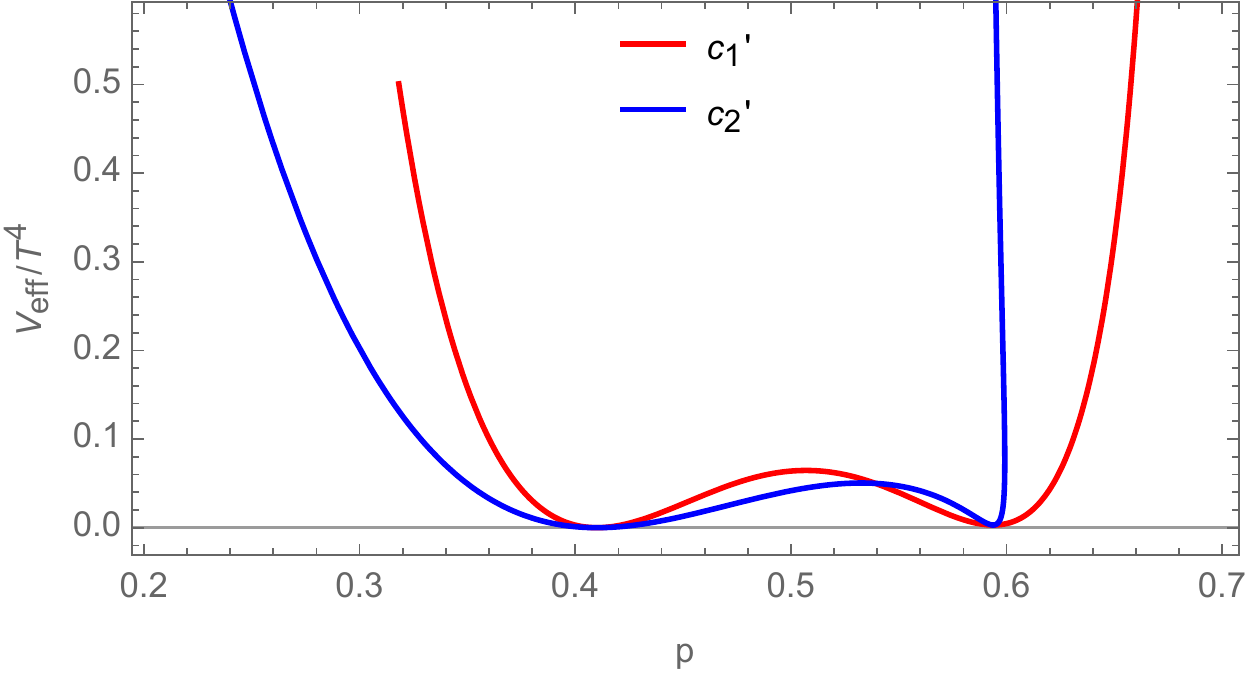}
\caption{ 
The potential $V_\text{eff}(L,\chi,T)/T^4$ at $T_{c,5}/\Lambda_H=3.615$ as a function of $\chi$ (left), $\ell$ (center) and $p$ (right) on the two different lines $c_1'$ (red) and $c_2'$ (blue) linking the two minimum points given in \eqref{min5-chiPol-1} and \eqref{min5-chiPol-2}.
}
\label{fig:VSandglue5} 
\includegraphics[width=5.9cm]{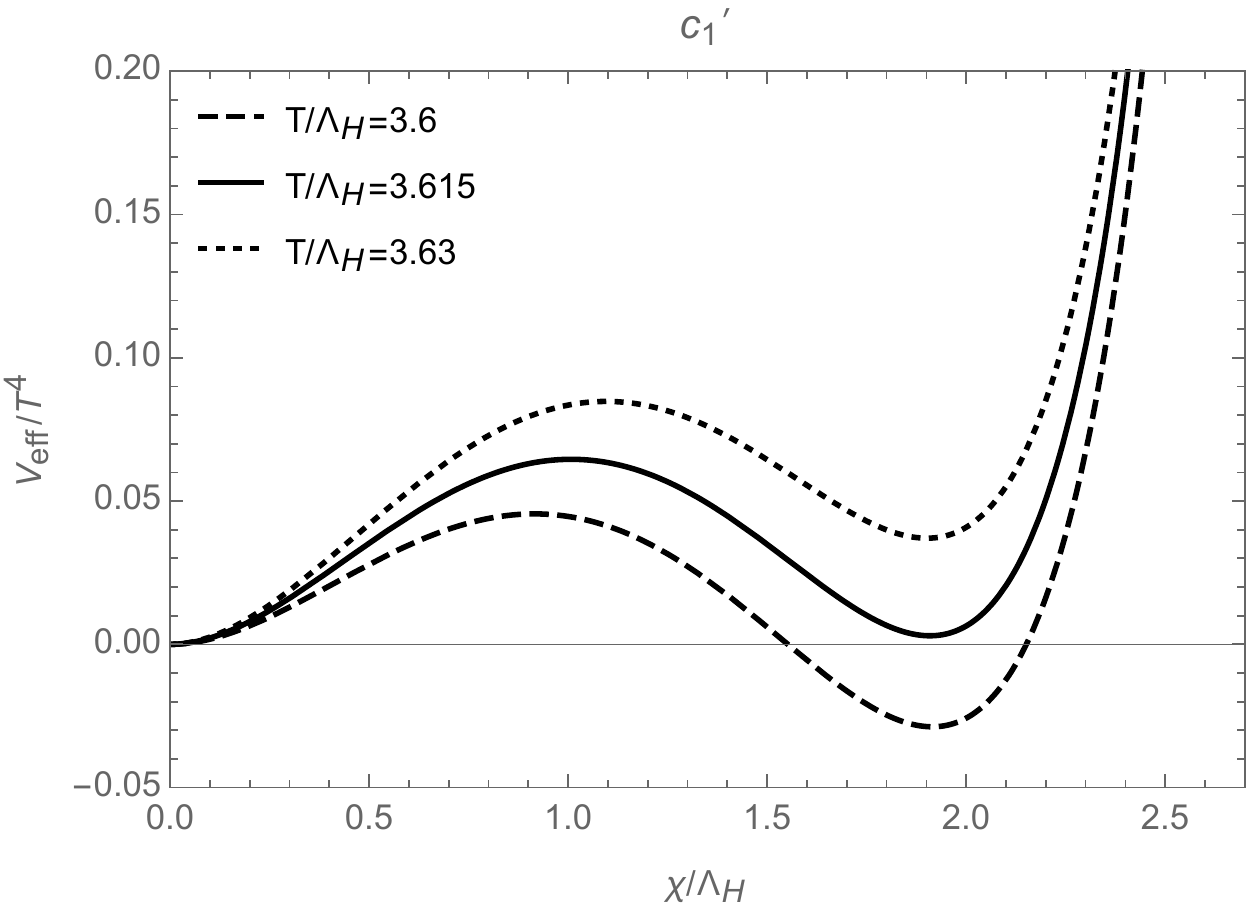}
\includegraphics[width=5.9cm]{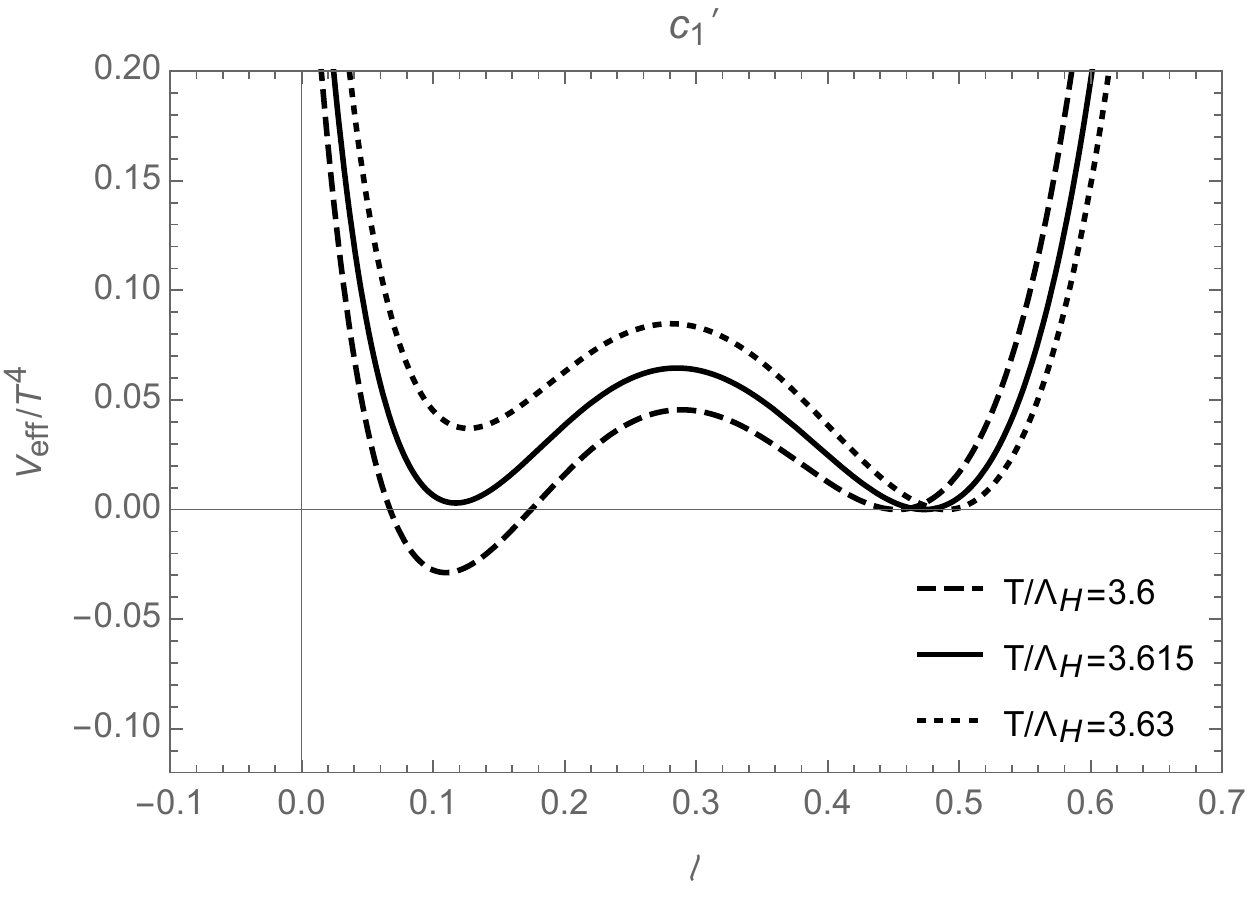}
\includegraphics[width=5.9cm]{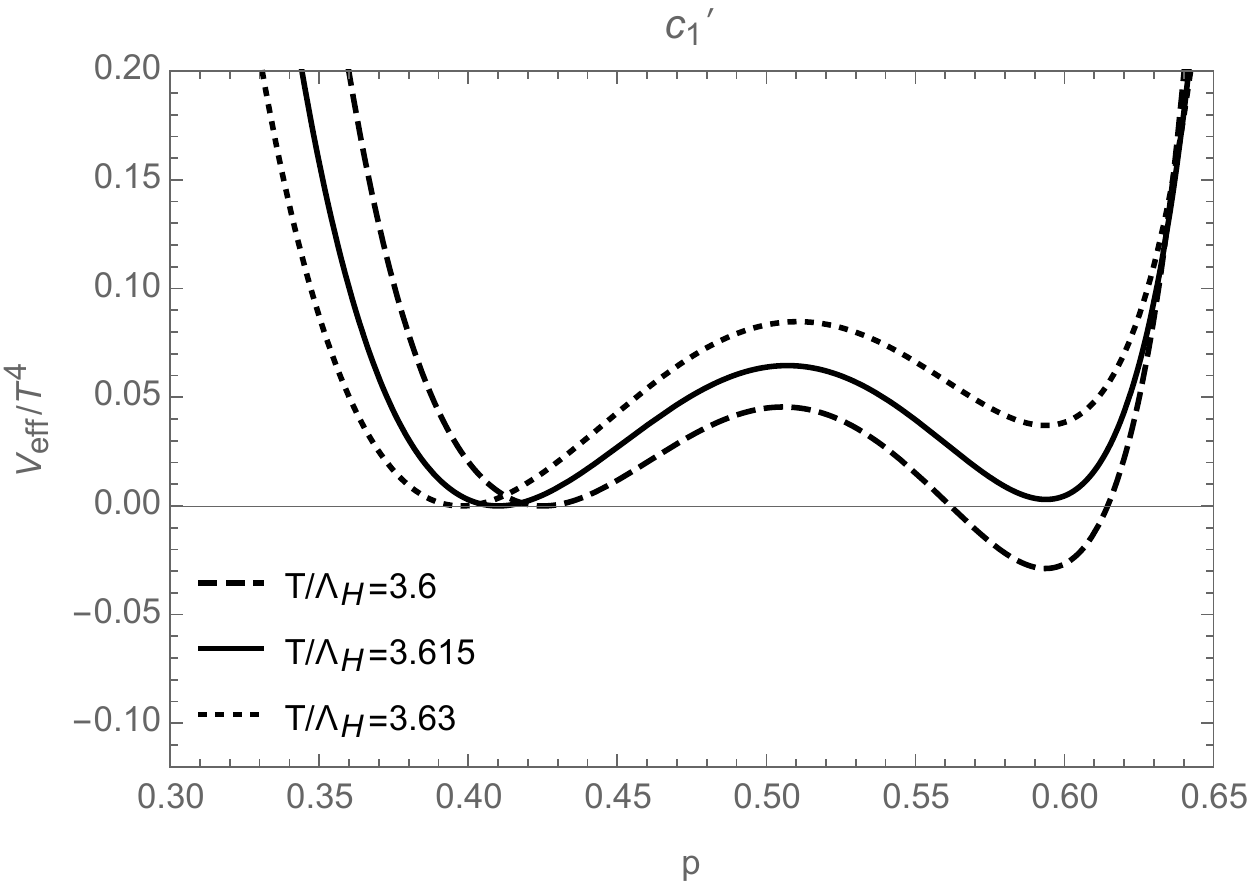}
\includegraphics[width=5.9cm]{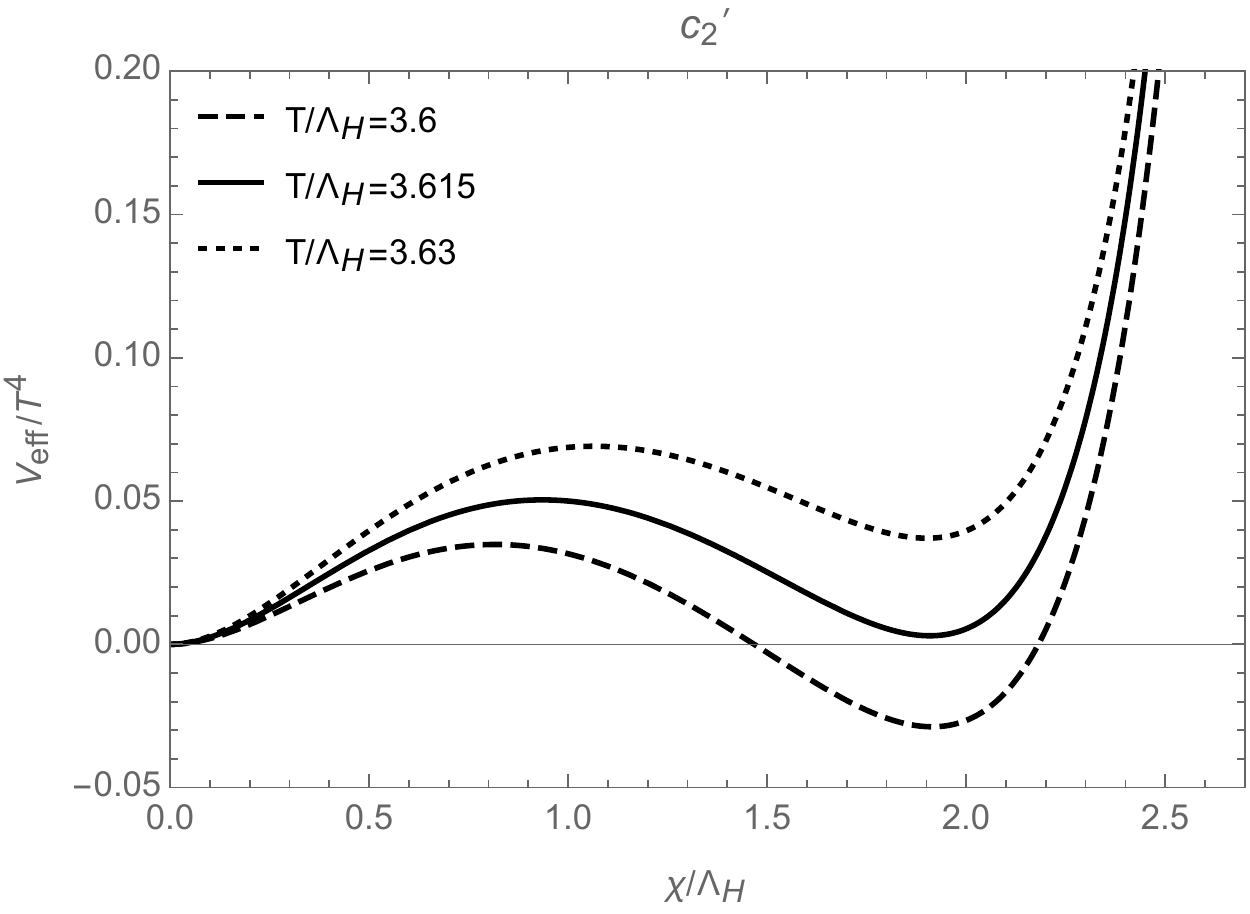}
\includegraphics[width=5.9cm]{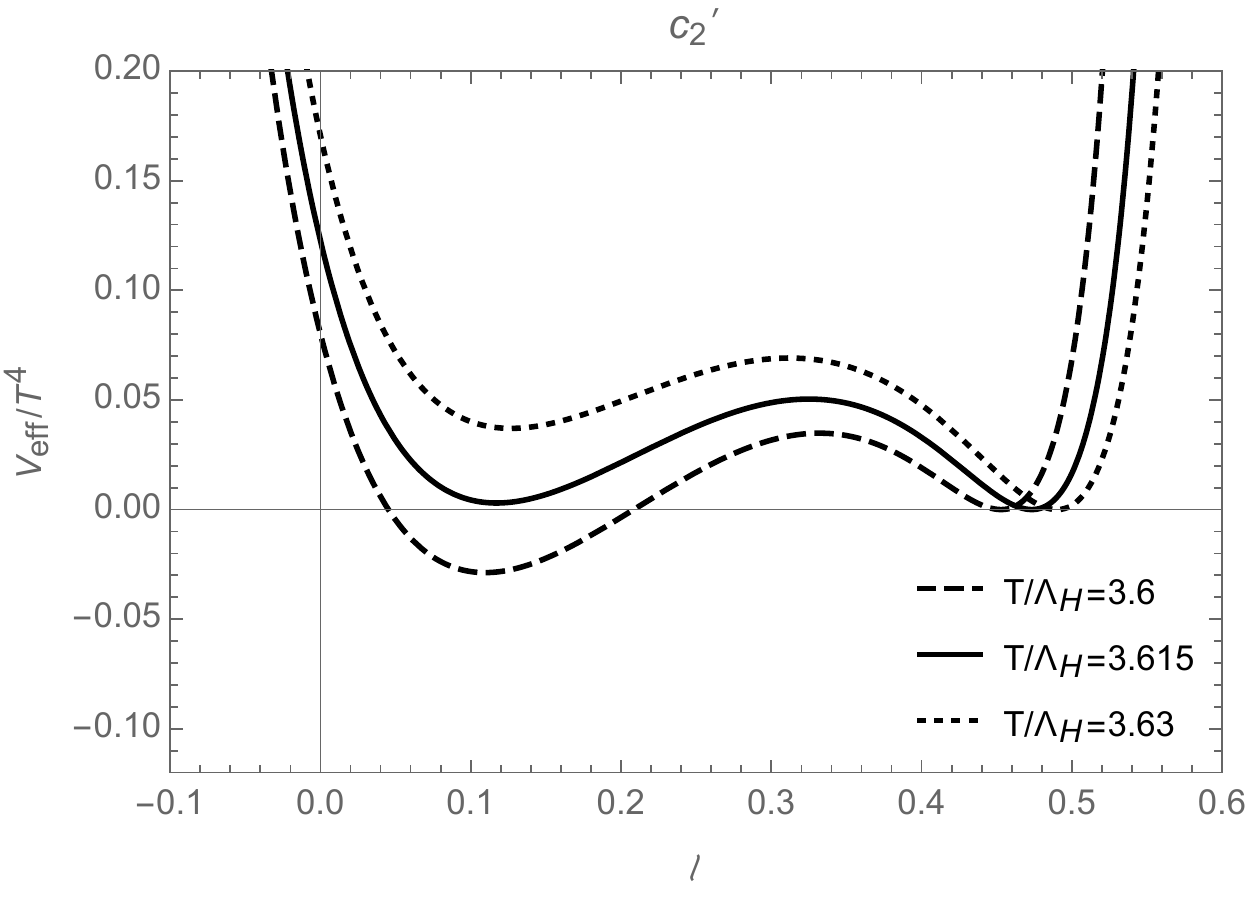}
\includegraphics[width=5.9cm]{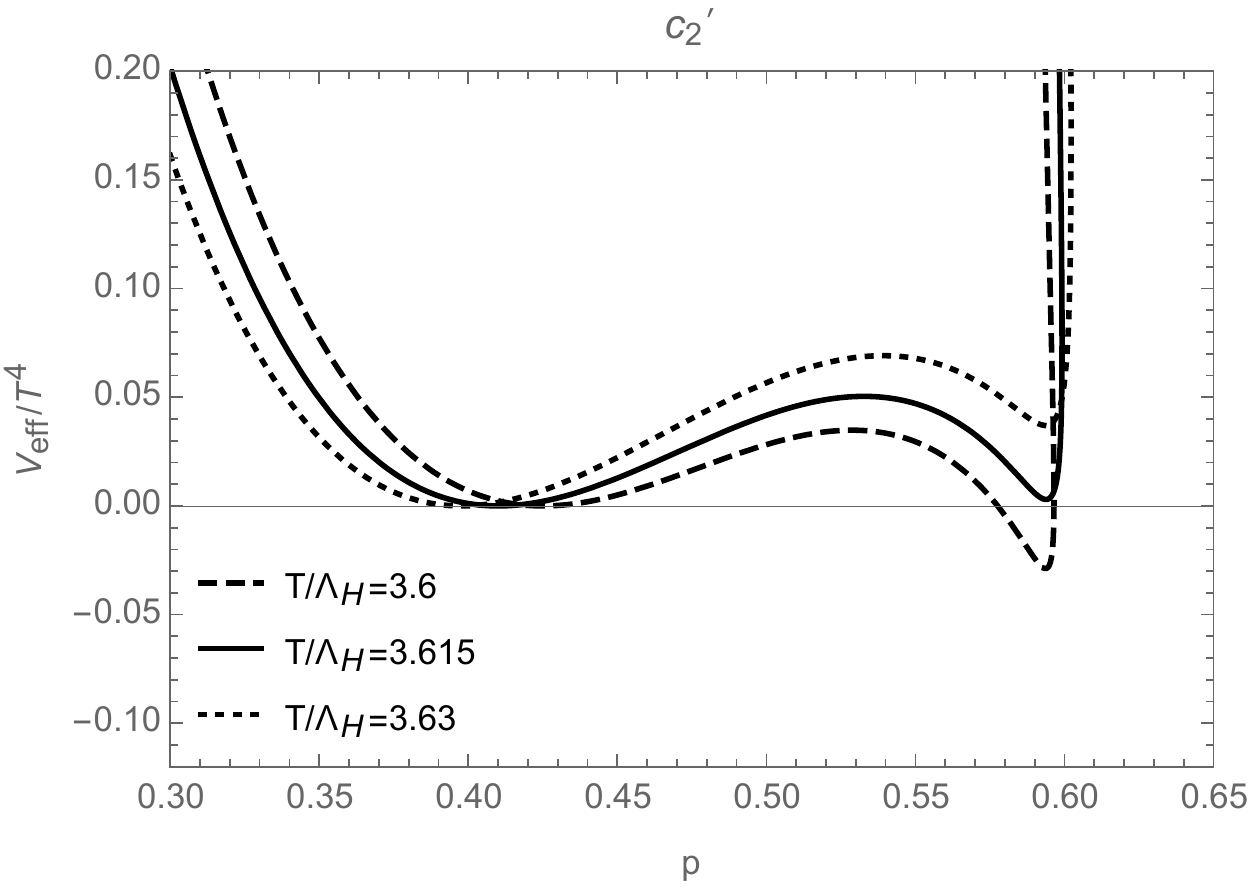}
\caption{ 
The potential $V_\text{eff}(L,\chi,T)/T^4$ around $T_{c,5}/\Lambda_H=3.615$ as a function of $\chi$ (left), $\ell$ (center) and $p$ (right) on the two different lines $c_1'$ and $c_2'$ linking the two minimum points given in \eqref{min5-chiPol-1} and \eqref{min5-chiPol-2}.
}
\label{fig:V5 change T} 
\end{figure}

With the parameter set \eqref{parameter set} and the following choice
\al{
T_{c,5}/\Lambda_H&=3.615,&
a_5/\Lambda_H&=8.0,&
\label{temperature and a parameter for n5}
}
we obtain
\al{
&\text{(i) broken phase}:&
&\chi/\Lambda_H=1.9088, &
&\ell=0.1171, &
&p=0.5938, &
\label{min5-chiPol-1}
\\
&&  &(x_1=0.4902,& &x_2=-0.6974)&\nn[10pt]
&\text{(ii) symmetric phase}: &
& \chi/\Lambda_H=0,&
&\ell=0.4733,&
&p=0.4101,&
\label{min5-chiPol-2}
\\
&&  & (x_1=0.7517,&  &x_2=-0.0684).& \nonumber
}
We make two lines connecting the vacua \eqref{min5-chiPol-1} and \eqref{min5-chiPol-2} as
\al{
&c_1':&
&\chi=1.9088 t,&
&x_1= - 0.2615 t+0.7517,&
&x_2= - 0.6290 t-0.06840,&\\
&c_2':&
&\chi=1.9088 t,&
&x_1=  - 0.2615t^2 +0.7517,&
&x_2= - 0.6290 t-0.06840.&
}
The effective potential along these lines is shown in Fig.\,\ref{fig:VSandglue5}.
In Fig.\,\ref{fig:V5 change T} we plot the effective potential around the critical temperature \eqref{temperature and a parameter for n5}.\\

\noindent
{\underline{$\bvec{N=6}$}}\\

\begin{figure}
\includegraphics[width=5.9cm]{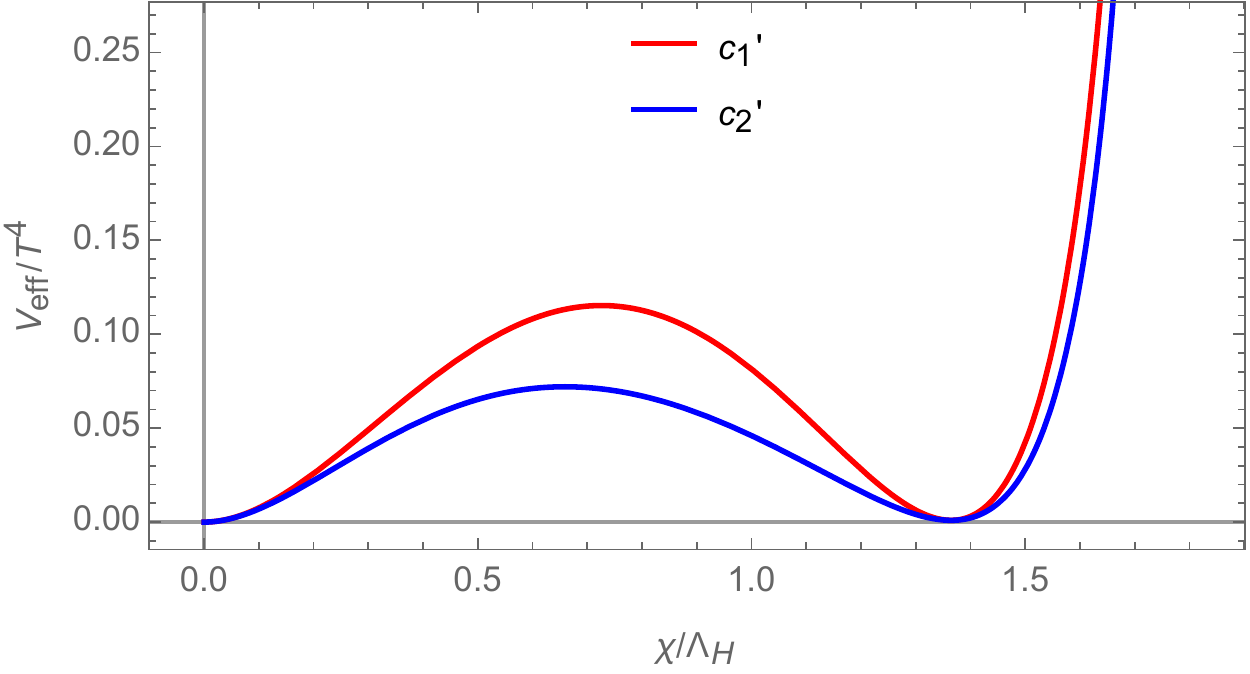}
\includegraphics[width=5.9cm]{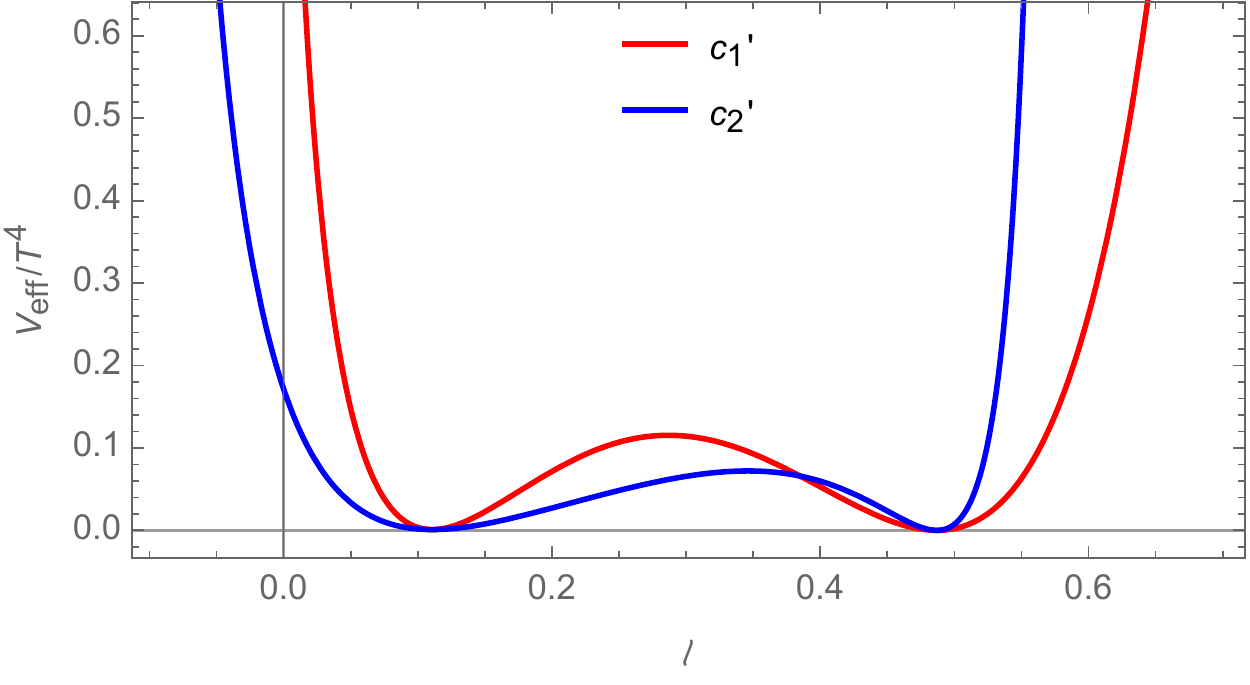}
\includegraphics[width=5.9cm]{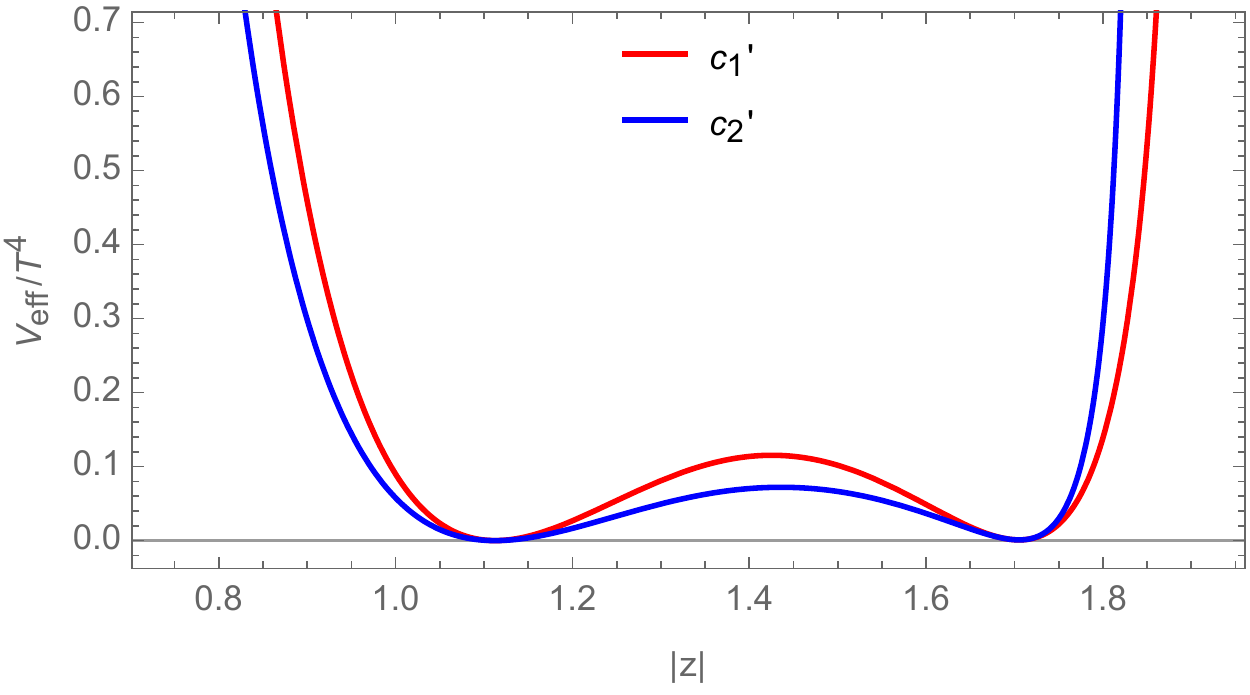}
\caption{ 
The potential $V_\text{eff}(L,\chi,T)/T^4$ at $T_{c,6}/\Lambda_H=2.918$ as a function of $\chi$ (left), $\ell$ (center) and $|z|$ (right) on the two different lines $c_1'$ (red) and $c_2'$ (blue) linking the two minimum points given in \eqref{min6-chiPol-1} and \eqref{min6-chiPol-2}.
}
\label{fig:VSandglue6} 
\includegraphics[width=5.9cm]{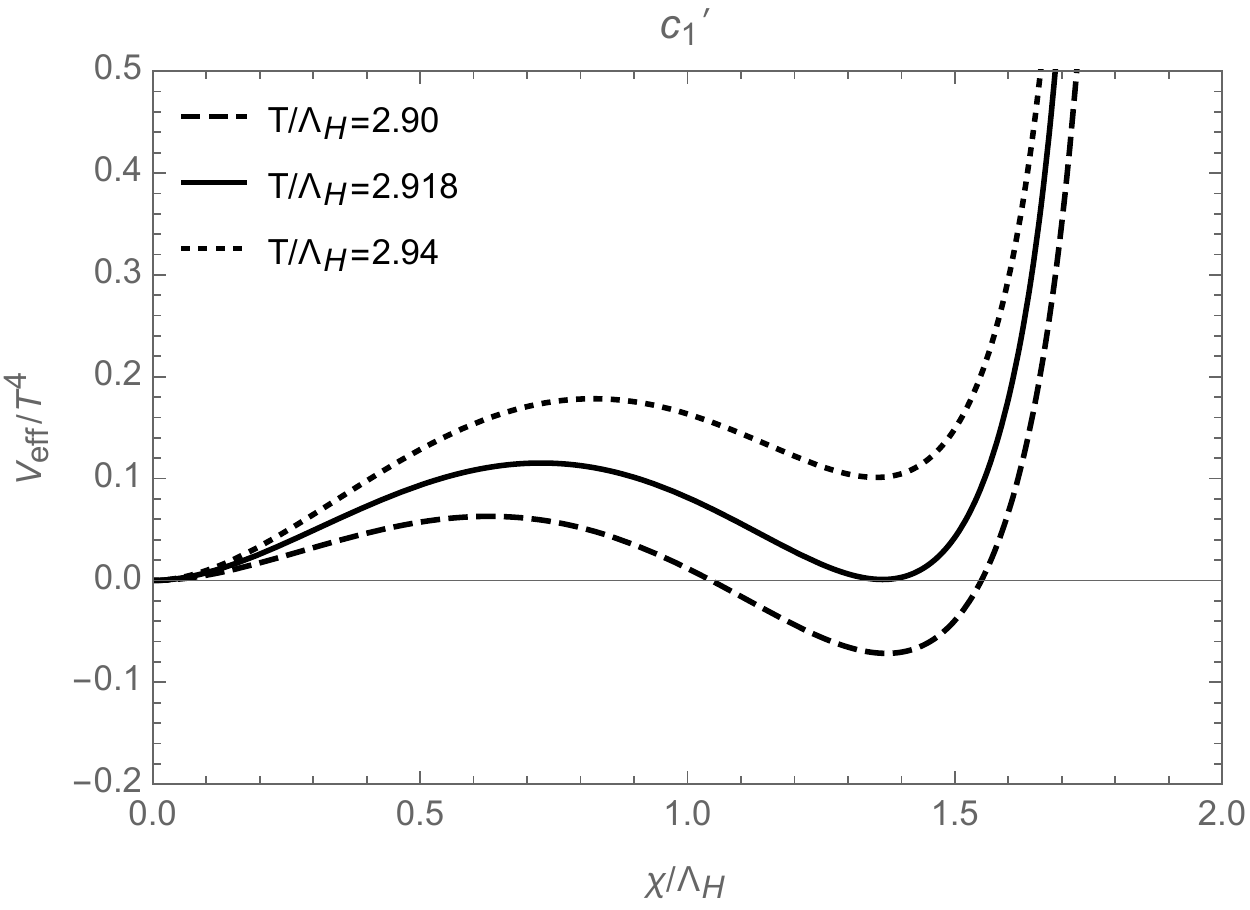}
\includegraphics[width=5.9cm]{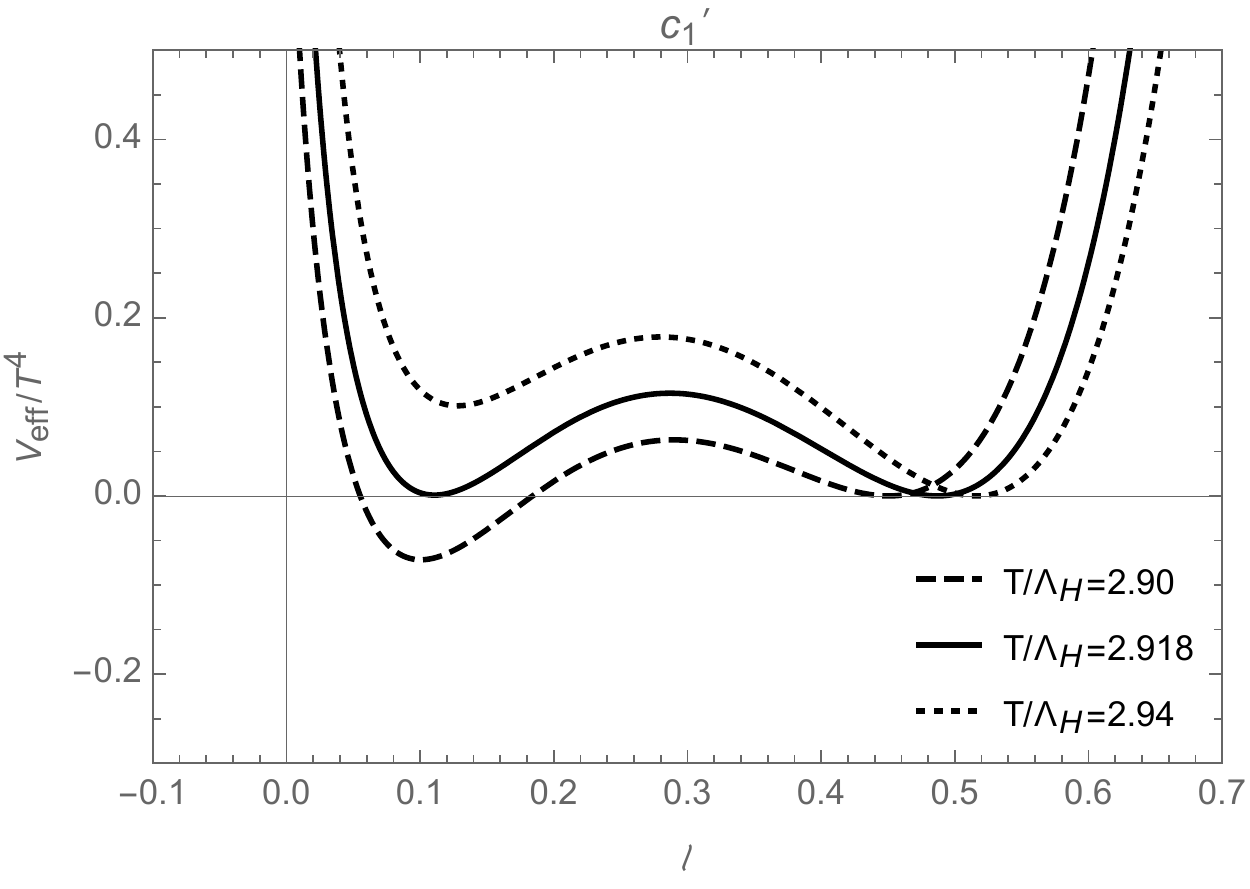}
\includegraphics[width=5.9cm]{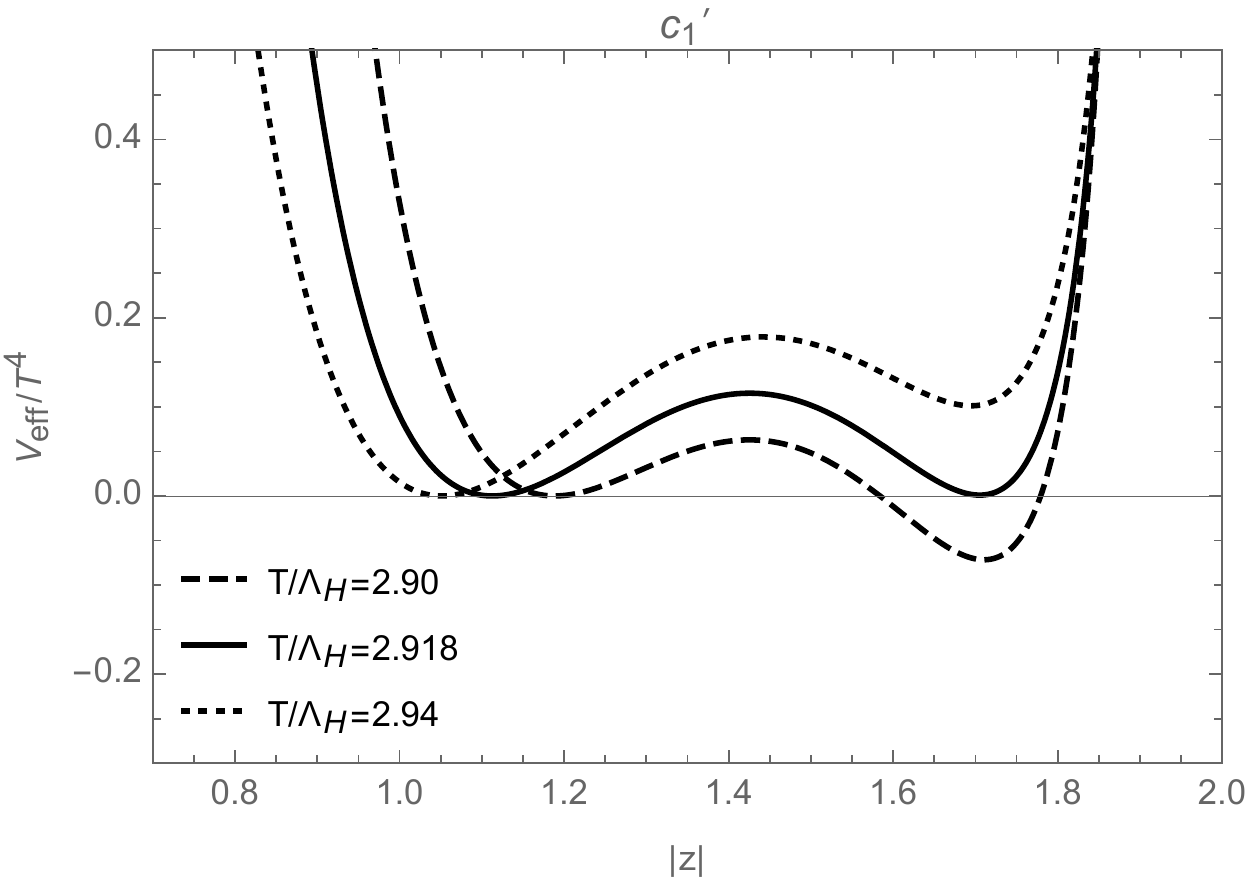}
\includegraphics[width=5.9cm]{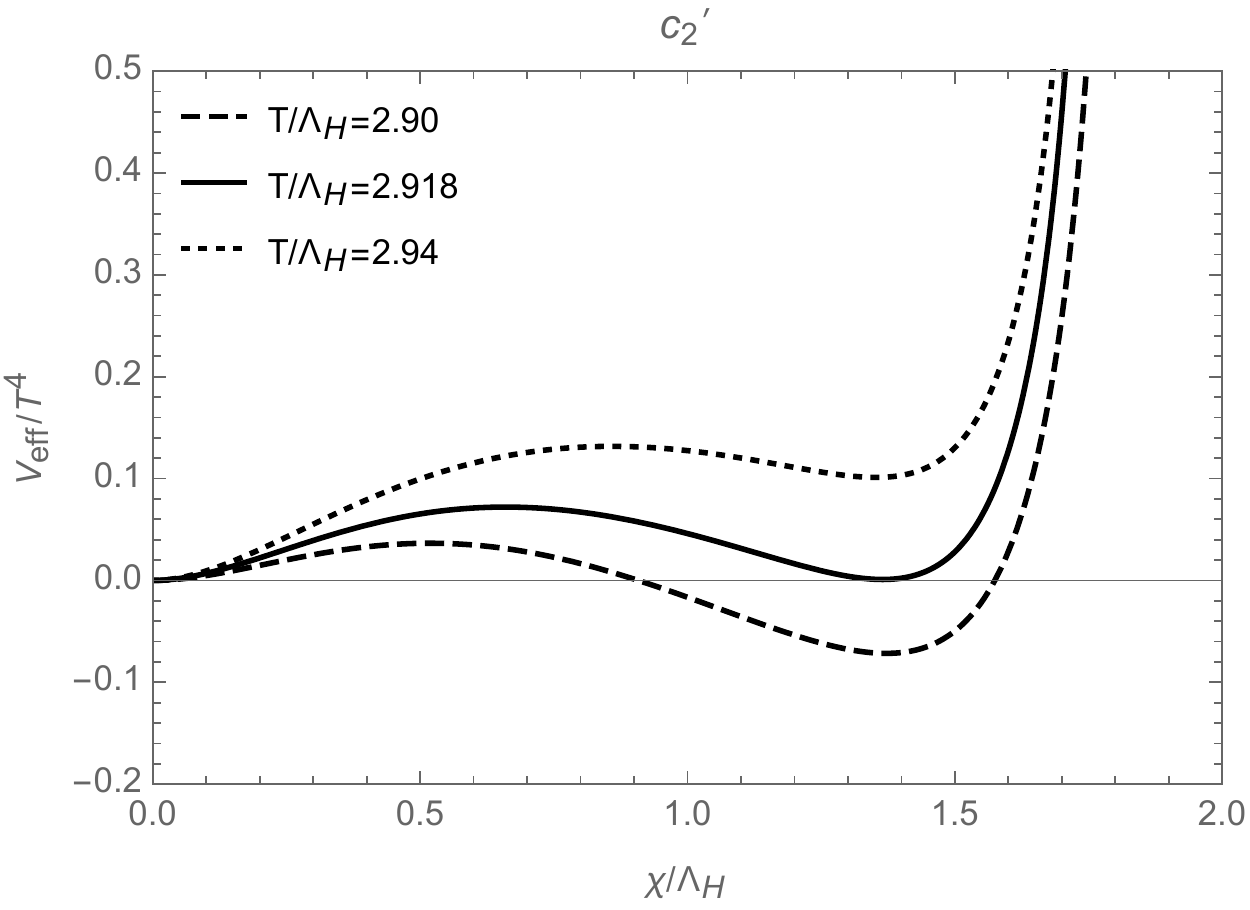}
\includegraphics[width=5.9cm]{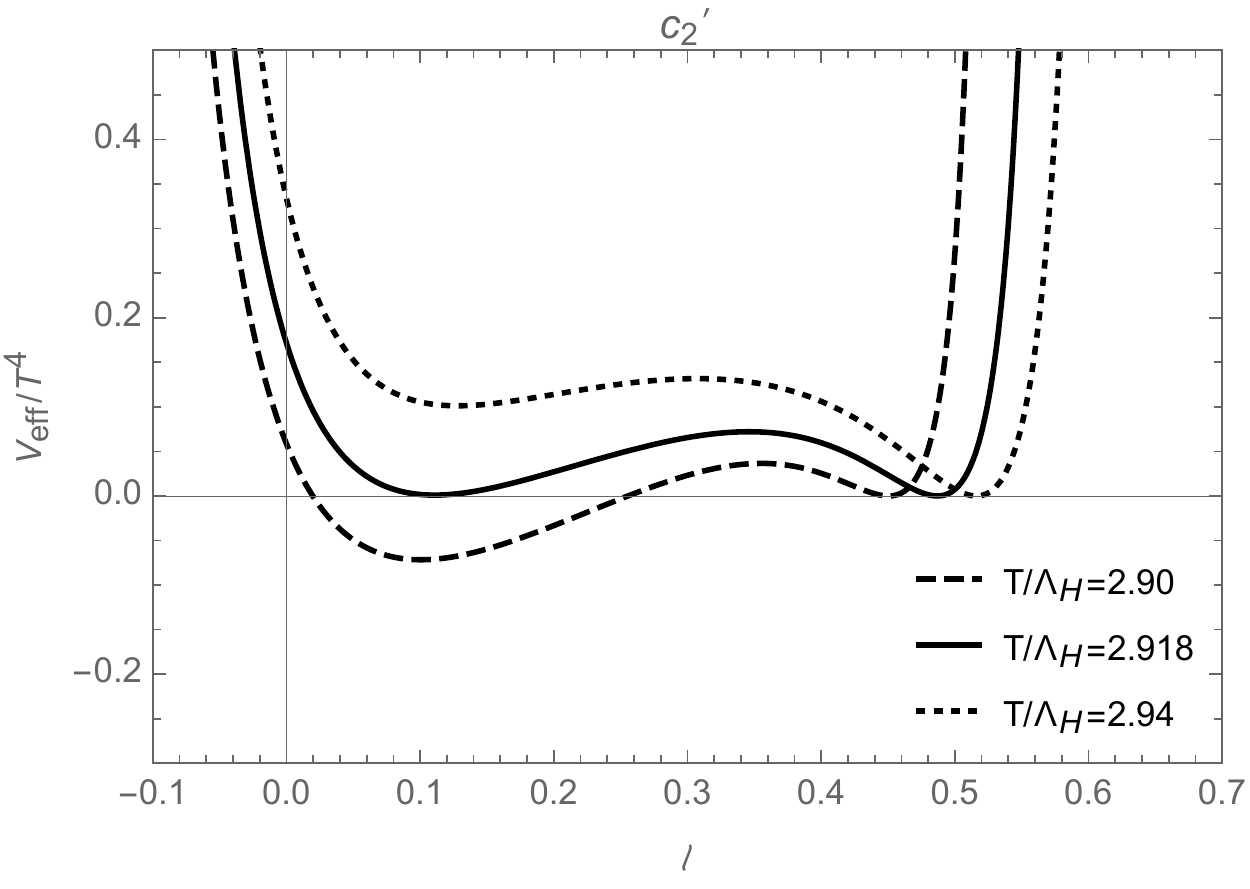}
\includegraphics[width=5.9cm]{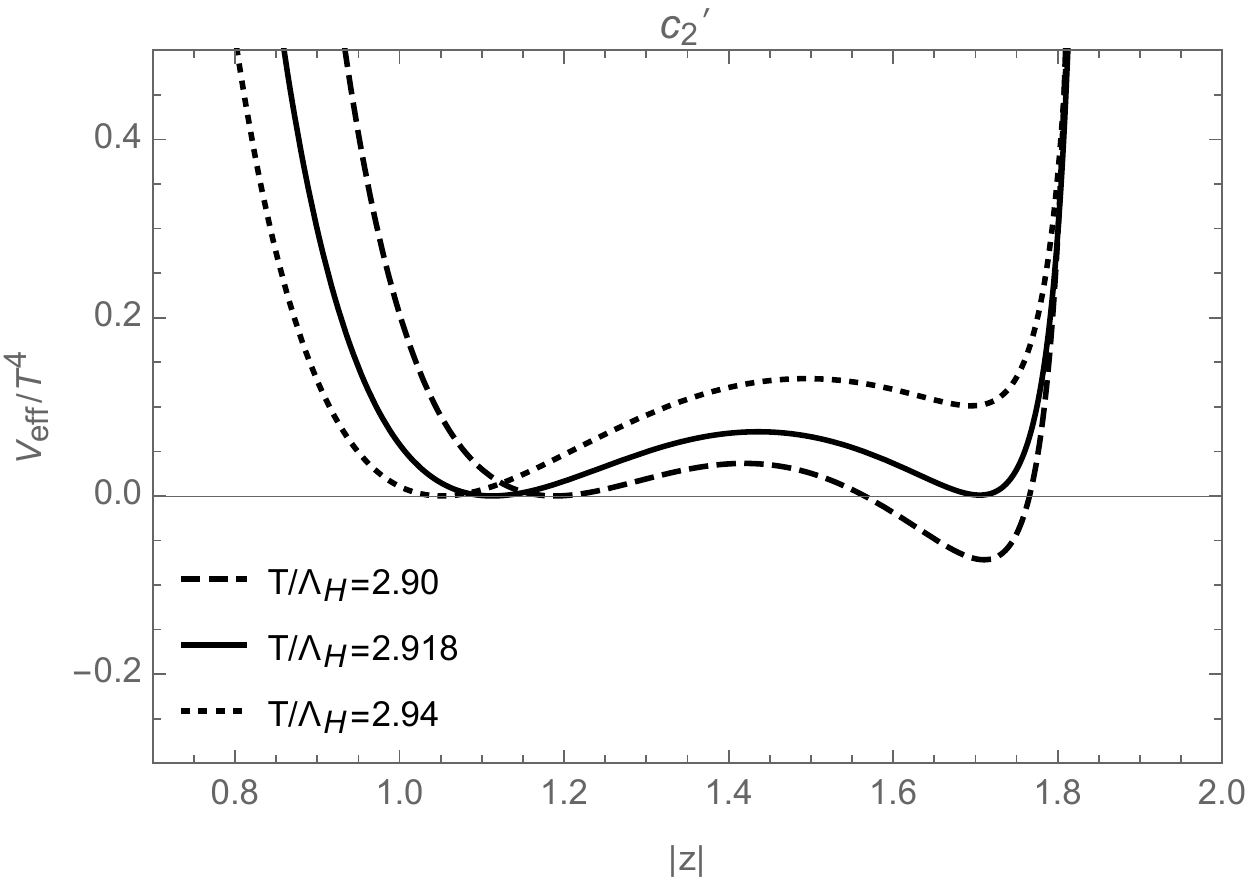}
\caption{ 
The potential $V_\text{eff}(L,\chi,T)/T^4$ around $T_{c,6}/\Lambda_H=2.918$ as a function of $\chi$ (left), $\ell$ (center) and $|z|$ (right) on the two different lines $c_1'$ and $c_2'$ linking the two minimum points given in \eqref{min6-chiPol-1} and \eqref{min6-chiPol-2}.
}
\label{fig:V6 change T} 
\end{figure}

A representative example of the critical value of $T$ and $a$ is
\al{
T_{c,6}/\Lambda_H&=2.918,&
a_6/\Lambda_H&=6.2.&
\label{temperature and a parameter for n6}
}
At this critical point two minima of $V_\text{eff}\fn{L,\chi,T}/T^4$ appear at:
\al{
&\text{(i) broken phase}:&
&\chi/\Lambda_H=1.3649, &
&\ell=0.1107, &
&z=1.0009- 1.3806 i, &
&z^*=1.0009+1.3806 i,&
\label{min6-chiPol-1}
\\
&& &&  &(x_1=0.2126,& &x_2=-0.78828,& &x_3=0.90782)&\nn[10pt]
&\text{(ii) symmetric phase}: &
& \chi/\Lambda_H=0,&
&\ell=0.48691,&
&z=0.7528 - 0.81933i,&
&z^*=0.7528 + 0.81933i,&
\label{min6-chiPol-2}
\\
&& &&  & (x_1=0.62679,&  &x_2=-0.12601,& &x_3=0.95995).& \nonumber
}
We plot $V_\text{eff}\fn{L,\chi,T}/T^4$ at the critical point (given in \eqref{temperature and a parameter for n6}) as a function of  $\chi$ (left), $\ell$ (center) and $|z|$ in Fig.\,\ref{fig:VSandglue6}, where we vary $\chi$, $\ell$ and $|z|$ along the two different lines $c_1'$ (red) and $c_2'$ (black), which link the two minimum points:
\al{
&c_1':&
&\chi=1.3649 t,&
&x_1= - 0.41419 t+0.62679,&
&x_2= - 0.66227 t-0.12601,&
&x_3= - 0.052131 t+0.95995,&\\
&c_2':&
&\chi=1.3649 t,&
&x_1=  - 0.41419t^2 +0.62679,&
&x_2= - 0.66227 t-0.12601,&
&x_3= - 0.052131 t^3+0.95995.&
}
Fig.\,\ref{fig:V6 change T} exhibits the effective potential around the critical temperature.
From these figures we conclude that the scale phase transition for $N=6$ with the Polyakov loop effect included is a first-order phase transition. 

\subsubsection{Latent heat}
Let us here evaluate the latent heat which is defined by
\al{
\epsilon\fn{T}=-\Delta V_\text{eff}\fn{L,\chi,T}+T\frac{\p \Delta V_\text{eff}}{\p T},
\label{definition of latent heat}
}
where $\Delta V_\text{eff}$ is the difference between those of broken phase (B.P.) and symmetric phase (S.P), namely, $\Delta V_\text{eff}\fn{L,\chi,T}=V_\text{eff}\fn{L,\chi,T}|_\text{B.P.}-V_\text{eff}\fn{L,\chi,T}|_\text{S.P.}$.
This is a key quantity of gravitational waves produced by a first-order phase transition.\footnote{
Note that in statistical physics, the ``latent heat" is defined as a critical temperature times the difference of the entropies between two phases.
That is, it corresponds to the second term on the right-hand side of \eqref{definition of latent heat} at the critical temperature.
On the other hand, phase transitions in the early Universe tend to take place below their critical temperatures, i.e., the supercooling, due to the expansion of the Universe, which is called the ``cosmological phase transition".
Therefore, we here call \eqref{definition of latent heat} the latent heat although it is the Helmholtz free energy in statistical physics.
}
The larger the latent heat becomes, the stronger spectra of gravitational waves become.

We first evaluate the latent heat in the pure gluon case.
Here, to this end, we set $a=1$ and $b_{N}=T^3$ in the effective potential of the Polyakov loop \eqref{VglN}.
In Table\,\ref{critical temperature and latent heat}, we show the critical temperature and the latent heat normalized by $T_c^4$.
We see that larger number of the color yields the larger latent heat,
which can be understood from the analytic expression
\al{
\epsilon(T_c)/T_c^4
=\frac{b_{N}}{T_c^4} e^{-1/T_c} \ln \ell_c^2~,
}
where $\ell_c$ is the traced Polyakov loop
in  the de-confining phase.\footnote{
The latent heat for  the $SU(3)$  pure gluonic case
has been computed in lattice  gauge theory 
with the result   $\epsilon/T_c^4=0.75\pm 0.17$
\cite{Shirogane:2016zbf}.
If we use $b_{3}=(0.69 T)^3$ instead of $b_{3}=T^3$, we can reproduce the lattice result.}

\begin{table*}[htb]
 \caption{Critical temperature and latent heat in pure gluon case with $a=1$ and $b_{N}=T^3$.}
 \label{critical temperature and latent heat}
 \begin{tabular}{@{\hspace{0cm}}l@{\hspace{1cm}}c@{\hspace{1cm}}c} \hline \hline
    $N$ & $T_c$ & $\epsilon\fn{T_c}/T_c^4$ \\ \hline
      3 & 0.40736  &  2.2857   \\ 
      4 & 0.44692  &  5.7278   \\ 
      5 & 0.47015 &   10.313   \\
      6 & 0.48529 &   15.917  \\ \hline
\end{tabular}
\end{table*}

Next, we consider the system where the scalar field is coupled with the Polyakov loop.
Here, before evaluating the latent heat numerically, let us describe what we could observe when the Polyakov loop effects are taken into account.
In the case without the Polyakov loop, the vacuum contribution from the scalar field loop at finite temperature is
\al{
V_{\rm FT}^\text{wo}(0,T)&=-2N_fN\times \frac{\pi^2}{90}T^4.
\label{vacuum contribution without Polyakov loop}
}
Since this term does not depend on the field $\chi$ and is subtracted in $\Delta V_\text{eff}\fn{\chi,T}$, it does not contribute to the latent heat \eqref{definition of latent heat}.
In contrast, when we take the Polyakov loop effects into account, as one can see from \eqref{Finite temperature loop effect}, the traced Polyakov loop is coupled to the vacuum contribution \eqref{vacuum contribution without Polyakov loop}:
\al{
V_{\rm FT}^\text{w}(L,0,T)&= -\frac{4N_f T^4}{\pi^2}\sum_{j=1}^\infty \frac{1}{j^4}\times
\begin{cases}
\left( \cos\fn{j\theta_1}+\cdots +\cos\fn{j\theta_{N/2}} \right)\\[10pt]
\left( \cos\fn{j\theta_1}+\cdots +\cos\fn{j\theta_{(N-1)/2}}+1/2 \right)
\end{cases}
~~\text{for}~~
\begin{cases}
\text{even}~~N\\[10pt]
\text{odd }~~N
\end{cases},
\label{vacuum contribution with Polyakov loop}
}
where we used the fact that $K_2\fn{x}=2/x^2-1/2+\cdots$.
Note that if we set all angles to zero ($\theta_1=\theta_2=\cdots =0$), \eqref{vacuum contribution with Polyakov loop} produces \eqref{vacuum contribution without Polyakov loop} using $\sum_{j=1}^\infty \frac{1}{j^4}=\zeta\fn{4}=\pi^4/90$.
Then, we see the relation between the vacuum contributions with and without the Polyakov loop:
\al{
V_{\rm FT}^\text{w}(L,0,T)=V_{\rm FT}^\text{wo}(0,T)\times \frac{90}{\pi^4}\sum_{j=1}^\infty \frac{1}{j^4} \Upsilon\fn{j,\{\theta_i\}},
\label{relation between the vacuum contributions with and without the Polyakov loop}
}
where
\al{
\Upsilon\fn{j,\{\theta_i\}}=\frac{2}{N}\begin{cases}
\left( \cos\fn{j\theta_1}+\cdots +\cos\fn{j\theta_{N/2}} \right)\\[10pt]
\left( \cos\fn{j\theta_1}+\cdots +\cos\fn{j\theta_{(N-1)/2}}+1/2 \right)
\end{cases}
~~\text{for}~~
\begin{cases}
\text{even}~~N\\[10pt]
\text{odd }~~N
\end{cases}.
}
Note that $\Upsilon\fn{j=1,\{\theta_i\}}=\ell$.
Since $\Upsilon\fn{j,\{\theta_i\}}$ takes different values between broken and symmetric phases, the vacuum term \eqref{relation between the vacuum contributions with and without the Polyakov loop} contributes to the latent heat.

\begin{figure}
\includegraphics[width=9cm]{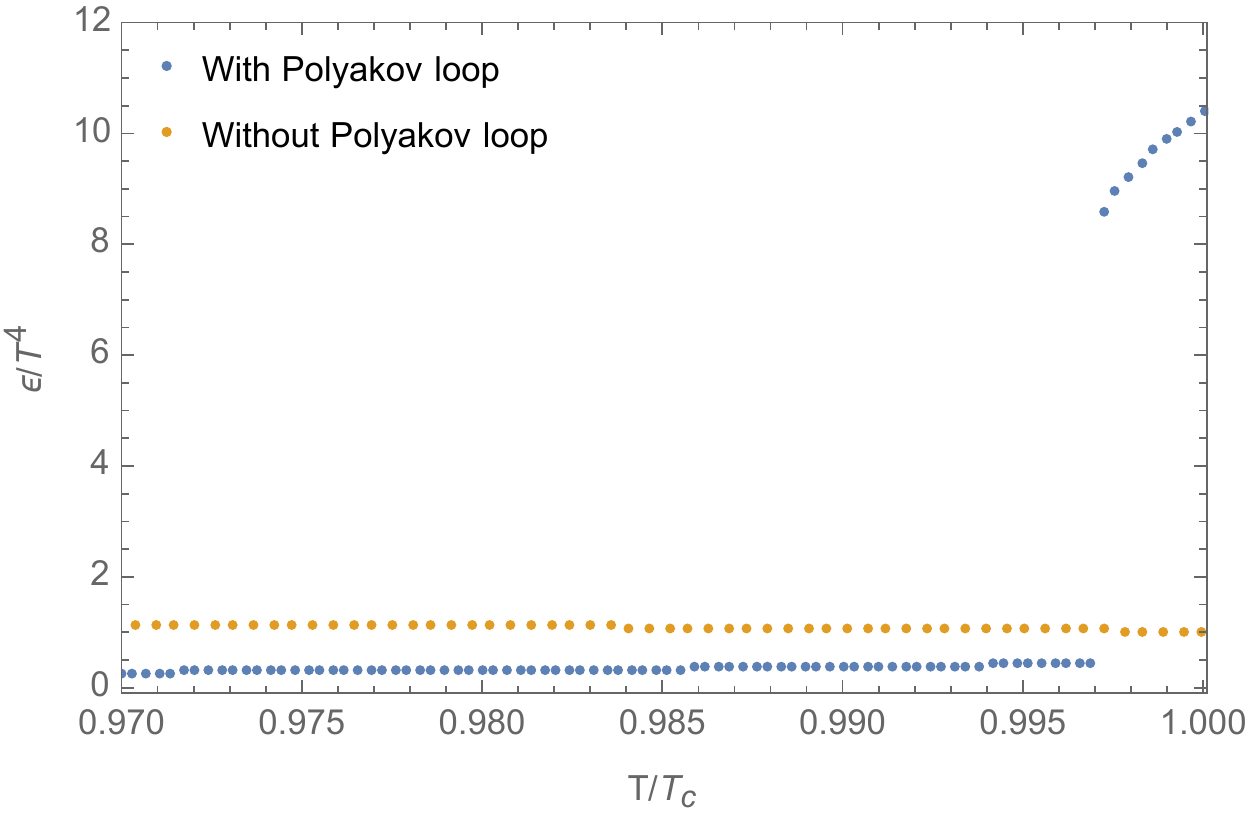}
\caption{
The latent heat \eqref{definition of latent heat} normalized by $T^4$ as a function of $T/T_c$.
}
\label{fig:latent heat} 
\end{figure}

We show the $N=6$ case with the parameters \eqref{parameter set} and \eqref{temperature and a parameter for n6}.
Fig.\,\ref{fig:latent heat} shows the latent heat normalized by $T^4$ slightly below the critical temperature.
At the critical temperature, we have vacua \eqref{min6-chiPol-1} and \eqref{min6-chiPol-2} at which the latent heat in the case with and without the Polyakov loop is
\al{
\frac{\epsilon\fn{T_c}}{T_c^4}=\begin{cases}
10.3733 & \text{with Polyakov loop}\\[10pt]
1.02791& \text{without Polyakov loop}
\end{cases}.
\label{total latent heat}
}
We see that the Polyakov loop effect increases the latent heat.
The vacuum contribution \eqref{relation between the vacuum contributions with and without the Polyakov loop} at the critical temperature becomes $\Delta \epsilon\fn{T_c^4}/T_c^4=2.4905$.
That is, this contribution accounts for about $24\%$ within the total contribution \eqref{total latent heat}.
In Fig.\,\ref{fig:latent heat}, we show the temperature-dependence of the latent heat slightly below the critical temperature.
We see that there is a jump at about $T/T_c=0.99724$.
Below this temperature, the scalar condensate takes different values between the true and false vacua, whereas the Polyakov loop does not.\footnote{
As one has seen below \eqref{minimum of pure Nc3 case}, in the $N=3$ case, the false vacuum appears at $a/T=2.48491$ which is very close to the critical temperature $a/T_c=2.45483$.
}
In other words, the false vacuum for the Polyakov loop in the effective potential appears above $T/T_c=0.99724$. 
Once the false vacuum of the Polyakov loop is generated, the latent heat could be large.

\section{Summary}\label{summary}
In this paper we have considered scalegenesis,
the spontaneous breaking of scale invariance, by the condensation
of the scalar bilinear in an $SU(N)$ gauge theory, where the scalar field is in the fundamental representation of 
$SU(N)$.
This non-perturbative effect has been studied by means of
an effective theory which we have developed in \cite{Kubo:2015cna,Kubo:2015joa,Kubo:2016kpb,Kubo:2017wbv}.
In the previous formulation of the effective theory  no confinement effect 
has been taken into account. Following the ansatz of \cite{Fukushima:2003fw}
we have included the Polyakov loop effect into the scale phase transition
at finite temperature, where we have assumed that the deconfinement
transition and the scale phase transition appear at the same critical temperature.
$N$ is not restricted to $3$ in phenomenological applications
of the scalar-bilinear condensation in a scale invariant extension
of the SM.  We therefore have studied the cases with
$N=3,4,5$ and $6$ and found that  in all these cases the phase transition is a first-order phase transition.
We could introduce a current scalar mass (which breaks the scale symmetry softly and investigate the change of the nature of the scale phase transition. As in QCD we expect the scale phase transition will become cross-over type above some current scalar mass. But this would go beyond the scope of our paper, and we would like to leave it to the next project.

Since the latent heat is an important quantity
to estimate  the strength of   the gravitational waves
background which is produced by a first-order phase transition in
the early Universe, we have
calculated it at and slightly below the 
critical temperature  and compared the results
with those obtained without the Polyakov loop effect. 
We have found that the Polyakov effect can indeed
increase the latent heat if the cosmological phase transition
occurs very close to the critical temperature of 
the first-order phase transition.  
This would mean a large  increase 
 in the energy density of  the gravitational waves background, if it  were produced by the scale phase transition.

\subsection*{Acknowledgements}
M.\,Y. thanks Jan.\,M.\,Pawlowski for valuable discussions.
The work of J.\,K. is partially supported by the Grant-in-Aid for Scientific Research (C) from the Japan Society for Promotion of Science (Grant No.16K05315).
The work of M.\,Y. is supported by the DFG Collaborative Research Centre SFB1225 (ISOQUANT).

\bibliography{refs2}
\end{document}